\documentclass{aa}  

\usepackage{natbib}
\bibpunct{(}{)}{;}{a}{}{,} % to follow the A&A style

\usepackage{graphicx}
\usepackage{txfonts}
\usepackage{amsmath}	% Advanced maths commands
\usepackage{amssymb}	% Extra maths symbols
\usepackage{arydshln} %for vertical dashed lines in tables
\usepackage{xspace}
\usepackage{textcomp}
\usepackage{hyperref}
\usepackage[normalem]{ulem}

\newcommand{\galagithub}{\href{https://github.com/MegaMorph/galapagos}{https://github.com/MegaMorph/galapagos}}
\newcommand{\mmweb}{\href{http://nottingham.ac.uk/astronomy/megamorph/}{http://nottingham.ac.uk/astronomy/megamorph/}}

\usepackage[usenames,dvipsnames]{xcolor}
\definecolor{mygray}{gray}{0.6}

\newcommand{\astromatic}{{\scshape AstrOmatic}\xspace}
\newcommand{\budda}{{\scshape Budda}\xspace}
\newcommand{\buddi}{{\scshape Buddi}\xspace}
\newcommand{\califa}{{\scshape Califa}\xspace}
\newcommand{\candels}{{\scshape Candels}\xspace}
\newcommand{\combo}{{\scshape Combo-17}\xspace}
\newcommand{\cosmos}{{\scshape Cosmos}\xspace}
\newcommand{\deVa}{de Vaucouleurs\xspace}
\newcommand{\euclid}{{\scshape Euclid}\xspace}
\newcommand{\galapagos}{{\scshape Galapagos}\xspace}
\newcommand{\galapagostwo}{{\scshape Galapagos-2}\xspace}
\newcommand{\galfit}{{\scshape Galfit}\xspace}
\newcommand{\galfitm}{{\scshape Galfitm}\xspace}
\newcommand{\gama}{{\scshape Gama}\xspace}
\newcommand{\gems}{{\scshape Gems}\xspace}
\newcommand{\gimtwod}{{\scshape Gim2d}\xspace}
\newcommand{\goods}{{\scshape Goods}\xspace}
\newcommand{\Haeussler}{{H\"au\ss ler}\xspace}
\newcommand{\HST}{{\scshape HST}\xspace}
\newcommand{\imfit}{{\scshape ImFit}\xspace}
\newcommand{\kmos}{{\scshape Kmos}\xspace}
\newcommand{\kids}{{\scshape KiDS}\xspace}
\newcommand{\lsst}{{\scshape LSST}\xspace}
\newcommand{\manga}{{\scshape MaNGA}\xspace}
\newcommand{\megamorph}{{\scshape MegaMorph}\xspace}
\newcommand{\omegacam}{{\scshape OmegaCAM}\xspace}
\newcommand{\pymorph}{{\scshape Pymorph}\xspace}
\newcommand{\profit}{{\scshape ProFit}\xspace}
\newcommand{\profound}{{\scshape ProFound}\xspace}
\newcommand{\psfex}{{\scshape PSFEx}\xspace}
\newcommand{\sami}{{\scshape Sami}\xspace}
\newcommand{\sdss}{{\scshape Sdss}\xspace}
\newcommand{\sersic}{S\'ersic\xspace}
\newcommand{\sex}{{\scshape SExtractor}\xspace}
\newcommand{\sigmakelvin}{{\scshape Sigma}\xspace}
\newcommand{\swarp}{{\scshape Swarp}}
\newcommand{\twoDPhot}{{\scshape 2DPhot}\xspace}
\newcommand{\twomass}{{\scshape 2Mass}\xspace}
\newcommand{\ukidss}{{\scshape Ukidss}\xspace}
\newcommand{\video}{{\scshape Video}\xspace}
\newcommand{\vircam}{{\scshape VIRCAM}\xspace}
\newcommand{\viking}{{\scshape VIKING}\xspace}
\newcommand{\wfcam}{{\scshape WFCAM}\xspace}
\newcommand{\wfirst}{{\scshape WFIRST}\xspace}
\newcommand{\sn}{S/N\xspace}

\begin{document} 

    \title{\galapagostwo/\galfitm/\gama\ -- multi-wavelength measurement of galaxy structure: separating the properties of spheroid and disk components in modern surveys}
%   \subtitle{I. Overviewing the $\kappa$-mechanism}
% authors
\author{Boris~\Haeussler\inst{1},
        Marina~Vika,
        Steven~P.~Bamford\inst{2},
        Evelyn J. Johnston\inst{3},
        Sarah Brough\inst{4},
        Sarah Casura\inst{5},
        Benne W. Holwerda\inst{6},
        Lee S. Kelvin\inst{7},
        Cristina Popescu\inst{8,9}
        }
%%%% affiliations 
   \institute{European Southern Observatory, Alonso de Cordova 3107, Casilla 19001, Santiago, Chile\\
        \email{Boris.Haeussler@eso.org}
    \and School of Physics and Astronomy, University of Nottingham, University Park, Nottingham, NG7 2RD, UK\\
    \and N\'ucleo de Astronom\'ia de la Facultad de Ingenier\'ia y Ciencias, Universidad Diego Portales, Av. Ej\'ercito Libertador 441, Santiago, Chile\\
    \and School of Physics, University of New South Wales, NSW 2052, Australia\\
    \and Hamburger Sternwarte, Universit\"at Hamburg, Gojenbergsweg 112, 21029 Hamburg, Germany\\
    \and Department of Physics and Astronomy, 102 Natural Science Building, University of Louisville, Louisville KY 40292, USA\\
    \and Department of Astrophysical Sciences, Princeton University, 4 Ivy Lane, Princeton, NJ 08544, USA\\
    \and Jeremiah Horrocks Institute, University of Central Lancashire, Preston, PR1 2HE, UK\\
    \and The Astronomical Institute of the Romanian Academy, Str. Cutitul de Argint 5, Bucharest, Romania\\
    }

   \date{Received December 2021}

% \abstract{}{}{}{}{} 
% 5 {} token are mandatory
 
  \abstract
  % context heading (optional)
  % {} leave it empty if necessary  
   {}
  % aims heading (mandatory)
   {To present the capabilities of \galapagostwo and \galfitm in the context of fitting 2-component profiles -- Bulge/Disk decompositions -- to galaxies, on the way to providing complete multi-band, multi-component fitting of large samples of galaxies in future surveys. We also release both the code and the fit results to 234,239 objects from the DR3 of the \gama survey, a sample significantly deeper than previous works.
}
  % methods heading (mandatory)
   {We use stringent tests on both simulated and real data, as well as comparison to public catalogues to evaluate the advantages of using multi-band over single-band data.}
  % results heading (mandatory)
   {We show that multi-band fitting using \galfitm provides significant advantages when trying to decompose galaxies into their individual constituents, as more data are being used, by effectively being able to use the colour information buried in the individual exposures to its advantage.\\
   Using simulated data, we find that multi-band fitting significantly reduces the deviations from the real parameter values, allows component sizes and \sersic indices to be recovered more accurately, and -- by design -- constrains the band-to-band variations of these parameters to more physical values.\\
   On both simulated and real data, we confirm that the spectral energy distributions (SEDs) of the 2 main components can be recovered to fainter magnitudes compared to using single-band fitting, which tends to recover \lq disks\rq\ and \lq bulges\rq\ to -- on average -- have identical SEDs when the galaxies become too faint, instead of the different SEDs they truly have.\\
   By comparing our results to those provided by other fitting codes, we confirm that they  agree in general, but measurement errors can be significantly reduced by using the multi-band tools developed by the \megamorph project.}
  % conclusions heading (optional), leave it empty if necessary 
   {We conclude that the multi-band fitting employed by \galapagostwo and \galfitm significantly improves the accuracy of structural galaxy parameters and enables much larger samples to be be used in a scientific analysis.}

    \titlerunning{Galapagos-2/GalfitM/GAMA -- multi-band Bulge/Disk decomposition in modern surveys}
    \authorrunning{H\"au\ss ler, et al.}
   \maketitle
%
%________________________________________________________________

\section{Introduction}
\label{sec_intro}
All information that we can gather from galaxies in order to constrain models of galaxy formation and evolution is encoded in the light that they emit.
From this single source we need to infer the physical processes which form these objects, and distinguish different evolutionary mechanisms, e.g. different quenching mechanisms, continuing star-formation, merger history, etc.
Squeezing as much information as possible from this limited resource is hence vital in order to understand how galaxies form and evolve.

Modern surveys and analysis have come a long way, partly through technical/hardware advances of instruments and telescopes, partly through advances in the analysis and software packages used.
One such advance in instrumentation is the gathering of Integral Field Units (IFU) data, which provides spectral information for each part of the sky/galaxy, which can be explored in detail in a sophisticated manner.
However, taking these IFU data is very expensive as it requires vast amounts of telescope time and delivers data only for individual of small samples of targeted galaxies or -- worse -- only the centres of those galaxies due to the limited field of view (FoV) of these instruments at present time.
Only recently, IFUs with a larger FoV have become available \citep[e.g. MUSE,][$\rm 1 arcmin^2$]{MUSE}, but even using those, it is -- for now -- impossible to gather statistically useful samples of galaxies in a cosmological volume.
The largest current IFU surveys have observed \textasciitilde600, \textasciitilde3000 and \textasciitilde210000 (\citealp[\califa,][]{califa}; \citealp[\sami,][]{sami21}; \citealp[ and \manga,][]{manga}), respectively.
Additionally, these surveys are limited to low redshifts of $\rm z\sim0.03$ (\califa) and $\rm z\sim0.1-0.15$ (\manga and \sami) and still lack a comparison sample at higher redshifts, vital for formation/evolution studies that span a significant fraction of the Hubble time.
This redshift issue might be somewhat solved using different wavelengths in different instruments, e.g. \kmos \citep{KMOS}, but this does not solve the issues with sample size and observing time required for statistically meaningful studies.

On the other hand, imaging of large survey areas can be -- and routinely is -- done relatively cheaply, e.g. by \sdss \citep{SDSS},\twomass \citep{2MASS}, \cosmos \citep{Cosmos}, \candels \citep{Grogin,Koekemoer}, \goods \citep{GOODS1,GOODS2}, \gems \citep{GEMS}, \video \citep{JarvisVideo}, \combo \citep{Wolf2003}, and countless others.
This cheap access to vast amounts of data and large galaxy samples, however, comes at the price of significantly poorer spectral/wavelength resolution.
For example the \cosmos field is now covered in \textasciitilde30 different filters with a sky coverage of \textasciitilde2 $\rm deg^2$, Alhambra \citep{Alhambra} observed \textasciitilde4 $\rm deg^2$ in 20 filters and JPAS \citep{JPAS} is planning to observe \textasciitilde8000 $\rm deg^2$ in 56 medium-band (and non-overlapping) filters, effectively creating high-quality spectral energy distributions (SEDs) at each image pixel.
However, while these datasets do deliver a huge amount of information, only part of it is generally exploited in present-day analyses, e.g. via aperture photometry.
Even these simple measurements do contain valuable information (e.g. one can turn the measured colours of a galaxy into estimates of global stellar masses, star formation rates, etc), which allow the development of a general picture of the merger and star-formation history of a given galaxy. 
Typically, however, a variety of star-formation histories can produce similar integrated properties, making this simple approach sensitive to measurement uncertainties, which are increased when measurements at different wavelengths are carried out independently.
This is further complicated by the fact that many of the tools used rely on aperture photometry, which -- while summing up all the light from an object -- entirely ignore the light distribution within the aperture radius. 
Different galaxy shapes, or internal distribution of light -- e.g. in different components -- are lost in such a simple measurement.

A different way to measure galaxy parameters is to measure the distribution of light using non-parametric codes, e.g. via Gini-M20 \citep{GiniM20} or CAS \citep{Conselice2003}.
To date, these techniques -- again -- only work independently in each band, which makes it difficult or impossible to make consistent measurements on images at different wavelengths.
Additionally, these and other non-parametric techniques generally do not take point spread function (PSF) effects into account and can hence bias the results as different fractions of the true light of a galaxy are measured at each wavelength, due to the different sizes and shapes seen in the PSFs at different wavelengths.

One widely used technique that makes it possible to take the image PSF into account is parametric light profile fitting.
Several codes for this work exist -- e.g. \budda \citep{Budda}, \gimtwod \citep{GIM2D}, \galfit \citep{Galfit,Galfit3}, \imfit \citep{imfit}, \profit \citep{PROFit}, \twoDPhot \citep{LaBarbera2008}, Profiler \citep{Profiler} -- and are more or less routinely used to measure galaxy parameters both on nearby galaxies and in an automated fashion on large samples of galaxies in large-scale galaxy surveys.
All these codes, however, suffer from the same problem, in that they -- once more -- only allow profile fitting in a single observed band \footnote{\profit allows multi-band fitting as of January 2021. \gimtwod allows 2-band fits in a limited manner in which structural parameters are not permitted to vary between the bands.}.

To improve on this situation, several authors have recently developed techniques in which a light profile is fit in one band and then applied to the other bands to derive magnitudes \citep{Bruce2014a,Bruce2014b,Head2014,Simard2011,Mendel2014}.
While this does paint a somewhat more consistent picture, it still loses information present in the imaging data itself, as it is not clear whether the profiles are actually the same at all wavelengths as has to be assumed for these analyses. 
For example, it is already known that a typical disk galaxy today generally contains an extended, blue disk and a compact, red bulge.
The same is also often found in S0 galaxies \citep[e.g.]{Bothun_1990,Peletier_1996,Head2014,Hudson_2010}, but exceptions to this general rule are known for individual S0 galaxies \citep{Johnston21} and dwarf galaxies.
A single profile fit to such a galaxy at different wavelengths would hence tend to fit something larger with low \sersic index \citep{Sersic} if one of the bluer bands is used in the fit, and something smaller and with a higher \sersic index if a one of the redder bands is used, making a consistent measurement hard to agree on between different science cases (e.g. at low and high redshift).
In fact, several authors \citep[e.g.][]{Kelvin2012,MMVulcani} have reported such variations of galaxy parameters with wavelength, which, using this approach, would stay undetected by design.
At least some of these studies have indeed attributed much of the change observed to the different mixing of bulges and disks at different wavelengths.
There is, however, a niche market for these kinds of analyses where the assumption of \lq no wavelength dependence\rq\ is valid, e.g. in first order for Bulge/Disk (hereafter: $B/D$) decomposition of high-signal-to-noise (S/N) galaxy images, where stellar population gradients within the individual galaxy components can be ignored.

For a full multi-wavelength analysis \citep[or what was called \textit{oligochromatic} by][]{deGeyert13}, it would however be preferred if all images were taken into account simultaneously in an equal manner, i.e. giving the same weight to each image while leaving any wavelength dependence possible to be accommodated for.
In the \megamorph\ project (Measuring Galaxy Morphology), we have developed and tested such a technique \citep[presented in][]{MMHaeussler1,MMVika1,MMVika2}, which allows the combination of some of the above techniques, while avoiding some of the obvious shortcomings. 
The software developed by the \megamorph\ team -- \galapagostwo and \galfitm\ -- are based on \galapagos and \galfit, two publicly available and well tested single-band codes\footnote{There is also single-band \galapagos version translated to C and optimized for speed \citep{galapagosc}}.
\galfitm, however, allows the simultaneous fitting of multiple images at different wavelengths, and \galapagostwo allows the automatic exploitation of this code on large scale surveys.
Not only does this enable more accurate measurements on even simple, single-component profiles -- e.g. when fitted values do not actually change with wavelength -- as it uses more data and effectively increases S/N, but it also allows structural parameters of galaxies to vary smoothly with wavelength, making additional measurements possible in a physically meaningful way.
This has been shown and exploited in
\citet{MMHaeussler1,MMVulcani,Kennedy15,Kennedy16a,Kennedy16b,Huertas,Dimauro,Mosenkov,Psychogyios,Nedkova20}.

While these single-component fits provide a great deal of useful information -- and are less challenging to perform -- there is further advantage of this multi-band approach when fitting multiple profiles to a galaxy in order to decompose the galaxy into its constituents.
The main (first-order) reason why galaxies look different at different wavelengths is that different galaxy components are mixed at different strengths at different wavelength.
Observing in blue bands generally observes much more of the disk light, while observing in red bands generally observes more light from the bulge component, explaining at least some of the trends found by \citet{MMVulcani,Kennedy15,Kennedy16a,Kennedy16b} and others.
Valuable information needed for a successful \textit{separation} of these components is hence stored in the colour information of each observed pixel, making the colour information vital in the decomposition process itself.
Although conceptually simple, $B/D$ decomposition remains a challenging task, due to the variety of structures that galaxies display, not to mention the usual observational limitations of resolution and S/N.
Any improvement and increase in parameter reliability is a big step forward to successfully separating bulges from disks so the individual galaxy components can be studied independently.

This successful decomposition of a galaxy light-profile -- or better: mass-profile -- into its individual constituents is the Holy Grail of Galaxy Profile fitting at least at redshifts outside the local universe, and is a vital step towards understanding galaxy formation and evolution.
However, this task is incredibly tricky to achieve, especially in an automated manner on a large sample of not-very-well-resolved galaxies spanning a wide range of redshifts, as are currently routinely imaged by large-scale surveys.
Previous approaches to this problem mostly worked on bright, well-resolved galaxies, where the profile shapes themselves contain enough information for a successful separation of the individual components even on single-band data.
Some papers have presented large catalogues of galaxies which were successfully separated into bulge and disk components \citep[e.g.][]{Simard2011,Huertas,Meert13}, allowing for further analysis, e.g. of their masses, stellar populations, etc \citep{Mendel2014,Dimauro}.
However, the sample used by \citet[][hereafter S11]{Simard2011} and \citet{Mendel2014}, is limited to $m_{\rm petro,r,corr} \le 17.77$ in \sdss data, \citet{Meert13} and \citet{Meert15} have presented a similar analysis on a sample using the same magnitude limit.

In this work, we present $B/D$ decompositions of galaxies which are nearly 2 magnitudes fainter than the ones discussed in S11 and \citet{Meert13} in order to show how well a multi-wavelength approach can separate different galaxy components and to quantify the advantages provided by multi-band fitting in the decomposition of distant galaxies in typical present-day surveys.
In agreement with the approach taken by \citet{MMHaeussler1} -- we use both simulated data to measure deviations from \lq true\rq\ values directly, and real data to carry out several consistency checks. 
We complement and expand on the work of \citet{MMVika2} by demonstrating the application of our technique to large surveys in an automated fashion, and with greater statistical power. 
We target a sample of real galaxies with spectroscopic information in \gama \citep{GAMA,Liske15,GAMADR3} with a limiting magnitude of $m_{\rm r} \le 19.8mag$,  using \swarp-ed \citep{terapix} data from \sdss and \ukidss LAS archival data \citep{ukidss}. We thus demonstrate how (and why) using multi-band fitting has advantages in terms of stability, improved accuracy and increased sample sizes, especially for the low \sn bands of a survey.

This paper is part of a series that investigates the benefits of this multi-wavelength approach to measuring galaxy structural properties. 
In \citet[hereafter H13]{MMHaeussler1} we presented an overview of \galfitm, with further details available on the project website\footnote{\galfitm can be obtained at \newline
\href{https://www.nottingham.ac.uk/astronomy/megamorph/}{https://www.nottingham.ac.uk/astronomy/megamorph/}, where some help in how to use the software is also provided.}. 
We demonstrated our approach by performing single-component fits on a large dataset from the \gama survey, automating both the preparation of the data and the fitting process itself. 
The resulting measurements -- in particular the variation of structural parameters with wavelength -- are studied further in \citet[][hereafter V14a]{MMVulcani} and \citet{Kennedy16b}. 
In \citet[hereafter V13]{MMVika1} we tested this new method by fitting single-\sersic models \citep[e.g.][]{Sersic,GrahamDriver} to original and artificially-redshifted images of 163 nearby galaxies.
In \citet{Kennedy15}, we used the results already discussed in V14a to analyse the colour gradients seen in galaxies vs. their colour, structure, and luminosity.

\citet[hereafter V14b]{MMVika2} showed that multi-band analysis on nearby galaxies allows a more accurate and more stable separation of 2 components, using the same galaxy sample as used in V13.
\citet{BUDDI} expanded the use of \galfitm\ to IFU data to separate cleanly  spectra of different components in individual galaxies, via the software \buddi.
A more complete approach to fit \textasciitilde1800 objects of the \manga\ survey is currently underway (Johnston et al., in prep).
This work will be based on the \sdss DR16 data \citep{SDSS-DR16} which contains \textasciitilde5000 galaxies and will use a sample cut based both on the fibre-bundle size used and pre-existing fits on \sdss\ data using \pymorph\ \citep{pymorphvikram} by \citet{pymorphfischer}.

The objective of the present paper is to present the ability of \galfitm to perform $B/D$ decomposition on galaxy images with a wide range of resolution and signal-to-noise, in a similar fashion as used by H13 for single-component fits.
\citet{Kennedy16a} already used the technique, code, catalogue and data sample discussed in this work to explain the wavelength dependencies seen by V14a to be mostly effects of mixing light from different galaxy components.
As such, the tests and results presented here are directly transferable to that paper.
Other works have already exploited \galapagostwo on other datasets, e.g. to examine Subaru SuprimeCam \citep{Kuchner} or \HST\ \candels data \citep{Dimauro,Nedkova20}.
Their results and code behaviours tie in well with the findings in this paper.

This paper is structured in the following way: \S~\ref{sec_software} explains necessary changes to \galapagos in comparison to H13, which are essential for the performance in this work.
This section also explains the setup of both codes -- \galapagostwo and \galfitm -- used throughout this paper (\S~\ref{sec_setup}), a description of the starting values used (\S~\ref{sec_setup_starts}) and constraints used during the fits (\S~\ref{sec_setup_constraints}). 
For further details about the technique itself, we refer readers to H13. \S~\ref{sec_sample} discusses the galaxy sample selection for the remaining part of the paper.
\S~\ref{sec_sims} shows tests from applying this software to simulated data, i.e. galaxies whose true intrinsic values are known. 
This comparison, while not containing any physical meaning about galaxy populations, allows us to show the improvement obtained by using multi-band fitting in more detail.
In \S \ref{sec_bad_habits} we will specifically highlight the dangers of using \sersic indices as a proxy for bulge-to-total (hereafter $B/T$ ratio).
In \S~\ref{sec_real} we show similar tests from applying this software to real \gama data and comparing their values to show in how much multi-band fitting improves the fitting results both on individual galaxies and on the galaxy population as a whole.
We will also show a comparison to other works in \S \ref{sec_comparison}, including S11, \citet{PROFit} and \citet{Casura}.
After briefly discussing the effect of dust in real galaxies on the fitting parameters in \S \ref{sec_dust}, we will discuss our results and Conclusions in \S \ref{sec_summary}.

In the Appendices, we present the technical details of how the simulated images were created in Appendix \ref{sec_app_sim_sims}.
Appendix \ref{sec_start_params} presents a brief analysis of the impact of varying starting parameters in \galapagostwo\ and \galfitm. 
Appendix \ref{App_gala_features} will discuss further improvements of the code in the current \galapagostwo\ version, compared to the version used in this work and discuss possible further improvements.
We will conclude with a description of the \gama catalogues released with this paper in Appendix \ref{sec_cat_release}, based on newer and deeper imaging data, which uses re-{\scshape Swarp}ed data from the \kids \citep[Kilo-Degree Survey,][]{KiDS1,KiDS2} and \viking \citep[VISTA Kilo-Degree Infrared Galaxy Survey,][]{Viking} surveys.

We should note at this stage that fitting additional components beyond a simple Bulge/Disk model might be advisable depending on the science case, e.g. see \citep{Kruk17,Kruk18,Davis19,Sahu19}.
Specifically, as the spatial resolution of images improves, extra components often need to be added to account for the detailed profile of a galaxy.
However, it should also be noted that overfitting of galaxy profiles is often possible if the spatial resolution of the data does not allow it, and as bad as under-fitting the data.
Care must hence be taken when selecting the appropriate number of components fit to a dataset.
As automating fitting of more and more components and subsequently selecting the \lq best-fit\rq\ model becomes more and more challenging, and in order to do one step at a time, we restrict ourselves to the analysis of $B/D$ models in this paper.
In Appendix \ref{App_gala_features}, we will show how this paper and the software used can be used as a setup for more in-depth analysis of additional components in selected objects.

%__________________________________________________________________
%%% SECTION CHANGES TO SOFTWARE %%%
\section{Changes to the previous software and setup}
\label{sec_software}
\galfitm-v1.4.4 \footnote{\galfitm is available for users online at \mmweb} has been used in this work, to incorporate some of the improvements of the code in terms of delivering robust measurements in special cases in comparison to H13.
Most improvements concern either \galfitm features not used in this analysis (e.g. implementation of non-parametric components), or bug fixes not affecting the results in this analysis due to the way the code is used by \galapagostwo and/or the data is set up for analysis, so improvements can be considered minor.\footnote{As an example it was found that \galfitm can misbehave in a bad way when the input images have different sizes or have a slight offset to each other. 
As by design of \galapagostwo all input images to \galfitm are created at the same size and with no offset between them, this does not impact the performance tested in this analysis.}
This means, however, that the results in this work are not \textit{directly} comparable to the work presented in H13, where \galfitm-0.1.2.1 has been used.
As that work presented single-\sersic fits only, no such comparison is carried out here.

Equally, we used a more recent version of \galapagos-v2.2.7 throughout this work.
This differs from the version used in \S7 of H13 (v2.0.3) mostly by being able to run $B/D$ fits and a few changes to optimise usages of CPU and disk space.
\galapagostwo has since been developed further, several required and convenient additional features -- e.g. as desired when running \galapagos on other datasets -- have been introduced in newer versions of the code.
However, none of the changes should have any impact on the parameters of the fits themselves.
optimise
For example, this more recent version of \galapagostwo can deal with surveys with different footprints in each band.
For the data used in this analysis -- both simulated and real -- all images in all bands show complete sky coverage in the areas used, so such a feature does not have any impact on the fitting results discussed in this work.
A more complete summary of these new features is given in Appendix \ref{App_gala_features}. 
The newest version of \galapagostwo can be derived as both a full github repository or a zipped package\footnote{\galagithub}, including example setup files and an extensive help file that includes practical help on how to properly set up both code and data.
The code is distributed under MIT license and can be used and edited by users.
We welcome any changes back into the repository if they provide new features, and if they are well tested.

The main purpose of this section is to present and discuss the setups for \galapagostwo and \galfitm used in this work.
Equivalently to H13, we use PSFs from the \gama survey as created by \sigmakelvin and used in \citet{Kelvin2012}.
For the simulated images, a mean of these PSFs is used for both images simulation and \galapagos fitting in each band, effectively removing any PSF uncertainty effects from this part of the analysis.
Images, object detection and analysis were identical to those used in H13, with the addition of a $B/D$ decomposition being carried out on all objects possible.
In fact, for practical reasons, the $B/D$ fits presented in this work are entirely based on the single-\sersic fits analysed in H13, including the postage stamps for each galaxy, the decision on masking or deblending, etc.
We hence assume the reader to be familiar with this paper.

\subsection{General code behaviour}
\label{sec_setup}

While most of the code setup and behaviour is analogous to the single-\sersic setup, some additional issues have to be considered when carrying out $B/D$ decompositions. 
Most importantly, we had to decide on the order/degrees of freedom (DOF) of the polynomial used in the $B/D$ decomposition and the starting values for the fit itself.
In this section, we discuss the general setup of the code and its logical flow in \S~\ref{sec_setup_code}, and will discuss possible improvements to the code in \S~\ref{sec_setup_improvements} in this context. 
In \S~\ref{sec_setup_starts}, we discuss how the starting values for the $B/D$ fits are derived in the code, while \S~\ref{sec_setup_constraints} will explain the parameter constraints used.
As \galfitm and \galapagostwo are closely linked in this work, separating them is difficult and we will present the general behaviour of \galfitm, and, where relevant, the specific implication for \galapagostwo.
In \S~\ref{sec_setup_choices}, finally, we explain the choices made for the analysis in this work in particular, e.g. the degrees of the polynomials chosen throughout this paper.
The main values from these sections are summarised in Table \ref{table_setup}.

\subsubsection{Code setup/features - logical flow}
\label{sec_setup_code}
Both codes, \galfitm and \galapagostwo, allow great flexibility with respect to the parameters used during the fit, most importantly, the degree of freedom allowed.
A user can trivially switch on wavelength variations in the setup files and define the degree of the polynomial used in these variations.
They can equally trivially hold the \sersic index fixed at $n==1$ for galaxy disks and $n==4$ for bulges, if they wished, see Appendix \ref{App_gala_features}.

The $B/D$  decomposition of a galaxy in \galapagostwo is set up by directly using the single-\sersic fit result.
\galapagostwo creates a new setup file for \galfitm, starting with the information available from the single-\sersic fit and the parameters specified in the \galapagostwo setup.
In this second fit, most of the settings are kept the same:

\begin{itemize}
\item The same postage stamps are used for all galaxies as well as the same masks.
\item The same PSF files are defined.
\item The wavelengths to be given to \galfitm\footnote{As a reminder, we used 3543\AA, 4770\AA, 6231\AA, 7625\AA, 9134\AA, 10305\AA, 12483\AA, 16313\AA, 22010\AA\ for $ugrizYJHK$-band, respectively.} are identical.
\item The sky values are directly set to the same values used of single-\sersic fits, they are not re-measured.
\item The same decisions regarding fitting (secondaries), masking (tertiaries) and deblending neighbouring galaxies are used as decided for the single-\sersic fits. Single-\sersic\ and 2-component fits are hence directly comparable regarding all effects of neighbouring objects.
\end{itemize}

However, there is one important difference regarding the last point.
Instead of fitting neighbouring galaxies (secondaries and tertiaries, see Barden et al., 2012\nocite{galapagos}) again, they are modelled with fixed values, i.e. are merely subtracted from the image as their parameters have already been optimized in the single-\sersic fit on the same primary target. 
It is a small, but important, point to note that the values used are not the \textit{true} best-fit values from when said neighbouring galaxy was the primary target, but from the associated single-\sersic fits of the \textbf{same} primary target.
The profile of the primary object is \lq\ simply\rq\ replaced by 2 profiles, a \lq bulge\rq\ and a \lq disk\rq\ (both characterised by \sersic profiles).
Starting values for the primary object are derived from the single-\sersic best fit in a way discussed in \S \ref{sec_setup_starts}.
As \galapagostwo automatically runs the single-\sersic fits as a first step, a user does not have to take care of such a \lq pre-analysis\rq.

This scheme has the big advantage that the single-\sersic fit and the $B/D$  fits are directly comparable, all values other than the ones of the primary object are identical, which makes any further analysis and a possible selection of the \lq better\rq\ fit easier. 
There are several additional practical advantages of this setup scheme that starts from the single-\sersic fits.

Firstly, due to its independence of any additional parameters e.g. from neighbouring objects, all $B/D$  fits can be carried out independently of each other.
No complicated queuing mechanism is needed to avoid simultaneous fitting of galaxies that might influence each other as was the case when fitting single-\sersic profiles.
As a result, fits can simply be carried out one by one in any order, minimising computing time by \galapagostwo itself.
\galapagostwo strictly goes from brightest to faintest objects in this step.

Secondly, and with much higher impact in crowded fields: as parameters of neighbouring galaxies are fixed, the \galfitm fit is carried out with fewer free parameters, reducing the time needed to carry out the fit itself (although the primary source itself of course shows a higher number of free parameters).
In the dataset used in this analysis this advantage is, however, minimal, due to the relatively low density of objects.
Galaxies typically have no or only one neighbour in the fit (mean value: 1.37 neighbouring objects per fit, median: 1), but this is expected to make a bigger difference in densely populated areas, reducing the fitting time by a factor of a few compared to the single-\sersic fits.
However, $B/D$ fits introduce one additional profile to each primary object, which slows the fit down somewhat.
In the dataset used here, with a median of 1 neighbouring galaxy, the DOF of a $B/D$ fit is comparable to that in the single-\sersic fits, leading to a similar total execution time.
The exact values, however, depend on the DOF allowed for bulge and disk parameters.

Additionally, there is a second reason why the overall speed of the $B/D$  fits can be faster than the single-\sersic fits as a whole.
In both steps, it is possible to submit a target list to \galapagostwo that includes objects that become 'primary objects'.
In the single-\sersic fits, it is advisable to not only target the galaxies of interest themselves, but also objects within a certain radius around them, but with higher luminosity.
The reason for including these additional objects is that those objects would, in a full \galapagos run (without a target list), be treated first.
As such, if/once they become neighbouring objects for a target of interest, their true best fit values are already known and they can be optimally \textit{removed} from the single-\sersic fit.
Not fitting these neighbouring objects first has the potential to bias the results of the objects one is interested in.
In the $B/D$  decomposition fits, these objects are not of interest anymore, as their best fit single-\sersic values are already known.
As they are not targets of interests themselves, running a $B/D$  fit on them can be avoided by removing them from the target list.
Not fitting those galaxies in the $B/D$  decomposition step hence speeds up the overall fitting process of that step significantly.

\subsubsection{Possible improvements}
\label{sec_setup_improvements}
Several possible improvements are obvious to this fitting scheme.
As this scheme depends on the existence of a single-\sersic output file, it means that if such a file does not exist for some reason -- e.g. the fit might have crashed or the fit has not been attempted as the object was not in the target list -- the $B/D$  fit can not be carried out, \galapagostwo will simply continue with the next object.
However, we have shown in H13 that these cases of crashed single-\sersic fits are rare as most fits \textbf{do} produce a result.

There is also the danger of biasing the results against certain objects, i.e. if there is a particular kind of object for which said single-\sersic fit would often fail/crash.
Such a bias, however, is unknown to us, although we can not exclude that it exists to a small degree.
The alternative would be that these objects with failed single-\sersic fits could run a $B/D$ fit using different starting values, e.g. directly from running a second single-\sersic fit with different starting values or from using \sex \citep{Bertin} values for the setup directly.
Due to the small number of objects which fail in the single-\sersic fit and the risks and complications involved in such a scheme, we have decided not to implement $B/D$  fitting from \sex values directly.
It is hence down to the user to make sure that any target lists provided to \galapagostwo are compatible, i.e. the $B/D$ targets should be equal to or a subset of the single-\sersic targets, to ensure that a $B/D$  fit is possible.

However, \galapagostwo\ is friendly in nature.
Individual, important cases where the fit crashed can be dealt with by hand.
All input files exist, they are easy to identify in the output catalogue, and it is straight forward to manually ensure that a fit for an object does produce a \lq good\rq\ result.
Once this is done on a single-\sersic fit, the missing $B/D$ fit on the same objects can be \lq filled in\rq\ by running the $B/D$ part of \galapagostwo again.
Equally, individual $B/D$ fits can be fixed in a similar manual manner.
After fixing such crashed fits manually where deemed necessary, \galapagostwo\ can simply be re-run to read out the all fit results as usual.

A second possible improvement would be if all neighbouring galaxies were always taken into account with their \lq real\rq\ fitting values, i.e. from the fit in which they were the primary target themselves, as those are the optimal values for them.
This is already true for the neighbours that are brighter than the primary target, as this is how they are dealt with in the single-\sersic fit in the first place.
For the galaxies fainter than the primary target, the improvement of such a scheme would be minimal and would make the single-\sersic and the $B/D$  fits less comparable, so we did not implement such a system in \galapagostwo.
Users are reminded that neighbouring galaxies are simply fit to improve the fit of the primary target, their fit values are not written into the output catalogue.

Similarly, one could obviously fit the secondary objects as multi-component systems to improve their fits.
However, this is expected to have very minor effect on the fitting results of the primary source as the important part in the fitting of neighbours -- the successful subtraction of the profile wings -- can be achieved by a simple single-\sersic fit.
The possible improvements are therefore expected to be small, while introducing a large amount of book-keeping.
We have hence decided against using this more complex scheme.

\subsection{Starting values}
\label{sec_setup_starts}

\begin{table*}
\centering
\caption{Starting parameters, constraints and DoF used by \galapagostwo}
\begin{tabular}{@{}l|l|l|l@{}}
\hline
Parameter & Starting Value  & Constraints & Choice for DOF in this work   \\
\hline
\hline
 & & & \\
Position x & $median(x_{SS})$ & $median(x_{SS})\pm0.5*r_{e,PB,SS}$ & 1 (constant with $\lambda$)\\
 & & $x_D=x_B$ & \\
Position y & $median(y_{SS})$ & $median(x_{SS})\pm0.5*r_{e,PB,SS}$ & 1 (constant with $\lambda$)\\
 & & $y_D=y_B$ & \\
Disk Magnitude $m_{D}$ & $m_{D} = m_{SS}+0.75$ & $m_{D,start}\pm 5$  (user defined) & 9 (full freedom)\\
 & (flux equally split at each $\lambda$) & & \\
Disk halflight radius $r_{e,D}$ & $r_{e,D}=1.2*median(r_{e,SS})> 1px$ & r$_{e,D}\leq 400px$ (user defined) &  1 (constant with $\lambda$)\\
 & & r$_{\rm{e,D}} > 0.3px$ (hard coded) & \\
Disk \sersic index $n_D$ & $n==1$ & $0.2 \leq n \leq 8$ (user defined) & 0 ($n_D==1$)\\
 & or $n_D=median(n_{SS}) < 1.5$ & & \\
Disk Axis Ratio $q_D$ &  $q_{SS}$ & $0.0001\leq q_D \leq 1$ & 1 (constant with $\lambda$)\\
 & & & \\
Bulge Magnitude $m_B$ & $m_{B} = m_{SS}+0.75$ & $m_{B,start}\pm 5$ (user defined) & 9 (full freedom)\\
 & (flux equally split at each $\lambda$) & & \\
Bulge halflight radius $r_{e,B}$ & $r_{e,B}=0.3*median(r_{e,SS})> 1px$ & r$_{\rm{e,B}}\leq 400px$ (user defined) &  1 (constant with $\lambda$)\\
 & & r$_{\rm{e,B}} > 0.3px$ (hard coded) & \\
Bulge \sersic index $n_B$ & $n==4$& $0.2 \leq n \leq 8$  (user defined) & 0 (in case of $n_B==4$)\\
 & or $n_B=median(n_{SS}) > 1.5$  & & 1 (constant with $\lambda$)\\
Bulge Axis Ratio $q_B$ &  $q_{SS} > 0.6$ & $0.0001\leq q_B \leq 1$ & 1 (constant with $\lambda$)\\
Position Angle $PA$ & $PA_{SS}$ & $-180\deg<\theta<180\deg$ & 1 (constant with $\lambda$)\\
\label{table_setup}
\end{tabular}
\tablefoot{
This table summarises the starting parameters in \galapagostwo, the constraints and the choices of DOF used throughout this work, as discussed in detail in sections \S \ref{sec_setup_starts}, \ref{sec_setup_constraints} and \ref{sec_setup_choices}. Please see Appendix \ref{App_gala_features} for changes in newer versions of \galapagostwo. Index \lq PB\rq indicates that the primary band is used for this parameter, i.e. the $r$-band value in this study.}
\end{table*}

Deriving good starting values for all fitting parameters in \galfitm is both difficult and important.
This is not very critical in single-\sersic fits where we have already shown that even rough estimates, e.g. on galaxy magnitudes (via a typical galaxy SED) are sufficient for a successful fit, but it becomes important when carrying out $B/D$ decompositions, due to the degeneracies that can be found between different profile parameters \citep[e.g.][]{Lange16}.
The more accurate the starting values are, the more likely one would expect the fit to converge to a physical solution rather than settling in a local minimum. 
However, getting accurate starting values is difficult in a fully automated code, as it should take care of all eventualities and possible galaxy types.
\galapagostwo has to employ a system that works well in most cases.
Below, we explain the behaviour of the code in case of the different parameters.

Before we discuss the general approach of \galapagostwo to derive starting values for the $B/D$ fits, we need to mention a specific case of how \galfitm handles the starting values it is given, as it explains some of our choices in \galapagostwo.
The degrees of variation in the parameters for each image/wavelength are defined by the user as Chebyshev polynomials, and the input values can be provided as the parameter values for each band (hereafter \lq band\rq\ values) or as the underlying Chebyshev parameters.
While it is irrelevant which input version is chosen, we find \lq band\rq\ values more practical for interaction/readability purposes and \galapagostwo makes use of this notation.
\galfitm behaves as naively expected when a degree of freedom $\rm DOF=0$ is used for the fit in any parameter, in that the parameter values are simply fixed at the input values at each band, no matter their values or shape of the chosen polynomial.
However, depending on the way the input values are being chosen, this is not necessarily true when another DOF is being used. 
In case of $\rm DOF \geq 1$ and giving the parameter values as \lq band\rq\ values (rather than the Chebyshev parameters themselves), the parameter values given might not resemble the desired polynomial degree or shape (i.e. is is not possible to fit the, e.g., 9 points with a polynomials of a lower DOF).
When $\rm DOF \geq 2$, \galfitm first fits a polynomial to the input values internally and starts the fit from this polynomial, which delivers the desired behaviour.

The behaviour, however, is different for the specific case of $\rm DOF==1$.
Here, the fit is technically a fit that allows the parameter to vary, but the values should be the same at all wavelengths.
As such, it would be possible that \galfitm simply calculates an average of the input values and uses this as a starting value at each wavelength, which would resemble the behaviour at higher $DOF$.
However, we have chosen that in this case, while \galfitm is allowed to vary/increase \& decrease the parameter as a whole, the \textbf{offsets} between the values for each image (i.e. the shape of the polynomial as defined in the input parameters) are being kept frozen as specified in the input values.
In other words, when $\rm DOF==1$, the shape of the final polynomial will be the same as that defined in the input parameters, but with a systematic offset to higher or lower values as derived by \galfitm.
The reason for allowing this behaviour (in \galfitm, not \galapagostwo) is that it makes it possible to assign a certain SED to an object -- e.g. a supernova or active galactic nuclei (AGN) where the SED are often reasonably well known -- but fitting its overall brightness only. 
A second case where this behaviour could be useful is to assign positions according to a known offset between images (i.e. all images from one telescope are slightly offset to all images from another telescope), so a consistent profile can be fit to all images simultaneously, while leaving its central position variable during the fit.
Ideally, such a case should be taken care of by shifting mismatched images to the same pixel-grid as all other images, and \galapagostwo strictly requires this, but \galfitm technically allows this.

This decision for the behaviour of \galfitm in case of $\rm DOF==1$, however, means that if one does desire to fit a value to be the same at all wavelengths, one has to make sure that the input values are indeed identical for all bands.
\galapagostwo takes care of this issue by homogenising the input values in this case if the DOF of any parameter -- as given in the setup file -- makes this necessary (specifically $\rm DOF==1$). 

The following list gives a brief overview over both the behaviour of \galapagostwo in general and the consequences for this work in particular.
Table \ref{table_setup} summarises the starting values used by \galapagostwo, as well as the constraints used, see \S \ref{sec_setup_constraints} and the degrees of freedom used in this work, which we will discuss in \S \ref{sec_setup_choices}.
Values indexed with \lq B\rq\ are bulge values, values indicated with \lq D\rq\ resemble disk values, with \lq SS\rq\ values of the single-\sersic fit.

\begin{description}
\item[\textbf{Position ($x$, $y$):}]
\galapagostwo uses the output values of the single-\sersic fit as the starting values for the position of both components.
Both components are started at (and in fact constrained to, see below) the same values.
If offsets between images are allowed in the single-\sersic fit, these will be reflected in the starting values for the $B/D$  fit.
Only if $\rm DOF=1$, the starting values are homogenised by taking the median value of the values at all bands from the single-\sersic output.
Generally, it is strongly advised to work with micro-registered images, so such a varying position can be avoided and $\rm DOF=0$ can be used.

\item[\textbf{Magnitude ($m$):}]
The magnitude starting values are also taken straight from the single-\sersic fits. 
The code simply divides the flux equally into the 2 components ($m_{\rm{D}} = m_{\rm{B}} = m_{\rm{SS}}+0.75$.\footnote{subscript indices here and in the remainder of the paper indicate the component or fits. D and B indicate disk and bulge parameters, respectively, SS indicates single-\sersic parameters/fit}). 
This means, critically, that both components are started at both the same brightness and the same SED. 
While it is known that generally speaking bulges are red and disks are blue, we did not want to bias the fits in this way as it might cover up interesting objects (e.g. objects with blue bulges and red disks). 
Any findings of a colour difference and SEDs discussed in this work as found by \galapagostwo are hence a pure results of the fit itself.
Again, only if $\rm DOF=1$, the starting values are homogenised by taking the median value of the values at all bands from the single-\sersic output.
We would advise code users to leave the SED of objects as freely variable as possible.

\item[\textbf{Size (half-light radius; $r_{\rm e}$):}]
We have decided on a slightly different scheme for half light radii. 
Generally speaking bulges are smaller than disks. 
From the fitting of very nearby galaxies carried out in V14b, we have found the fits to occasionally converge slightly better when the starting values reflect this, i.e. starting the disk profile to be bigger than the bulge profile. 
Using subsets of relatively bright objects that we re-fit with \textasciitilde 10 different sets of starting values, we have found that -- at least at the typical resolution and redshift of the objects used in this work -- these do not significantly change the fitting outcome.
Qualitatively, the fitting results showed the same trends and -- on average -- same values, although values for individual galaxies can vary, please see Appendix \ref{sec_start_params}.

Nonetheless, in this work we decided to start the sizes at $r_{\rm{e,D}}=1.2*median(r_{\rm{e,SS}})$ and $r_{\rm{e,B}}=0.3*median(r_{\rm{e,SS}})$, with lower limits on the sizes of 1 and 0.5 pixels respectively.
The general behaviour of \galapagostwo is such that the fit -- independent of the DOF of the individual components and the single-\sersic fit -- will always start at a value that is constant with wavelength, i.e. is the same at all wavelengths, and can vary as a constant value during the fit.
This is for simplicity and reflects the fact that at this stage in the program any changes of the sizes within the individual components would be unknown, as the single-\sersic fit would not be able to recover or predict those.
This automatically takes care of any variations that might be frozen when $\rm DOF=1$ is used (see discussion above). 
If a user chooses $\rm DOF \ge 2$, such variations will be allowed during the fit, and any different $r_{\rm{e}}$ in the fitting results at different wavelengths would indeed be a result of the fitting, not a bias from starting values.

\item[\textbf{\sersic index ($n$):}] 
The \sersic index $n$ is a somewhat special parameter in that it is possible that a user would want to hold it fixed during the fit \textbf{at a specific value}, e.g. several authors have fit disks with $n==1$ and bulges with $n==4$, instead of using free values. 
As we did not want to restrict the use of the code to either, the code allows both free and fixed (to specific values) \sersic indices, depending on the setup specified by the user. 
If the \sersic index is not fixed, the code always starts the fit at values constant with wavelength, as was the case with half-light radii.
It then uses $n_{\rm{D}}=median(n_{\rm{SS}})$ and $n_{\rm{B}}=median(n_{\rm{SS}})$, but imposes an upper limit 1.5 for the disk \sersic index starting value, and a lower limit of 1.5 for the bulge \sersic index starting value, respectively.
The limits used here reflect the \textit{general} case that bulges show higher \sersic indices than disks.
During the fit, however, these limits are not imposed.

Additionally -- and in contrast to any other parameter -- it is possible to fix the \sersic index to $n_{\rm{D}}=1$ and $n_{\rm{B}}=4$ for disks and bulges, respectively, if preferred, by simply setting the $\rm DOF=-1$ of the \sersic index for the respective component.
This allows fitting a true exponential disk and/or a \deVa bulge as seen in the \lq classic\rq\ case. 
In those cases, the \lq starting value\rq\ of the parameter becomes irrelevant, and is of course set to the fixed value.

\item[\textbf{Axis ratio ($q$):}] For both components, the values $q_{\rm{SS}}$ are taken as a starting value directly (including any possible variations with wavelength, if allowed in the single-\sersic fit). 
As usual, if $\rm DOF=1$, the starting values are homogenised by taking the median of the values at all bands from the single-\sersic output.
Additionally, a minimum starting value for $q_{\rm{B}}>0.6$ is introduced to reflect that galaxy bulges are generally round-ish.
However, the fit itself is allowed to go below this value.

\item[\textbf{Position angle ($\theta$):}]  For both components, the value $\theta_{\rm{SS}}$ is taken as a starting value directly, including any possible variations if allowed in the single-\sersic fit.
The case of $\rm DOF=1$ is treated in the usual fashion.
\end{description}

This means that apart from $r_{\rm{e}}$, $n$ and possibly $q_{\rm{B}}$ (depending on $q_{\rm{SS}}$), bulge and disk in $B/D$ fits in \galapagostwo are always started at identical parameter values, i.e. specifically the SEDs of the bulge and disk are identical at the start of the $B/D$ fit.
Any deviations in their final/fit magnitudes (and hence the SED of the components) are a result of the \megamorph method to use multiple images at different wavelengths to fit one consistent galaxy profile across all wavelengths, while allowing the magnitudes of each component and at each wavelength to vary freely.
The colour information embedded in these images at different wavelengths makes it easier for the fit to distinguish different components as we will show in \S \ref{sec_sim_results} and \ref{sec_real_results}.

A radically different approach to 2-component fits has been presented by \citet{Lange16}.
Instead of starting all $B/D$ fits at commonly derived starting values, they fit each object multiple times, starting on a fixed grid of starting values, i.e. different starting values for $r_{\rm{e}}$ and $n$ are being used.
They then analyse the convergence of these multiple fits to derive average/reliable \lq best-fit\rq\ parameters.
This method allows a better handle on the robustness of fitting parameters, but requires every galaxy to be fit multiple times, i.e. 40 fits are carried out for each objects.
For obvious reasons on sample size and required CPU time, we avoided such an approach and instead, using simulated data, show that our approach allows satisfactory/good results.

\subsection{Constraints}
\label{sec_setup_constraints}
Generally speaking, the same constraints are being used in the $B/D$ fit that have already been used during the single-\sersic fits.
These have been discussed in detail in H13 and are only quoted here for completeness with additional comments.

\begin{description}
\item[\textbf{Position ($x$, $y$):}] Positions are constrained to lie within a box of size $0.5*r_{\rm{e,PB,SS}}$ around the object centre as defined by the single-\sersic fit.
\footnote{In our \galapagostwo setup, we chose the $r$-band as the primary band PB, but as we keep position constant with wavelength in this particular paper, any other band would enforce the same constraints.}

Additionally, the position of the disk and the bulge are constrained to be the same. 
While in nature, the bulge and the disk can in principle be slightly offset, especially in peculiar objects (e.g. post mergers which have not yet entirely relaxed), this offset should be much smaller than 1 pixel, given the resolution and the distance of the galaxies typically fit with \galapagostwo.
On larger, nearby galaxies with visible dust lanes which could affect disk and bulge differently, such an offset might make sense.
However, these galaxies are not ideally dealt with using \galapagostwo, so a more flexible implementation seemed too complicated and not useful.
The advantage of constraining the $B/D$ positions to be the same, by avoiding many issues with clumpy galaxies or neighbouring objects, outweighs the disadvantages and limitations introduces by such a choice, as discussed in \S \ref{sec_sample}. 

\item[\textbf{Magnitude ($m$):}]
The constraint on magnitudes is a user specified values, we use $5 \leq m_{\rm{D/B,fit}}-m_{\rm{D/B,input}}\leq 5$ in this work for each band. 
Such a wide $\pm 5$ magnitude offset has to be allowed as the real brightness of bulge and disk, respectively, are unknown.
This limit generally resembles a limit well beyond the brightness/faintness of a component which one would still trust in a fit, so does not impose any significant effect on the output values that should be used in an analysis.

Additionally, we use the same basic constraint of $0\leq m\leq 40$ that has been used in the single-\sersic fitting already.
Other than most constraints discussed here, this is currently a limit hard coded into \galapagostwo (although trivial to change), as it easily covers all current galaxy surveys. 
Given the above constraint, however, this will only be violated in un-physical fits and catastrophically failed fits, and is only mentioned here for completeness.

\item[\textbf{Size (half-light radius; $r_{\rm{e}}$):}]
The upper value is a user specified value, we use $r_{\rm{e,D/B}}\leq 400px$ throughout this work, i.e. the same as in single-\sersic fitting mainly to prevent the fit from returning unphysical results.
A lower constraint of $0.3px \leq r_{\rm{e,D/B}}$ is currently hard coded into \galapagostwo.

\item[\textbf{\sersic index ($n$):}] 
Both upper and lower limits are user defined values in \galapagostwo.
In this work, we use $0.2 \leq n \leq 8$, i.e. the same as in single \sersic fitting, but for each component individually. 
Please note that $n_{\rm{D}}==1$ in this work (see \S \ref{sec_setup_choices}), so this constraint only has an effect on $n_{\rm{B}}$ during the $B/D$  fit examined here.

\item[\textbf{Axis ratio ($q$):}] $0.0001\leq q \leq 1$, for the reasons given in H13.

\item[\textbf{Position angle ($\theta$):}] $-180\deg<\theta<180\deg$, in order to prevent numerically different but otherwise identical fits with unnecessarily large values for $\theta$.
No other constraints are set, i.e. bulge and disk are allowed to have different $\theta$ values.
\end{description}

Neighbouring objects are used identically to the single-\sersic fits and with fixed values, so no constraints have to be employed for these objects.

\subsection{Choices of setup and degrees of freedom}
\label{sec_setup_choices}
In single-\sersic fitting, we allowed variation of certain parameters -- size $r_{\rm e}$ and \sersic index $n$ -- with wavelength and have exploited this in detail in H13, V13 and V14b.
This decision effectively allowed colour gradients within galaxies.
These gradients are largely a result of different mixing of the different stellar populations in the galaxy bulge (generally older and redder) and galaxy disk (generally younger and bluer) at different radii \citep{Kennedy16a}.

While, in principle, such gradients could also exist within individual galaxy components, the colour differences are expected to be small compared to the colour differences seen between the 2 components as a whole, at least in the vast majority of cases, making them very hard to measure.
As a result, we have decided to \textbf{not} allow any wavelength variation of any component parameters other than magnitude in this work (magnitude is a free parameter in all bands, as in our previous work).
We are aware that this is only a first step towards understanding galaxy components in detail, but we find it necessary to understand this step in detail before taking the next step, e.g. allowing colour gradients within galaxy components in more suitable datasets with brighter and better resolved galaxies, e.g. the ones analysed in V14b.

However, both codes, \galfitm and \galapagostwo, are written in a way that the user can trivially switch on wavelength variations in the setup files, if they so wished.
In order to support our choice, we have run $B/D$ decomposition with linear variation allowed for $r_{\rm e}$ of disks and bulges for a subset of 4000 galaxies, and have found that their values in fact \textbf{do} change slightly with wavelength, but by much smaller factors than reported by \citet{MMVulcani} for single-\sersic fits and, in fact, within the measurement uncertainty in most cases.
We have found that $r_{e,\rm D}$ and $r_{e,\rm B}$ change by \textasciitilde14\% and \textasciitilde23\%, respectively, from u- to K-band. 
Similarly, we have fitted the 163 nearby galaxies analysed in the V13 sample with linear variation of disk and bulge size (but not \sersic index) and have found little variation from u- to z-band covered by this data, with $r_{e,\rm D}$ changing by \textasciitilde-10\% and $r_{e,\rm B}$ by \textasciitilde+10\%, the exact value also depending on galaxy classification/type.
We conclude from this that keeping these parameters constant with wavelength is a justifiable assumption in case of $B/D$ decomposition of \gama galaxies.
In the following, we discuss the choices for each parameter.

\begin{description}
\item[\textbf{Position:}]
As images at different observed bands are accurately registered (although see \citealt{Kelvin2012} and \S 4.1), we keep the position constant with wavelength in the $B/D$ fits.

\item[\textbf{Magnitude:}]
As the polynomial shape of the SED of each component is unknown, we keep the magnitudes with full degree of freedom for both bulge and disk.
Users should be reminded of Runge's phenomenon, whereby a polynomial function can oscillate excessively between data points, particularly at the edges of the considered interval if the degree of freedom is similar to the number of points fit.
As such, it is in theory dangerous to use the Chebyshev polynomial of the fit directly to derive values at intermediate wavelengths.
While Nedkova et al. (in prep.) have found that such oscillations are rare in their fits on \candels data, great care has to be taken when interpolating the magnitudes.

\item[\textbf{Sizes:}]
For both component, bulge and disk, we use a size which is constant with wavelength, for the reasons given above.

\item[\textbf{\sersic index:}]
We use $n_{\rm D}=1$ (fixed) and a $n_{\rm B}$ that is constant as a function of wavelength (but variable during the fit).
This is also what we used for the simulated galaxies in \S \ref{sec_sim_sims} (see Appendix \ref{sec_app_sim_sims}), so no biases are introduced in these tests.

\item[\textbf{Axis Ratio:}]
Axis ratios are chosen to be wavelength independent, for each component, but can vary during the fit.

\item[\textbf{Position Angle:}]
Position Angles are also chosen to be wavelength independent, i.e. no rotation is allowed between images at different wavelengths. The value can vary during the fit.
 
\item[\textbf{List of target objects:}]
As was mentioned before, the list of target objects for these 2-component fits should be equal or a subset of the targets used in the single-\sersic step of the code.
For our runs, we made sure that this was the case.
In case of the simulated images, we simply fit all objects at both stages.
When fitting real galaxies in \S \ref{sec_real}, we target all \gama galaxies at this stage.
In the single-\sersic fits, we had also targeted bright neighbours to these objects in order to be able to take these properly into account when they become neighbouring objects for a fit.
Given how these neighbours are being dealt with in the $B/D$  fits, it is no longer necessary to deal with these objects.
\end{description}

%__________________________________________________________________
%%% DATA & SAMPLE SELECTION %%%
\section{Data \& rejection of bad fits}
\label{sec_sample}

As usual before any analysis, the catalogues created by \galapagostwo have to be cleaned of bad fits.
In this chapter, we explain only the principles used to define the samples presented in the following chapters.
As catalogues used are by design very different (simulated vs real data), we leave the discussion of the exact samples and object numbers to the respective Sections \S \ref{sec_sim_samples} and \S \ref{sec_real_samples}.

To identify \lq good\rq\ fits, we use the same parameter limits used in H13.
More precisely, we identify and discard those components with one or more parameters lying on (or very close to) a fitting constraint, as described in Section~\ref{sec_setup_constraints}. 
Such a fit is unlikely to have found a global minimum in $\chi^2$ space, but is rather constraint by the shape of $\chi^2$ space along the boundary box of the allowed parameter space, and is a good indication that the parameter values given can not be trusted.
While \galfit and \galfitm have ways to try to avoid such local minima in the final fit solution, these measures do not work in extreme cases of very deep/steep local minima.
The absolute $\chi^2$ (or reduced $\chi^2$, $\chi^2/\nu$) value itself is also not a good indicator on whether a fit is a good representation of the true galaxy profile, as it depends strongly on the image properties, especially the precise way the pixel-to-pixel noise is correlated.
It is also artificially increased in the case that the object of interest can not precisely be modelled with the model used in the fit.
Real galaxies often have features (bars, spiral arms, faint components, merger features) that prevent a perfect fit, and the contribution of neighbouring objects also adds to the absolute $\chi^2$ value, bright stars being a particularly bad example, so such a mismatch is obvious in real galaxies.
We hence avoid using absolute $\chi^2$ values in this analysis as a measure to define \lq good fits\rq.
Instead, we avoid any object where we already know that the fit was not a free fit.

For the analysis throughout this paper, we only keep objects which fulfil the following criteria:
\begin{itemize}
\item $m_{\rm input}-5<m<m_{\rm input}+5$,
\item ${\rm abs}(x_{\rm pos}-x_{\rm pos,SS})<0.5*re_{\rm ss}$
\item ${\rm abs}(y_{\rm pos}-y_{\rm pos,SS})<0.5*re_{\rm ss}$
\item $0<m<40$,
\item $0.205<n<7.95$,
\item $0.305\ [pix]<r_{\rm{e}}<395.0\ [pix]$,
\item $0.001 < q \leq 1.0$.
\end{itemize}

where the magnitude input values $m_{\rm input}$ are derived from the single-\sersic fit result as described in \S \ref{sec_setup_starts}.
For the individual single-band fits, this is obvious, but for the multi-band fits, these constraints are equally checked on each band, removing the component from the subsequent analysis if even a single one of these parameters fails this test in any band.
The limits on position make sure that the fit stayed on the primary target and did not run away to fit a neighbouring object.
As position of bulge and disk are by design fixed to be the same, this limit is in practice never violated (e.g. by the disk fitting one target and the bulge fitting a neighbour).
The criteria for $n$ and $r_{\rm e}$ are slightly more restrictive than the fitting constraints used.
When we use limits more tightly around the constraint values ($0.201<n<7.99$ and $0.301\ [pix]<r_{\rm{e}}<399.0\ [pix]$), the number of \textit{successful} fits in Tables \ref{tab_sim_numbers_all} are naturally higher as less galaxies are excluded, i.e. both bulges and disks in multi-band fitting are 98\% \lq successful\rq.
While this might seem like a preferable cutoff to use, it merely changes the definition of 'success' and would, in fact, include many galaxies where the fit was in practice not a free fit. 
This indicates that even values \textit{close} to a fitting constraint (rather than on the limit precisely) are potentially not the result of a free fit.
The numbers in Table \ref{tab_sim_numbers_bright}, however, would not change much, which shows that especially faint objects are removed by these cutoffs, which seems obvious as these objects will be hardest to fit.
These are hence the most critical limits we use to clean the galaxy samples and a balance between rigorously avoiding bad fits while allowing large samples must be found.

The obvious difference of this work compared to H13 is that these limits need to be checked for each component \textbf{individually}, instead of each galaxy as a whole.
The reason for this is simple: If a faint galaxy component sits within a bright component (e.g. faint bulge within a bright disk or vice versa), it is irrelevant whether the fainter component is fit well in order to decide whether we would believe the values of the brighter component.
We might believe the fit values of a disk even if we do not believe the fit result of the bulge in the same galaxy.
This can easily be seen in the numbers given in \S \ref{sec_sim_samples} and \S \ref{sec_real_samples}, where the bulge and disk samples contain different numbers of objects and do not fully overlap.
For such a scheme, it is also wise to include a further brightness limit below which one would not believe the parameters of the fainter component, despite it being \lq well\rq\ fit given the above limits.
During this work, we choose to use the fainter component only if it is within 1.5 magnitudes of the brighter component in the $r$-band, which was chosen as the main band in this analysis and within \galapagostwo in our setup.
This value was somewhat empirically established and will be discussed in \ref{sec_sim_samples}.
Realistically, such a choice could additionally be a function of magnitude of the galaxy as a whole, i.e. in brighter objects a larger magnitude difference between the components might be acceptable.
Such a scheme is hard to define universally, however, and beyond the scope of this paper.

%__________________________________________________________________

%%% APPLICATION TO SIMULATED DATA %%%
\section{Application to simulated imaging}
\label{sec_sims}

The advantages and disadvantages of using simulated vs. real data have already been discussed in H13.
In order to follow the same approach, we have decided to, once again, use both simulated and real data in our analysis.
Simulated data has the advantage that one knows the input galaxy parameters exactly, allowing for a direct comparison of recovered galaxy parameters to their true values, while assuming perfect profile match, i.e. in that galaxy components actually {\bf are} well described by \sersic profiles.
For these tests, the choice of simulated parameters is obviously an important one.
If the simulated galaxies do not resemble typical $B/D$  systems, not much can be learned from this analysis.
If, on the other hand, parameters are wisely chosen and resemble real objects (and potentially include interesting objects one would like to identify in the real data, e.g. galaxies in which the bulge is larger than the disk or with inverted colours), one can conclude that the parameters found in real galaxies are likely to be well measured and a good description of the galaxies found in nature.
As in H13 and similar work \citep[e.g.][]{Meert15}, simulations are used as an idealised case here and, while allowing a good comparison of single-band to multi-band fitting, they can only give a lower limit on the fitting uncertainties for real data.

In this section, we describe the galaxy simulations and the choice of parameters in detail. 
We will then present and discuss our finding when using the galaxy images to test the performance of \galapagostwo and \galfitm in \S \ref{sec_sim_results}.

\subsection{Creating the simulations}
\label{sec_sim_sims}

The simulations used in this work are created using the same codes/methods used and described in previous papers, so we refer the reader to \S~5.1 of H13 and \S~2 of \citep[][hereafter H07]{Haeussler2007} for technical details on how the images are created in detail.

In H07, we have presented analysis and testing for single-component, single-band datasets.
In H13, we have extended this analysis to multi-band fits, but have still restricted ourselves to single-component objects.
In this work, instead of creating single-\sersic objects, we aim to simulate galaxies with several components.
To achieve this, we have simply put 2 objects at the same position, using the same scripts, but with different parameters, one generally representing a bulge, one generally representing a disk.
As the details of the simulated data might be important to some readers, we give extensive details in Appendix \ref{sec_app_sim_sims} and refer to that section for the details in these simulated data.
The most important points in the resulting objects are
\begin{itemize}
    \item We simulate less faint objects compared to H13, where we used galaxies up to 4 magnitudes below the peak in the magnitude histogram in a size bin. Their main purpose in H13 was to push the single-\sersic fits to their limits on galaxies along the detection limit. As those objects are extremely unlikely to be decomposed by any code, we have restricted the objects simulated here to 1 magnitude below the peak in the magnitude distribution. In the following sections, we further restrict our analysis to objects at $m_{\rm r,B+D}<19.5$ only, which would avoid analysing fainter objects. This imposed magnitude limit somewhat matches the magnitude limit of the \gama survey, which provides spectroscopic redshifts down to a magnitude of $m_{\rm r} <19.8$. While results presented here still hold at this fainter limit qualitatively, it is at significantly lower significance, which is why we chose a somewhat brighter magnitude limit to present our results. Objects at $m_{\rm r}\sim19.8$ are very much on the edge of what could possibly be decomposed in these \sdss/\ukidss data and the samples derived from overlap with single-band fits are very small.
    \item All disks and all bulges show the same, somewhat extreme, SED, respectively, with disks being bluer and bulges being redder. This choice was taken in order to make it easier to analyse and present the results as they are likely colour dependent, disk and bulge with similar SEDs being much harder to decompose. However, initial tests with more general data showed the same trends, but with less statistical significance. Some - 0.1 mag - noise is added to these SEDs, in order to simulate some variation found in real galaxies, however.
    \item Bulge \sersic indices $n_{\rm B}$ shows a variety of values, centreed around 4, but with a wide spread. This allows bulges in the simulated data to explicitly \textbf{not} be classical \deVa bulges.
\end{itemize}

The resulting images are realistic looking at all wavelengths with thousands of galaxies that span a large range of parameters, but -- over all -- show similar parameter distributions to real galaxies with realistic noise properties, and for which we know the true parameter values, including -- and especially -- their subcomponents. 

It should be noted that these simulated images do not contain any stars. 
While this removes a potential source of error from the analysis, this source has already been tested in previous work (e.g. H07) and should not influence the $B/D$ decompositions more strongly than single-\sersic fits.
Overlapping galaxies are naturally included by our simulation method, and so the effects of blending several objects are still included in our results.

The resulting images are then fed through \galapagostwo and \galfitm as described in \S~\ref{sec_setup}, i.e. the same pipeline is used for both the simulated and the real data (discussed in \S~\ref{sec_real_results}). 

\subsection{Samples}
\label{sec_sim_samples}
In total, 95143 galaxies have been simulated in the survey area analysed in this work.
75998 of these objects were simulated as 2-component galaxies, the remaining $\rm \sim$20\% were one-component systems.
While a detailed analysis of these objects, and an attempt to identify them in an automated fashion within our dataset, are beyond the scope as this paper, we had a quick look at the 2-component fits of these one-component objects.

These objects seem to often fall into one of three qualitative categories.
\begin{itemize}
\item[\textbf{1)}]
The objects in the first category are objects in which the fits for one of the components runs into a fitting constraint.
For these objects, this is very often the \sersic-index of the bulge, which hits the upper limit of $n_{\rm B}==8$, often in combination with fitting a very small size of the bulge.
As fitting constraints are violated, these objects would be removed from the analysis when cleaning the catalogues for a 2-component analysis.
As the violating component can also be faint, there is a big overlap of this category to the next.
\item[\textbf{2)}]
The second category contains objects for which \galfitm returns a large magnitude difference between the components.
This behaviour is what one would naively expect in that one component fits the galaxy profile well (which component this is depends on the overall \sersic-index of the galaxy) while the other tries to fit a small correction to this profile, e.g. even by trying to fit a group of high-flux pixels in the noise pattern.
These objects can be identified by a large magnitude difference between the two components in the $B/D$ fit, and often a very small size in the fainter component.
This is why we introduce a limit on the magnitude difference in our analysis.
We only use the fainter component if it is not more than 1.5 mag fainter than the brighter component, as measured in the $r$-band.
\item[\textbf{3)}]
The third category seems to contain objects for which the fit behaves such that both profiles mimic each other and the overall profile.
For many of these objects, we find magnitudes (and hence SEDs) of both components to be very similar, as well as similar sizes of the components.
As the $B/D$ fits are started at $r_{\rm{e,D}}=1.2*median(r_{\rm{e,SS}}) (>1px)$ and $r_{\rm{e,B}}=0.3*median(r_{\rm{e,SS}}) (>0.5px)$, this means that the sizes in the fit actually converged to be the same value as one would expect if they fit the same profile.
Effectively, the flux of the galaxy is simply divided into 2 components, without any physically meaningful separation, making the SEDs of these components largely unconstrained, so it comes as somewhat as a surprise that the flux often seems to be shared equally by the two components.
Unfortunately, this does not always seem to coincide with $n_{\rm B}\sim1$ as one would naively expect as these are the galaxies where the profiles would mimic each other best.
The galaxies in this category are the hardest to find in large datasets, but we will discuss such an effect in \S \ref{sec_sim_results}.
\end{itemize}

There seems to be a big overlap between all these categories, especially between the first and second.
It should be noted that we only mention these general categories here for completeness.
While these categories can be found in our analysis, they are not pronounced enough to actually use them to separate out the single-\sersic objects from the 2-component galaxies as they are, and additional development and testing would be required for this purpose.
Such an attempt, however, will have to use a more sophisticated method to separate the object classes and is beyond the scope of this paper.
For example, it has been found that the classification scheme presented by \citet{Allen2006} works well on \candels\ data \citep[e.g.][Nedkova in prep]{Nedkova20}, but it is unclear whether the same scheme would work equally well on the \gama data used here.

In the following, we restrict our analysis to the 2-component galaxies.
Of these, not all galaxies are recovered by \sex, some are too faint to be detected.
Given our analysis limit of $m_r<19.5$, these missed galaxies are unlikely to be presented in this work.
Depending on the band used for object detection, this fraction is very different, the detection and fitting numbers can be found in Table \ref{tab_sim_numbers_all}.
At best -- when running detection in a band-combined/multi-colour stacked image -- $\rm \sim70\%$ ($53448/75998$) of the objects can be recovered. 
This detection completeness is not part of the analysis in this work and will hence not be discussed here. 
The important part for this work is that we can analyse galaxies all the way down to the detection limit.
In this work, we merely analyse what fractions we can successfully \textbf{fit}.

Object numbers shown in Tables \ref{tab_sim_numbers_all} and \ref{tab_sim_numbers_bright} show how much multi-band fitting improves the sample size for scientific studies, especially once measurements from several bands are required. 
Each row shows the numbers of objects in each -- single-band and multi-band -- \galapagostwo run, and the percentages of objects which deliver a \lq successful\rq\ fit for single-\sersic (on the left) and $B/D$ fits (on the right), in case of $B/D$ fits split up for bulges and disks.
Additionally, we show the object numbers and success rates when combining several single-band fits (in black), e.g. in case a science case requires values from more than 1 band ($ugrizYJHK$ and $griYHK$, respectively), which drastically reduces the available sample size that we would recommend to use.
These numbers also include the cut as the fainter component being within 1.5 magnitudes of the brighter component, based on the fitting values.
In light grey, we give the same values when using simulated values to make this decision.
As one can see, multi-band fits are far more likely to produce a valid fit result.
For all galaxies at $m_{\rm r}<19.8$ (\gama spectroscopy limit, Table \ref{tab_sim_numbers_bright}), only 2369 objects (14\% of objects of the 16941 objects for which a $B/D$  decomposition was attempted) have \lq good\rq\ single-band fit in all $griYHK$ bands in both components (366, 2.2\%, in case one adds $uzJ$ bands), immediately reducing the sample size for any analysis that requires these parameters, e.g. magnitudes in several bands when a SED of a component is required.
In comparison, 14732 (85.5\%) of objects have \lq good\rq\ multi-band fits, increasing a possible science sample by more than a factor of 6.
In the following sections, in order to estimate the effect of multi-vs single-band fits, we try to use the largest sample possible in most cases, in order to maximise statistical significance of our findings and allowing a fair comparison of the 2 fit performances, i.e. on the same objects.
Unfortunately that drastically reduces the sample size available in most plots.

\begin{table*}
\centering
\caption[]{Simulated object numbers and success rates, full sample}
\scriptsize
\begin{tabular}{@{}l|rrr|rrr:rr:rr@{}}
\hline
Band    &  \multicolumn{3}{c|}{\#objects SS fit} &  \multicolumn{7}{c}{\#objects BD fit} \\
            & \#detected &  \#successful & Success rate & tried $B/D$  & success B   & rate  & success D  & rate  & success BD  & rate\\
\hline 
\hline
%table comes automatically from plotting IDL script!
u & 1915 & 1412 & 73.7\% & 1915 & 1182 & 61.7\% & 1869 & 97.6\% & 1143 & 59.7\%\\
g & 34746 & 27387 & 78.8\% & 34693 & 21460 & 61.9\% & 33540 & 96.7\% & 20394 & 58.8\%\\
r & 48711 & 41197 & 84.6\% & 48610 & 31787 & 65.4\% & 47021 & 96.7\% & 30336 & 62.4\%\\
i & 48457 & 41491 & 85.6\% & 48351 & 31655 & 65.5\% & 46823 & 96.8\% & 30267 & 62.6\%\\
z & 22926 & 17934 & 78.2\% & 22912 & 13939 & 60.8\% & 22299 & 97.3\% & 13392 & 58.4\%\\
Y & 33879 & 28439 & 83.9\% & 33848 & 22209 & 65.6\% & 32748 & 96.8\% & 21190 & 62.6\%\\
J & 25544 & 20896 & 81.8\% & 25513 & 16476 & 64.6\% & 24774 & 97.1\% & 15793 & 61.9\%\\
H & 37471 & 32013 & 85.4\% & 37415 & 24752 & 66.2\% & 36493 & 97.5\% & 23923 & 63.9\%\\
K & 31655 & 26566 & 83.9\% & 31612 & 20779 & 65.7\% & 30857 & 97.6\% & 20078 & 63.5\%\\
 & & & & & & & & & & \\
\textbf{combined single band} $ugrizYJHK$ & \textbf{50780} & \textbf{1295} & \textbf{2.6\%} & \textbf{50780} & \textbf{406} & \textbf{0.8\%} & \textbf{1681} & \textbf{3.3\%} & \textbf{366} & \textbf{0.7\%}\\
of those bright enough (r-band S1 fit) & & & & &  (376) & & (1458) & & (253) & \\
\textcolor{mygray}{additionally bright enough (simulated)} & & & & &  \textcolor{mygray}{(305)} & & \textcolor{mygray}{(1430)} & & \textcolor{mygray}{(191)} & \\
additionally with valid mwl fits & & (1284) & & &  (369) & & (1453) & & (245) & \\
additionally bright enough (r-band M fit) & & & & &  (346) & & (1340) & & (201) & \\
\textbf{combined single band} $griYHK$ & \textbf{50771} & \textbf{15772} & \textbf{31.1\%} & \textbf{50771} & \textbf{3900} & \textbf{7.7\%} & \textbf{22299} & \textbf{43.9\%} & \textbf{3120} & \textbf{6.1\%}\\
of those bright enough (r-band S1 fit) & & & & &  (3524) & & (21088) & & (2559) & \\
\textcolor{mygray}{additionally bright enough (simulated)} & & & & &  \textcolor{mygray}{(3063)} & & \textcolor{mygray}{(17369)} & & \textcolor{mygray}{(1738)} & \\
additionally with valid mwl fits & & (14580) & & &  (3343) & & (20908) & & (2377) & \\
additionally bright enough (r-band M fit) & & & & &  (3045) & & (18792) & & (1896) & \\
 & & & & & & & & & & \\
\textbf{mwl} & \textbf{53443} & \textbf{39524} & \textbf{74.0\%} & \textbf{53439} & \textbf{46793} & \textbf{87.6\%} & \textbf{52289} & \textbf{97.8\%} & \textbf{45674} & \textbf{85.5\%}\\
\textcolor{mygray}{of those bright enough (simulated)}  & & & & & \textcolor{mygray}{(37172)} & & \textcolor{mygray}{(42506)} & & \textcolor{mygray}{(27851)} & \\
of those bright enough (r-band M fit) & & & & & (36937) & & (46663) & & (31465) & \\
\end{tabular}
\tablefoot{Object numbers in the simulated dataset and fraction with successful fits for single-band fits (Mode S1, as defined in H13 as single-band fits with single-band detection) and multi-band fits (Mode M, multi-band fits with detection in co-added image). 
Please note that the success rate in this simulated sample resembles an actual fitting success rate and does not include the effect of missing galaxies in the detection step. Sample sizes are given in brackets to show the general samples used in this analysis.
}\label{tab_sim_numbers_all}
\end{table*}

\begin{table*}
\centering
\caption{Simulated object numbers and success rates, bright sample}
\scriptsize
\begin{tabular}{@{}l|rrr|rrr:rr:rr@{}}
\hline
Band    &  \multicolumn{3}{c|}{\#objects SS fit} &  \multicolumn{7}{c}{\#objects BD fit} \\
            & \#detected &  \#successful & Success rate & tried $B/D$  & success B   & rate  & success D  & rate  & success BD  & rate\\
\hline
\hline
%table comes automatically from plotting IDL script!
u & 1899 & 1401 & 73.8\% & 1899 & 1172 & 61.7\% & 1854 & 97.6\% & 1133 & 59.7\%\\
g & 14956 & 13603 & 91.0\% & 14925 & 10186 & 68.2\% & 14321 & 96.0\% & 9598 & 64.3\%\\
r & 16535 & 15604 & 94.4\% & 16489 & 11982 & 72.7\% & 15869 & 96.2\% & 11389 & 69.1\%\\
i & 16535 & 15622 & 94.5\% & 16488 & 12144 & 73.7\% & 15810 & 95.9\% & 11500 & 69.7\%\\
z & 13406 & 11816 & 88.1\% & 13395 & 8716 & 65.1\% & 12949 & 96.7\% & 8290 & 61.9\%\\
Y & 15207 & 14019 & 92.2\% & 15190 & 11072 & 72.9\% & 14542 & 95.7\% & 10451 & 68.8\%\\
J & 14031 & 12656 & 90.2\% & 14007 & 9928 & 70.9\% & 13500 & 96.4\% & 9442 & 67.4\%\\
H & 15650 & 14648 & 93.6\% & 15612 & 11568 & 74.1\% & 15107 & 96.8\% & 11093 & 71.1\%\\
K & 15045 & 13837 & 92.0\% & 15015 & 10896 & 72.6\% & 14529 & 96.8\% & 10433 & 69.5\%\\
 & & & & & & & & & & \\
\textbf{combined single band} $ugrizYJHK$ & \textbf{16943} & \textbf{1292} & \textbf{7.6\%} & \textbf{16943} & \textbf{406} & \textbf{0.8\%} & \textbf{1668} & \textbf{9.8\%} & \textbf{366} & \textbf{2.2\%}\\
of those bright enough (r-band S1 fit) & & & & &  (376) & & (1446) & & (253) & \\
\textcolor{mygray}{additionally bright enough (simulated)} & & & & &  \textcolor{mygray}{(305)} & & \textcolor{mygray}{(1418)} & & \textcolor{mygray}{(191)} & \\
additionally with valid mwl fits & & (1282) & & &  (369) & & (1441) & & (245) & \\
additionally bright enough (r-band M fit) & & & & &  (346) & & (1328) & & (201) & \\
\textbf{combined single band} $griYHK$ & \textbf{16941} & \textbf{10881} & \textbf{64.2\%} & \textbf{16941} & \textbf{3012} & \textbf{17.8\%} & \textbf{11796} & \textbf{69.6\%} & \textbf{2369} & \textbf{14.0\%}\\
of those bright enough (r-band S1 fit) & & & & &  (2714) & & (10801) & & (1861) & \\
\textcolor{mygray}{additionally bright enough (simulated)} & & & & &  \textcolor{mygray}{(2352)} & & \textcolor{mygray}{(9187)} & & \textcolor{mygray}{(1248)} & \\
additionally with valid mwl fits & & (10395) & & &  (2616) & & (10701) & & (1759) & \\
additionally bright enough (r-band M fit) & & & & &  (2406) & & (9723) & & (1408) & \\
 & & & & & & & & & & \\
\textbf{mwl} & \textbf{17231} & \textbf{15029} & \textbf{87.2\%} & \textbf{17232} & \textbf{15254} & \textbf{88.5\%} & \textbf{16698} & \textbf{96.9\%} & \textbf{14732} & \textbf{85.5\%}\\
\textcolor{mygray}{of those bright enough (simulated)}  & & & & & \textcolor{mygray}{(11809)} & & \textcolor{mygray}{(13941)} & & \textcolor{mygray}{(8868)} & \\
of those bright enough (r-band M fit) & & & & & (12173) & & (14996) & & (10334) & \\
\end{tabular}
\tablefoot{Same table as table \ref{tab_sim_numbers_all}, but for \lq bright\rq\ objects at $m_{\rm r}<19.8$, the spectroscopic limit of the \gama survey. This gives a handle on how successful we would be able to decompose galaxies observed by \gama. As can be seen, single-band fitting is more successful to return fit values for brighter objects, as expected, but still on a much lower level than multi-band fitting, which is basically unchanged between the samples.}
\label{tab_sim_numbers_bright}
\end{table*}

In simulated data, it is possible to define galaxy samples using 2 different methods: using measured values, or using simulated values, e.g. when identifying the $B/T$ ratio of a galaxy (recovered $B/T$ vs true $B/T$).
The first would allow a more direct comparison to real, observed objects, as the selection would/could be entirely identical.
The latter defines slightly different samples, but allows a cleaner comparison to simulated values.
In this paper, we decided to define galaxy samples using the simulated/true values where possible, as the comparison to those is the main purpose of this paper.
However, we have carried out the same analysis using the measured/recovered values instead, which qualitatively leads to the same results.
However, plots are somewhat harder to read using this approach as effects can be less pronounced and different effects are harder to separate from each other.
The conclusions in this paper have not been significantly altered by this choice.
In tables \ref{tab_sim_numbers_all} and \ref{tab_sim_numbers_bright}, we give numbers for both definitions.
From these numbers one can see that the sample size increases when using observed values.
This indicates that the fainter component in a galaxy often ends up accounting for some flux from the brighter component, boosting its magnitude and hence pushing it into an observed sample.

\begin{figure}
\begin{center}
\includegraphics[width=0.48\textwidth,trim=25 30 8 15, clip]{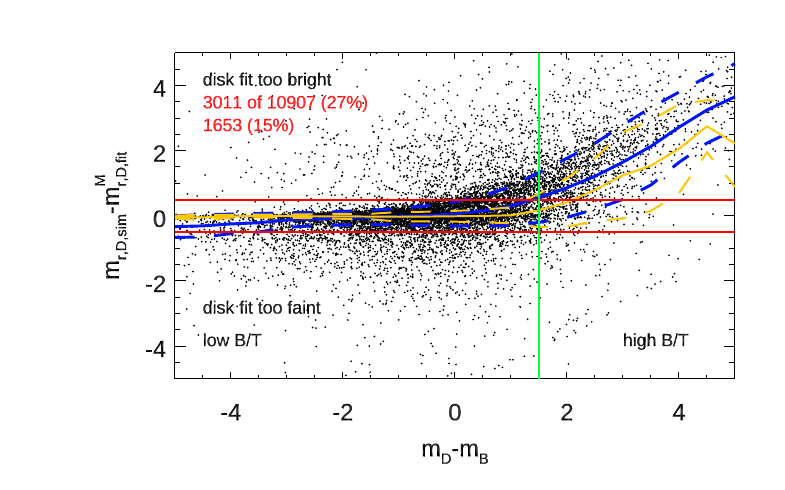}
\includegraphics[width=0.48\textwidth,trim=25 5 8 15, clip]{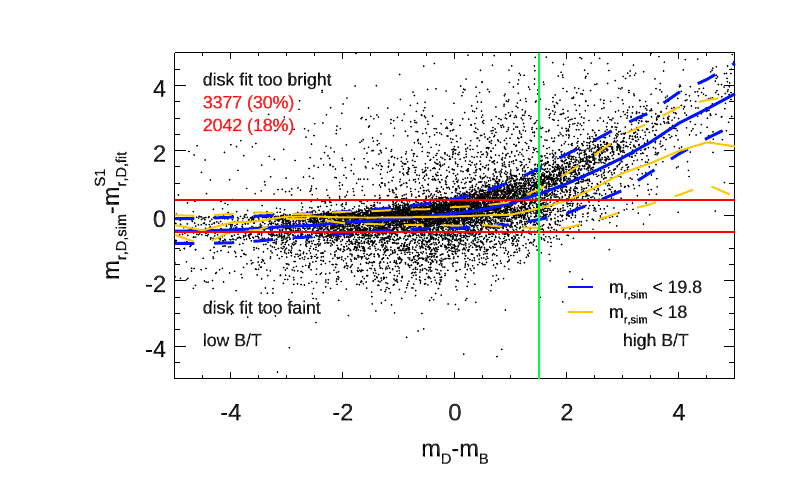}
\caption{For objects with $m_{\rm r}<19.8$, we show the deviation of fitting to simulated disk $r$-band magnitude $\delta m_{\rm r}=m_{\rm r,D,sim}-m_{\rm r,D,fit}$ as a function of the faintness of the disk (Top: multi-band fit. Bottom: single-band fit). Here, and in all successive plots, the subscripts of the axis labels indicate [band] plotted where relevant, as well as [B/D] to indicate bulge or disk component and [fit/sim] to indicate fit or simulated value. Where necessary, we also indicate superscripts M or S1 to indicate multi-band or single-band fitting according to the definition in H13. 
These plots show that the fainter a disk is within a given bulge, the harder it is to fit, as one would expect. Red horizontal lines show an (empirically chosen) allowed deviation of $\rm \pm 0.5 mag$, red numbers in the top left corner indicate the number of objects violating these limits. The systematic trend reaches this limit when the disk is $\rm \sim 1.5$mag fainter than the bulge. Blue lines show mean deviation and sigma of the relation (robust mean, clipped at 3 sigma), with the same values for objects at $m_{\rm r}<18$ indicated in orange.
}
\label{fig_disk_mag_acc}
\end{center}
\end{figure}

\begin{figure}
\begin{center}
\includegraphics[width=0.48\textwidth,trim=25 30 8 15, clip]{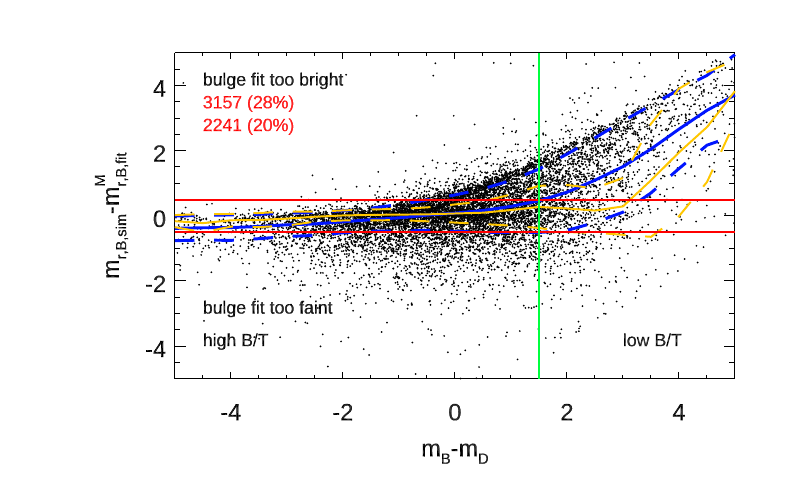}
\includegraphics[width=0.48\textwidth,trim=25 5 8 15, clip]{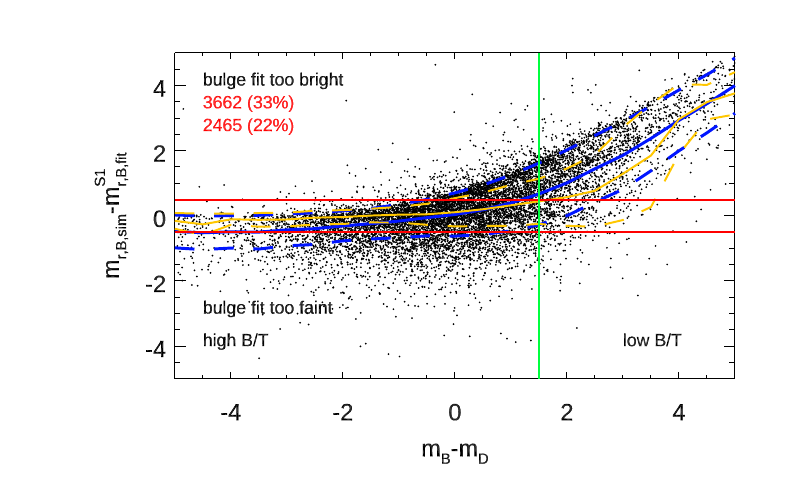}
\caption{For objects with $m_{\rm r}<19.8$, we show the deviation of fitting to simulated bulge $r$-band magnitude $m_{\rm r,B,sim}-m_{\rm r,B,fit}$as a function of the faintness of the bulge (Top: multi-band fit, mode M. Bottom: single-band fit, mode S1). The fainter a bulge within a given disk, the harder it is to fit. Red horizontal lines show a (empirically chosen) allowed deviation of $\rm \pm 0.5 mags$. The systematic trend reaches this limit when the bulge is $\rm \sim 1.5$mag fainter than the disk.
}
\label{fig_bulge_mag_acc}
\end{center}
\end{figure}

\begin{figure}
\begin{center}
\includegraphics[width=0.48\textwidth,trim=25 30 8 15, clip]{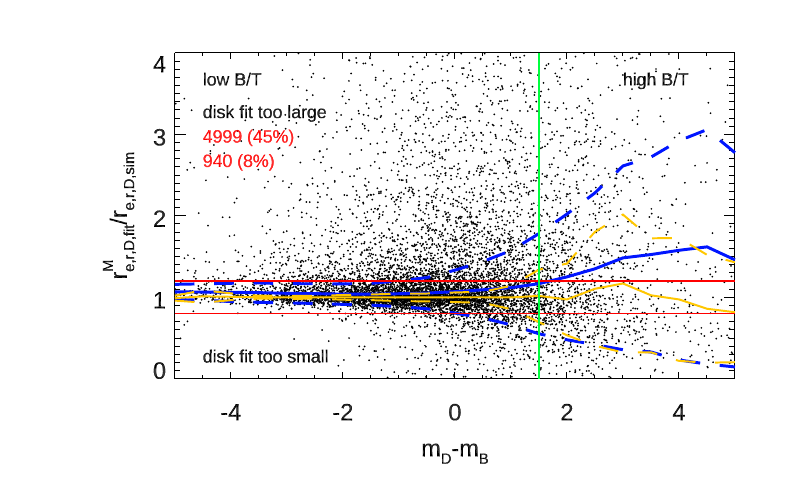}
\includegraphics[width=0.48\textwidth,trim=25 5 8 15, clip]{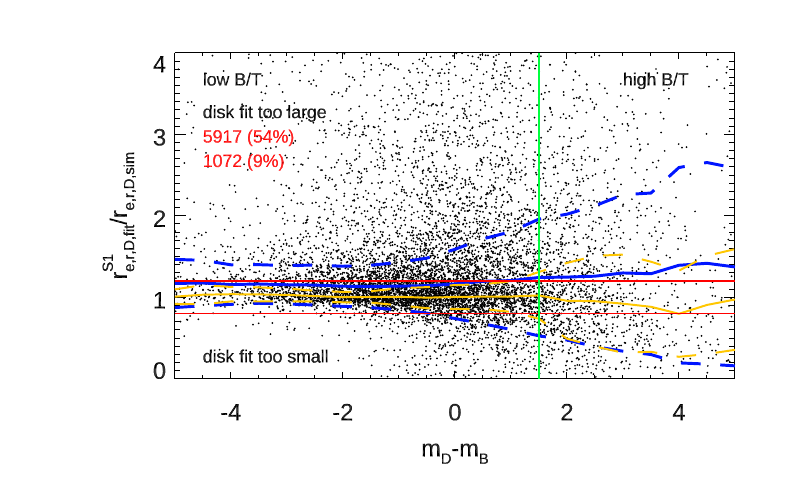}
\caption{For objects with $m_{\rm r}<19.8$, we show the deviation of fitting to simulated disk size $r_{\rm e,D,fit}/r_{\rm e,D,sim}$ as a function of the faintness of the disk (Top: multi-band fit. Bottom: single-band fit). The fainter a disk within a given bulge, the harder it is to fit. Horizontal lines show a (randomly chosen) allowed deviation of $\rm \pm 20\%$.
}
\label{fig_disk_re_acc}
\end{center}
\end{figure}

\begin{figure}
\begin{center}
\includegraphics[width=0.48\textwidth,trim=25 30 8 15, clip]{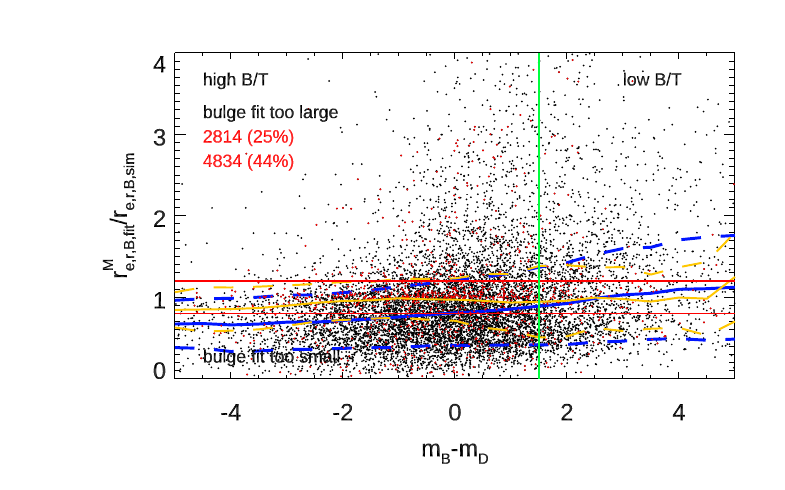}
\includegraphics[width=0.48\textwidth,trim=25 5 8 15, clip]{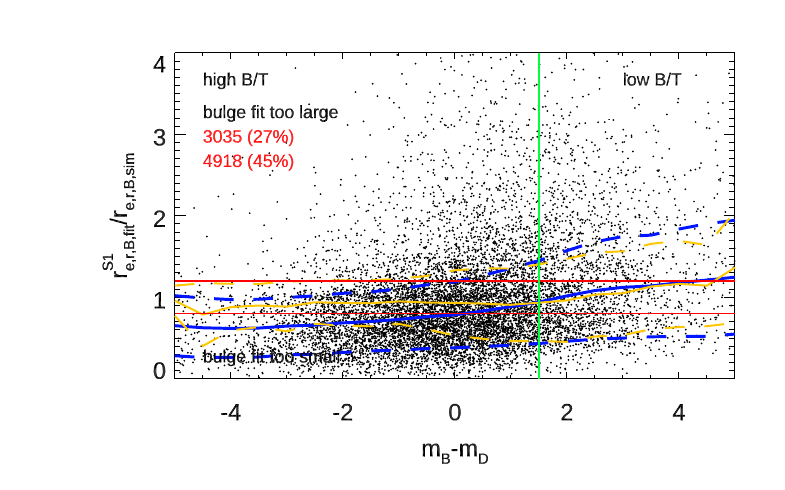}
\caption{For objects with $m_{\rm r}<19.8$, we show the deviation of fitting to simulated bulge size $r_{\rm e,B,fit}/r_{\rm e,B,sim}$ as a function of the faintness of the disk (Top: multi-band fit. Bottom: single-band fit). The fainter a bulge within a given disk, the harder it is to fit, represented in larger scatter. Horizontal lines show a (randomly chosen) allowed deviation of $\rm \pm 20\%$.
}
\label{fig_bulge_re_acc}
\end{center}
\end{figure}

\begin{figure}
\begin{center}
\includegraphics[width=0.48\textwidth,trim=25 30 8 15, clip]{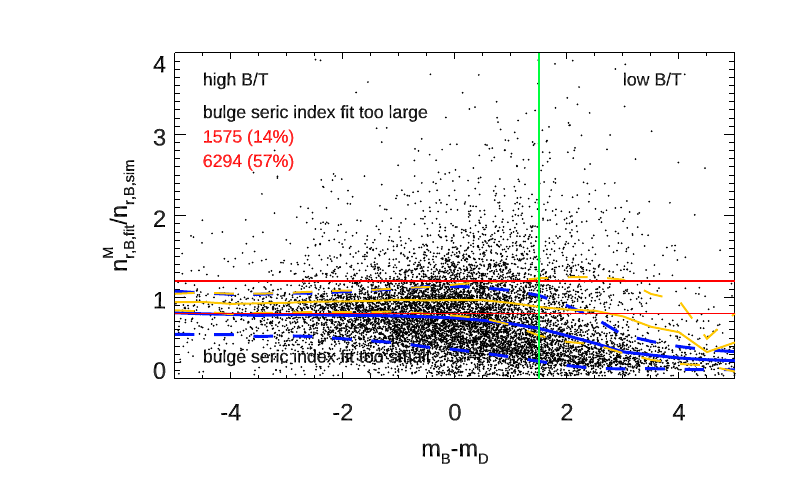}
\includegraphics[width=0.48\textwidth,trim=25 5 8 15, clip]{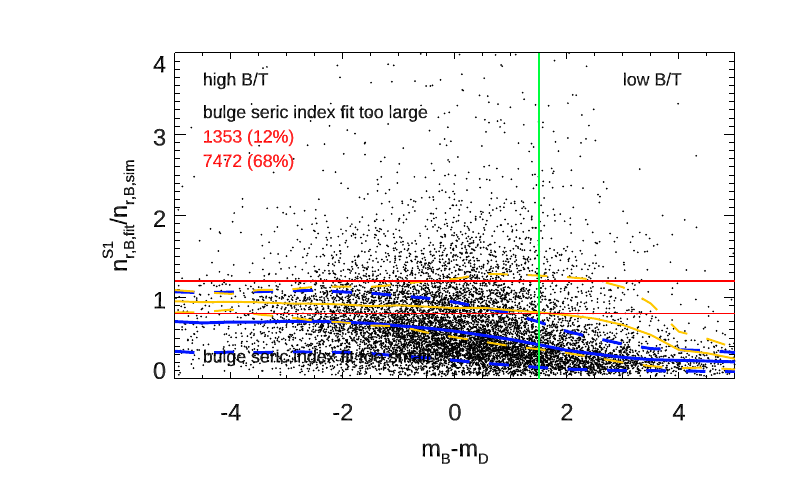}
\caption{For objects with $m_{\rm r}<19.8$, we show the deviation of fitting to simulated bulge \sersic index $n_{\rm D,fit}/n_{\rm D,sim}$ as a function of the faintness of the disk (Top: multi-band fit. Bottom: single-band fit). The fainter a bulge within a given disk, the harder it is to fit. Horizontal lines show a (randomly chosen) allowed deviation of $\rm \pm 20\%$.
}
\label{fig_bulge_n_acc}
\end{center}
\end{figure}

In \S \ref{sec_sample}, we have mentioned that we only believe the fainter of the 2 galaxy components if it is less than 1.5 magnitudes fainter than the brighter component ($\delta m_{\rm r}<1.5$).
The reason for this becomes obvious in Figures \ref{fig_disk_mag_acc} to \ref{fig_bulge_n_acc}.
Figure \ref{fig_disk_mag_acc} shows the deviation of the disk $r$-band magnitude as a function of the difference between disk and bulge magnitude (i.e. $B/T$ ratio) for the $\rm \sim11000$ galaxies that have successful fits in both multi-band and $r$-band fits, with a running mean and scatter over-plotted as blue lines.
Somewhat arbitrarily, we define $\rm \pm0.5~ $mag deviation from the true value as acceptable (red horizontal lines).
A clear trend is visible such that faint disks are badly fit, showing systematically brighter fit values.
This is understandable as they are embedded in a much brighter component and the fit compensates fitting residuals of this component.
The $\rm \pm0.5 $mag is reached when the disk is more than 1.5 magnitudes fainter than the bulge within a galaxy (blue median line crosses the red vertical line), slightly more for brighter galaxies (orange median line).
The red numbers in the top left corner indicate the numbers and fraction of objects with deviations larger than 0.5 mag above (top number) and below (lower number). 
However, it should be noted that these numbers should be taken with a grain of salt, as it is obvious that they are dominated by fits that we already know are bad and refer to relatively small deviations, which explains the high fraction of objects.
For comparison, in the bottom panel of this figure, we show the same plot for single-band fitting in the r-band.
There are a somewhat larger number of outliers (as indicated also by the numbers), and systematic deviations are typically a little larger in this case. 
The 0.5 magnitude deviation is reached at about 1 magnitude difference between bulge and disk.
This exercise can obviously be made in any of the bands used, and it is important to point out that the $r$-band is the deepest and best behaved of the single-band fits, some of the other band looking significantly different, especially the bands with shallower images, e.g. $uzJ$.
We chose the $r$-band here, as it serves as our main band throughout this work.

Figure~\ref{fig_bulge_mag_acc} shows the deviation in the fits to the simulated bulge magnitudes in the same way as Fig. \ref{fig_disk_mag_acc}.
The fraction of outliers is higher here as would be expected for bulges, and we find the same value of 1.5 magnitude difference between bulges and disks, above which their values are \textit{reliably} fit.
This limit restricts the objects in which we believe the fitting values for both components to targets with $0.2<B/T<0.8$, which incidentally are values also found in the literature, i.e. in that objects with $B/T>0.8$ are often called \lq pure spheroids\rq. 
There are other trends visible in these plots, which need to be explained.
Firstly, it can be seen that very pure/bright components, both in case of disk and bulge, are somewhat under-fit, i.e. the fit returns a fainter magnitude value.
This can be somewhat understood with the same argument used above.
If the fainter component takes away some of the flux of the brighter component, the brighter component will need to compensate and hence return a fainter magnitude.
Secondly, there are a large number of outliers in this plot.
Points and blue lines show all these galaxies with $m_{\rm r}<19.8$ to match the science sample discussed in \S \ref{sec_real}.
When restricting the sample to $m_{\rm r}<18$ (mean value and scatter shown as orange lines in all plots), both under-fitting and extreme outliers appear less frequently and the plots appear cleaner.
For these bright objects, a fainter limit of $\delta m_{\rm r} \sim 2$mag could in principle be used in multi-band fits, but it is important to point out that several of the brightness bins contain only a few objects.
However, for multi-band fits, this improvement is larger than for single-band fits.
Additionally, as bulges have been simulated with a range of \sersic indices, a trend with bulge \sersic index is to be expected.
High-$n$ objects are generally harder to fit (see H07 and H13), so one would assume the same to be true for bulges embedded in a different object.
Indeed, when restricting the plot to high-$n$ objects, trends are emphasised in all bulge plots, in that offsets are generally larger for high-$n$ bulges.

Figures \ref{fig_disk_re_acc} and \ref{fig_bulge_re_acc} show the deviations in the fits to the disk and bulge sizes respectively, and show that neither multi-band or single-band fits display large systematic trends for galaxy disks. 
Red horizontal lines indicate a $\pm20\%$ deviation, red numbers indicate the numbers and fractions of objects above and below these lines.
Deviations are somewhat more pronounced in single-band fitting of disks, but the bigger difference can again be seen in the scatter of the distribution.
For bulges in Fig. \ref{fig_bulge_re_acc}, the same effects can be seen, although both fits seem to fit bright \textit{pure} spheroids too small, which is true even in bright object.
Even a limit of more than $\rm \sim 1.5$mag would work for both components when large, bright samples are being used.

For the bulge \sersic-indices, however, this is not true, as can be seen in Fig. \ref{fig_bulge_n_acc}.
Both multi-band and single-band fits have trouble recovering the \sersic-index if the bulge is too faint, as one would also naively expect.
In multi-band fits, $\rm \sim20\%$ accuracy is reached at most $\delta m_{\rm r}$ values and about constant until $\delta m_{\rm r} = 0$, i.e. $B/T=0.5$.
However at $\delta m_{\rm r} = 0$mag the single-band \sersic indices are already badly underestimated by about 50\%.
In general, fainter bulges become more and more disk-like, as it becomes more and more likely that the bulge profile fits small residuals of the galaxy disk instead of the actual bulge.
There is also an effect at very bright bulges in that \sersic-indices are already $\rm \sim20\%$ under-estimated.
This effect is enhanced to $\rm \sim30\%$ in single-band fits. 
We will confirm these trends in \S \ref{sec_sim_results_n}.
Again, these trends weaken significantly when the samples are restricted to brighter galaxy samples (orange lines), where the S/N in the individual components is also higher and multi-band fits can still recover $n_B$ at $\rm \lesssim10\%$ at $\delta m_{\rm r} = 1.5$ ($\rm \sim20\%$ for single-band fits).
As was expected, and as has been reported by other authors before, parameters of high-n objects indeed seem harder to recover accurately.
The \sersic index is the hardest parameter to fit, but its known degeneracies with galaxy sizes and magnitudes mean that those parameters are also less accurate than in the case of galaxy disks.

Over all, we can establish from these plots that a magnitude limit for the fainter component has to be used and values are generally recovered within 20\% (and 0.5mag) when limiting the samples to those components with $\delta m_{\rm r} \leq1.5$mag. 

\subsection{Results}
\label{sec_sim_results}
In this section, we discuss the results from the simulations and will establish how well we can recover the $B/D$ parameters that we put into the simulated images using both single-band and multi-band fitting.
Throughout this section, unless stated otherwise or trivially visible from the axis range, we restrict ourselves to show results of galaxies at $m_{\rm r}<19.5$ in order to match the sample of real galaxies (see \S \ref{sec_real_results}) for which \gama provides spectroscopic redshifts at $m_{\rm r}<19.8$, and to avoid very faint galaxies where analysis suffers badly from low-number statistics of \lq successful fits\rq in several single-band fits. 
Where a different/brighter sample cut is being used, this will be indicated.
Many of our conclusions still hold qualitatively for fainter galaxies, i.e. the advantage of multi-band over single-band in recovering the profile parameters and of the sample size being available for a scientific analysis (see numbers in Tables \ref{tab_sim_numbers_all} and \ref{tab_sim_numbers_bright}) are expected to become even more pronounced. 
However, we do not present them in this work in detail.
We highly recommend to any user of \galapagostwo and/or \galfitm to run similar tests on their own data in order to establish the accuracies and issues, especially on faint galaxies.

We also only examine the \lq pure\rq\ fitting modes and what we called \verb+Mode_M+ (multi-band fitting with object detection in a co-added image) and \verb+Mode_S1+ (single-band fitting with object detection in the according single-band image) in H13.
We do not discuss \verb+Mode_S2+ fitting (single-band fitting with object detection in a co-added image), as this was already shown to be counter-productive as many galaxies are detected in the co-added (and hence deeper) image, but are too faint to be sensibly fit in single-band images, confusing the single-band fits even of the visible targets.

\begin{figure*}
\begin{center}
\includegraphics[width=0.48\textwidth,trim=23 30 8 15, clip]{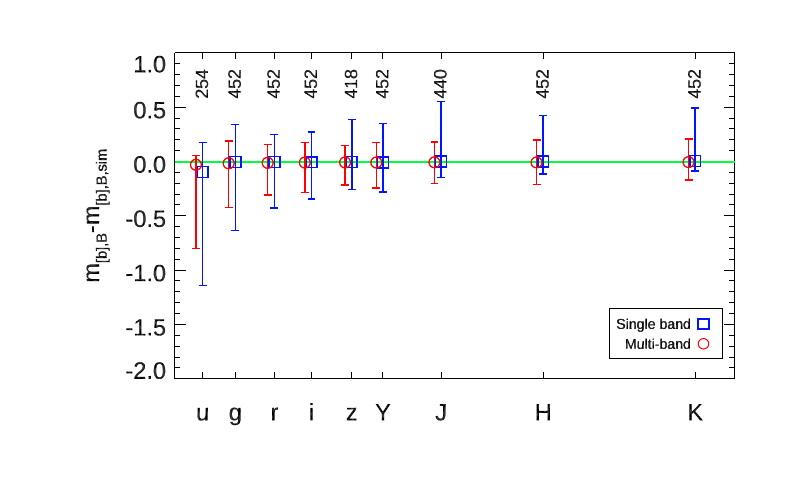}
\includegraphics[width=0.48\textwidth,trim=23 30 8 15, clip]{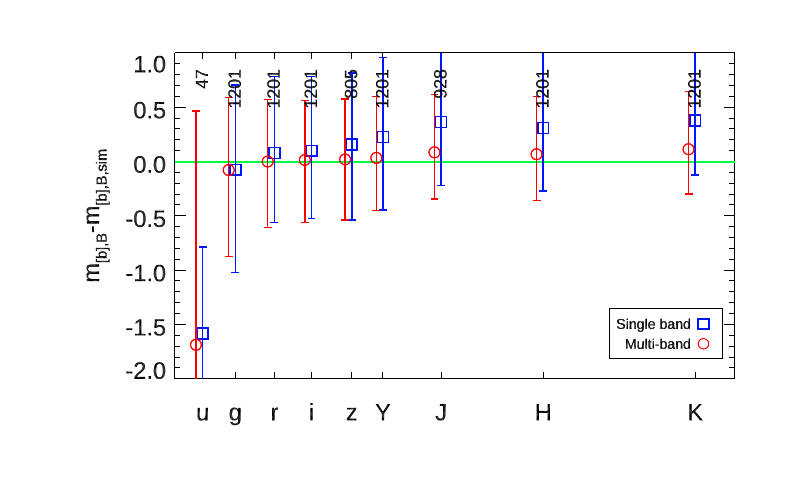}
\includegraphics[width=0.48\textwidth,trim=23 5 8 15,clip]{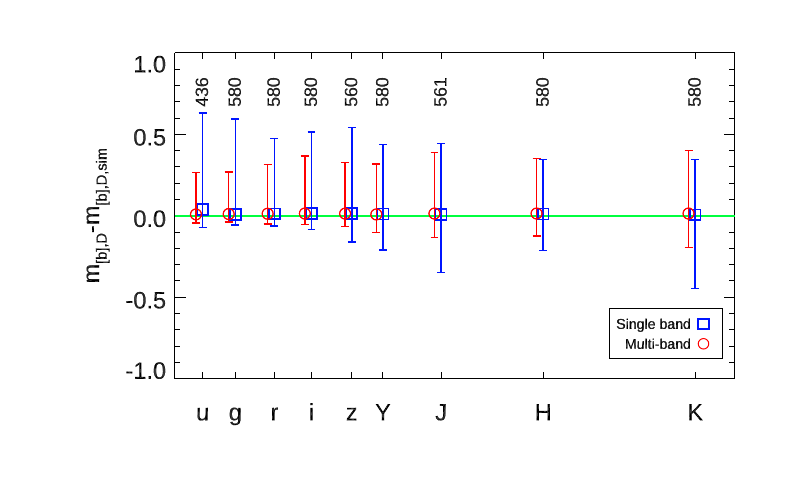}
\includegraphics[width=0.48\textwidth,trim=23 5 8 15,clip]{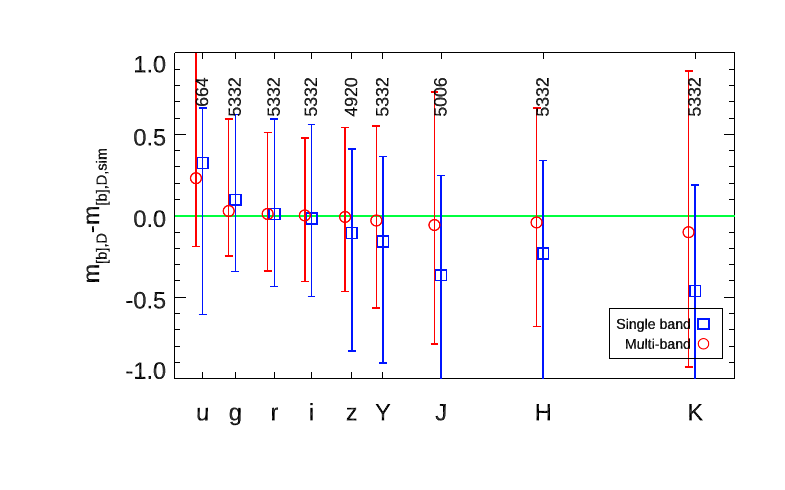}
\caption{Recovery of magnitude values for bulges (top) and disks (bottom) as a function of wavelength at each band $[b]$, for a bright ($m_{\rm r,B+D,sim}<17.5$mag, left) and faint ($18.5<m_{\rm r,B+D,sim}<19.5$mag, right) sample. Symbols show median values, errorbars indicate 16 and 84 percentiles, respectively. Numbers at the top of the panels indicate the sample size for each data point, e.g. the 580 bright disk galaxies for which both single-band (in all of $griYHK$-band) and multi-band fitting produces a \lq good\rq\ fit, and in which the disk is bright enough ($m_{\rm D} < m_{\rm B}+1.5$). For  $uzJ$-bands we plot the biggest possible subsample of these objects. Deviations towards the top indicate that a component is fit too faint. Note the different scales on the Y-axis for bulges and disks.
}
\label{fig_sim_mag_comp}
\end{center}
\end{figure*}

\subsubsection{Component magnitudes}
\label{sec_res_sim_mag}
In Figure \ref{fig_sim_mag_comp}, we show how well magnitudes can be recovered using single-band and multi-band fitting for bulges (top) and disks (bottom), respectively.
The left plots show the analysis of bright galaxies at $m_{\rm r,B+D,sim}<17.5$mag from their simulated magnitudes, the right column for faint objects at $18.5<m_{\rm r,B+D,sim}<19.5$mag.
The plots show the magnitude deviation compared to the simulated value of the multi-band fit in red and the single-band fits in the respective band in blue.
This forms the most basic magnitude comparison for each of the components and is equivalent to the way results were presented in H13, where errorbars represented symmetric standard deviations.
In this work, we use 16 and 84 percentiles in most plots to indicate not only the errors of the fit, but simultaneously allow an estimation of how symmetric the distributions are.

Several effects are visible in this plot.
Firstly, for most bands, it becomes apparent that multi-band indeed does improve the recovery of magnitude values, as expected.
The user should be reminded that the magnitudes in our multi-band fitting are not constrained by any assumption on a polynomial, the improvement seen here is a result of constraining the profile shape itself, i.e. the other fit parameters follow certain functional forms.
We already reported in H13 that the improvements of fitting performance on magnitudes is not as strong as on other parameters for this reason.
Secondly, it becomes apparent how asymmetric some of the distributions are.
Both single-band and multi-band fits tend to fit disks too faint, and bulges too bright, but this effect is much more pronounced in the single-band fits, as is visible from the larger, and asymmetric dispersion of the objects.
While neither of the methods show large systematic offsets for these bright galaxies (left) and recovers bulge and disk magnitudes -- on average -- nearly perfectly (offsets generally below 0.05 mag), the multi-band fits recover the magnitudes of bulges and disks more accurately at all wavelengths, showing smaller error bars, highlighting the advantage of multi-band fits.
This is especially true for the bulge component, which is known to be harder to fit.

In the right column of Fig \ref{fig_sim_mag_comp}, we show the same plot, but for fainter galaxies with $18.5<m_{\rm r,B+D,sim}<19.5mag.$. 
As expected, at these magnitudes fitting uncertainties are in general much larger, indicated by the larger error bars.
It can also be seen that especially single-band fitting returns values with significant systematic/median offsets.
Systematic offsets in multi-band fits are much smaller where present, with systematic offsets of the bulge/disk magnitudes below \textasciitilde0.1 mag, but larger dispersion.
There is, however, a significant effect in the $u$-band of the bulge magnitudes (top right panel), where a substantial systematic offset can be measured for both multi-band and single-band fits. 
We emphasise here that the very small number of bulges with \lq good fits\rq\ in galaxies at these magnitudes and number statistics are a problem.
This is largely due to the $u$-band single-band fit itself being unsuccessful in providing a fit value, as the image is much shallower compared to the images in other bands (compare detection numbers in Table \ref{tab_sim_numbers_all}).
Given the SEDs used in the simulations, the $B/T$ ratio in the $u$-band is also significantly smaller than the $B/T$ ratio in the $r$-band which we use to define out samples, increasing the difficulty to derive a good value, and resulting in many of the bulges of these galaxies indeed being so faint that neither single-band nor multi-band fits can recover their magnitudes accurately.
The opposite effect, albeit smaller, can be seen in the disk magnitudes (bottom right panel).
As the offsets seen in the single-band fits go in different directions in different bands, they become especially critical when using colours in a scientific analysis, as colours derived from single-band fitting would be especially affected.
Disks are generally still fit too faint, bulges too bright (exception: $u$-band), and this effect is strongly enhanced for these faint objects.
This has a catastrophic effect when $B/T$ ratios in a specific band are required (see Figure \ref{fig_sim_BT_comparison}).

\begin{figure*}
\begin{center}
\includegraphics[width=0.48\textwidth,trim=23 30 20 0,clip]{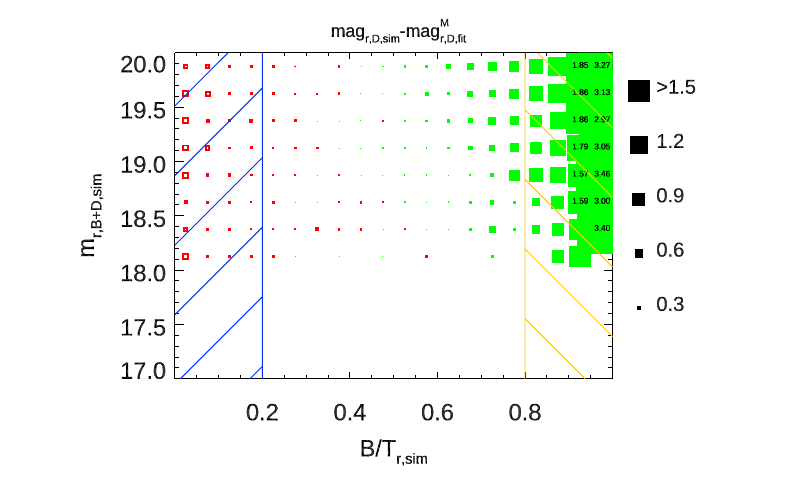}
\includegraphics[width=0.48\textwidth,trim=23 30 20 0, clip]{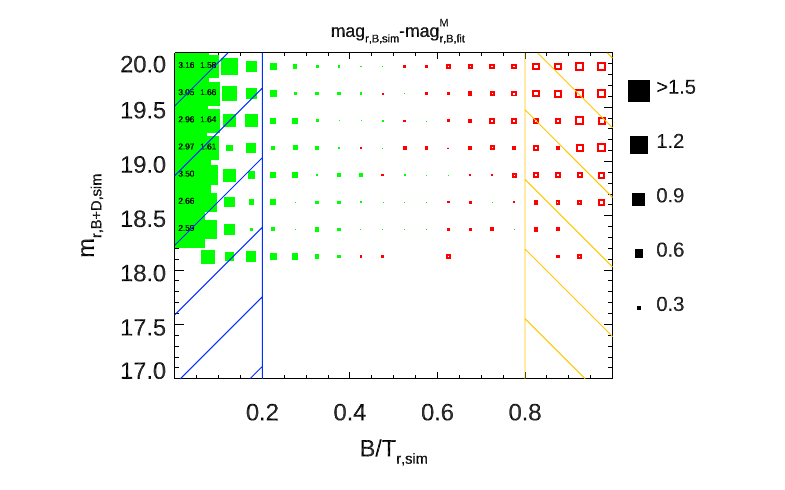}
\includegraphics[width=0.48\textwidth,trim=23 5 20 0, clip]{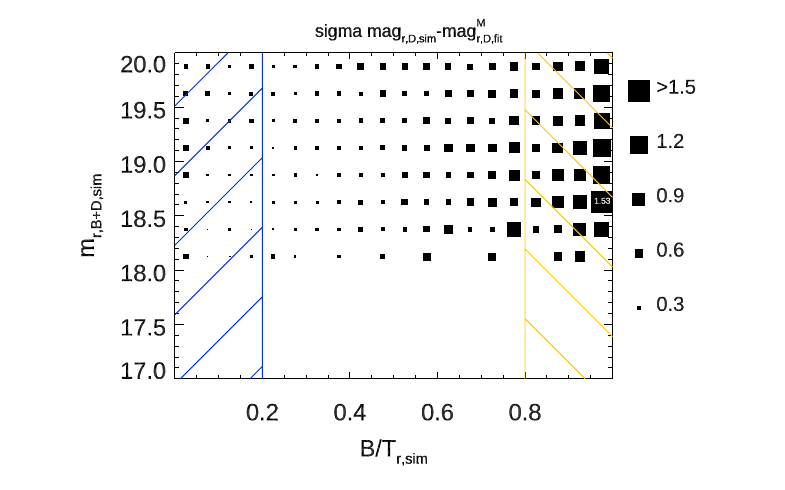}
\includegraphics[width=0.48\textwidth,trim=23 5 20 0, clip]{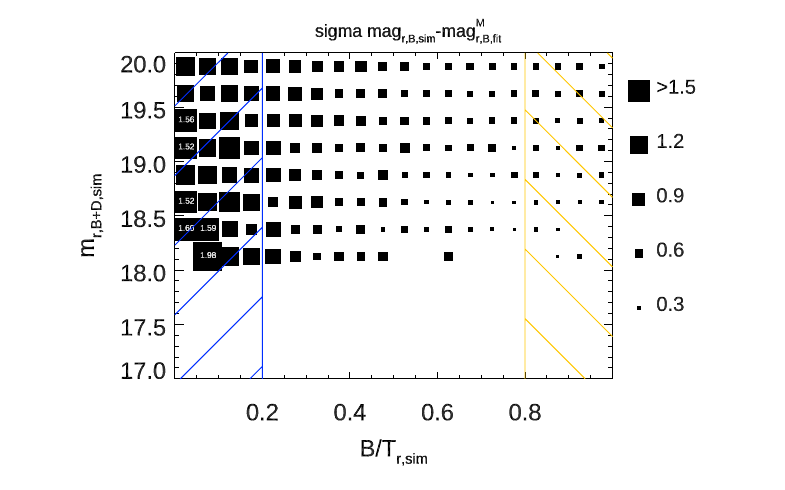}
\caption{Top left: Disk magnitude, $m_{\rm r,D,sim}-m_{\rm r,D,fit}$ average offsets in multi-band fits as a function of $B/T$ and $m_{\rm r,B+D,sim}$. 
The symbol size indicates the magnitude of the offset according to the legend on the right, where small symbols indicate small offsets and large symbols indicate large offsets, with the solid green and open red squares reflecting positive and negative values respectively.
Bottom left: standard deviation of the same value (smaller symbols represent less scatter). As discussed in the text, the $m_{\rm r,D}$ can be well recovered as long as $(B/T)_{\rm r,sim}<0.8$.
In both plots, the blue hashed area indicates galaxies with $(B/T)_{\rm r,sim}<0.2$, i.e. \lq pure disks\rq, and the orange hashed area indicates galaxies with $(B/T)_{\rm r,sim}>0.8$, i.e. \lq pure spheroids\rq.
Right column: Same plots for bulge magnitudes, $m_{\rm r,B,sim}-m_{\rm r,B,fit}$. These can be well recovered as long as $(B/T)_{\rm r,sim}>0.2$.
Small numbers indicate the actual value in those bins with a systematic offset larger than 1.5 magnitudes.
We only show bins with more than 20 objects in this and all similar plots, to assure reasonable number statistics in each bin.
}
\label{fig_sim_mendel_mag}
\end{center}
\end{figure*}

We can develop a more detailed picture of recovering the disk and bulge magnitudes in Fig. \ref{fig_sim_mendel_mag}, where we show the deviations of the respective values from the simulated values as a function of $(B/T)_{\rm sim}$ and $m_{\rm r,sim}$ for multi-band fits, in order to understand their performance in more detail.
In each bin of this 2D plane, we show how well the component magnitude is recovered on average via a 2 sigma clipped robust median, with larger squares reflecting larger differences between the magnitudes derived from the fits and those used to simulate the galaxies. 
Please note that in order to ensure reasonable number statistics in each bin, we only plot bins which contain more than 20 objects.
Green solid squares in these plots indicate that the magnitude is under-estimated, i.e. the component is fit too bright, while red open squares indicate that the fit to the component is on average too faint.
We show the scatter in the distributions in the respective bottom panel.
In each panel, the hashed areas indicate the area in which we would not trust the fit of the fainter components, and we consider objects \lq pure disks\rq\ and \lq pure spheroids\rq.
In these areas, we would suggest the use of the single-\sersic fit values instead, as we will argue in the following sections.
We will show below why even the fit to the brighter component is not ideal in the $B/D$ fits in these areas.

The first effect visible in these plots is that, as expected, the fainter component in the extreme areas can indeed not be recovered well.
Disks (left panels) are recovered with large systematic offsets in objects with high $B/T$ ratios, bulges show especially large systematic offsets in objects with low $B/T$ ratios.
In both cases, the fainter component is recovered too bright by more than 1 magnitude, 3 magnitudes in extreme cases.
The same effect can be seen in the lower panels that show the scatter of the recovered values, the values are highest in the area where the component is faintest.
This shows that indeed the fits of these fainter components are not accurate and reliable.
Similar to the analysis in Section \ref{sec_sim_samples}, these plots suggest that the choice of $\rm \pm 1.5$mag between the components is a good choice, corresponding to $B/T<0.2$ and $B/T>0.8$.
In between these values, both bulge and disks magnitude seem to be recovered relatively well.
There is also a (weak) trend with galaxy brightness, in that fainter galaxies are harder to separate, as one would expect.
This is especially true in the case of galaxy disks and will be more pronounced in other parameters.

However, there is a second effect, which is somewhat unexpected at first glance.
Disks in \lq pure disk\rq\ objects and bulges in \lq pure spheroidal\rq\ objects are also fit less well, and show systematic offsets to fainter magnitudes.
This might come as somewhat of a surprise, as one would naively assume that e.g. a disk in a \lq pure disk\rq\ galaxy should be easy to fit.
However, this effect can be explained by the fitting process itself and is, in fact, a consequence of the first effect discussed above.
We are forcing a $B/D$ fit on an object that is effectively a 1-component model with a smooth \sersic profile and no internal structure.
With the constraints on object positions etc, this makes it likely that the fainter component fits some flux of the brighter component. 
This is indeed the effect described and seen above in that disks in \lq pure spheroidal\rq\ galaxies are over estimated in brightness.
However, as a result, in the bulge, this flux is missing such that the bulge brightness would be underestimated, which is the effect seen here.
This forced 2-component fit also affects the galaxy size and \sersic index, which we will discuss further in Figs.~\ref{fig_sim_mendel_re_bd} and \ref{fig_sim_mendel_n_b_v}.

\begin{figure}
\begin{center}
\includegraphics[width=0.48\textwidth,trim=23 30 8 15, clip]{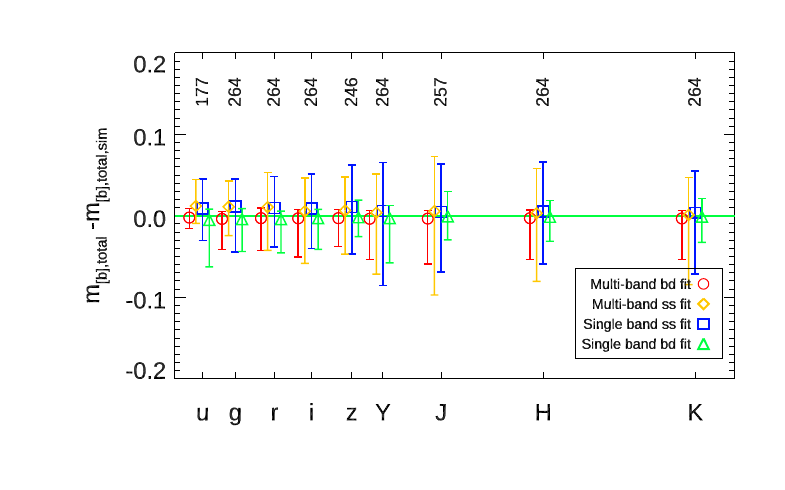}
\includegraphics[width=0.48\textwidth,trim=23 5 8 15,clip]{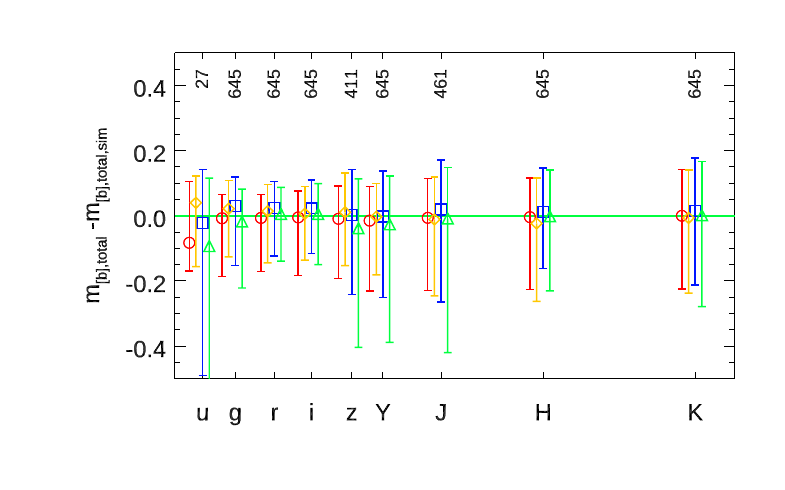}
\caption{Comparison of total magnitude values in each band $[b]$ from both SS and $B/D$ fits for single-\sersic and $B/D$ fits at $m_{\rm r,sim}<17.5$mag (top) and $18.5<m_{\rm r,sim}<19.5$mag (bottom). Please note the different ranges on the y-axis. Offsets to the top indicate fits being too faint. Objects used in the analysis are the maximally large sample with $griYHK$ fits with $B/D$ fits (additionally $u$,$z$ and $J$ in those bands respectively) and the single-\sersic fits of each band. 
}
\label{fig_sim_magbd_comparison}
\end{center}
\end{figure}

\subsubsection{Total magnitudes}
While in these objects with extreme $B/T$ ratios the flux of the brighter component is under-estimated and that of the fainter component over-estimated, these 2 effects cancel out nearly completely, as we analyse in Figure \ref{fig_sim_magbd_comparison}, where we show how well both single \sersic profiles and $B/D$  fits in both single- and multi-band fits can recover the \textit{total} magnitude of a galaxy. 
For each band, we show a group of 4 values, best visible in the reddest bands, as they are more spread out.
The inner 2 (orange and blue) points show the recovery of single-\sersic fits of the total magnitude of a $B/D$ object in each band using multi- and single-band fits respectively.
The outer 2 (red and green) data points show the same values for the $B/D$ fits in multi- and single-band fits, respectively.
For bright objects (upper panel) all 4 methods measure the total magnitude of a galaxy accurately (please note the different ranges on the Y-axis in comparison to the previous plots) and with small dispersion.
This behaviour can be understood by the nature of fitting techniques, in that any residual from the one component would be minimised by the other component, i.e. balancing out any error in the estimation of the total flux.
However, while offsets in recovering the total magnitude are in general small, they are minimised in multi-band fitting, especially in the low S/N bands, e.g. $u$.
The difference in the scatter -- multi-band showing asymmetric errorbars towards brighter magnitudes, while single-band show more symmetric errors -- is interesting, but has minimal effect on the median values.
For faint galaxies (bottom panel) this effect is enhanced and one can see that single-band fits show larger systematic offsets and larger dispersion, especially in case of $B/D$ fits (please note the low number statistics).
Multi-band fits recover total magnitudes more accurately, although small systematic offsets can still be seen.

It is also visible from the errorbars shown in this plot that if one only requires a total magnitude of faint galaxies -- instead of bulge and disk magnitudes separately -- single-\sersic (multi-band) fits seemingly provide equally good measurements, although the top panel suggests that for bright galaxies, a $B/D$ fit actually recovers the total flux of these objects better.
This is possibly an effect of a single-\sersic profile not being able to perfectly describe the profile of these 2-component objects.

\begin{figure*}
\begin{center}
\includegraphics[width=0.48\textwidth,trim=23 30 20 0,clip]{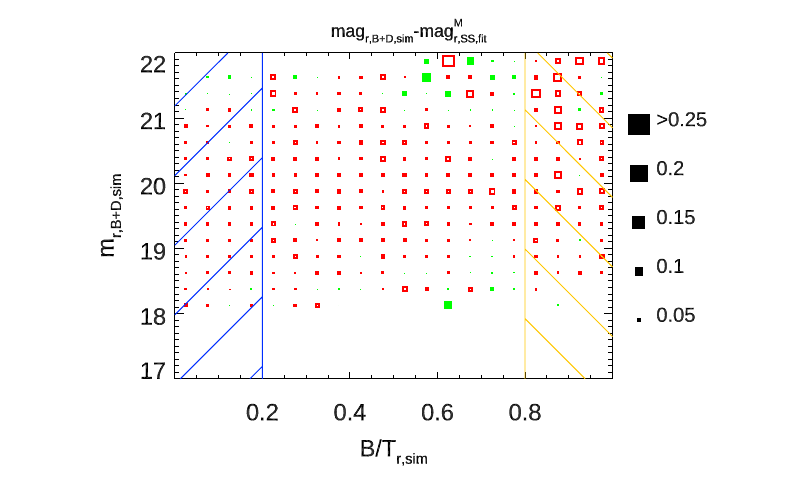}
\includegraphics[width=0.48\textwidth,trim=23 30 20 0,clip]{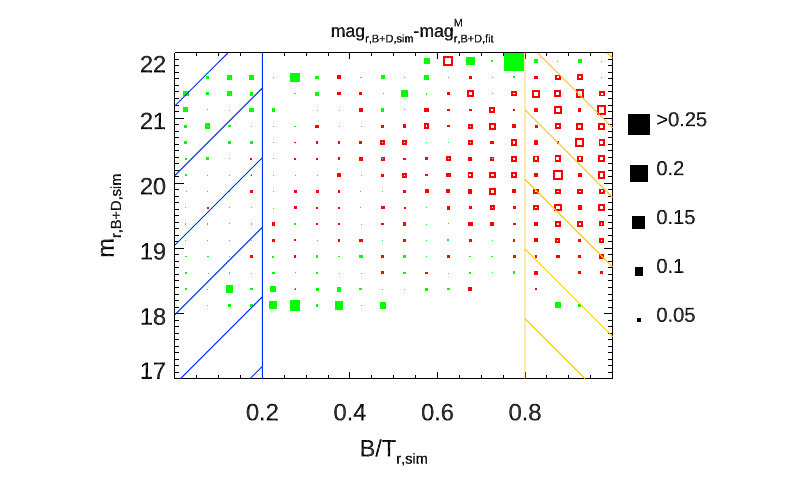}
\includegraphics[width=0.48\textwidth,trim=23 5 20 0, clip]{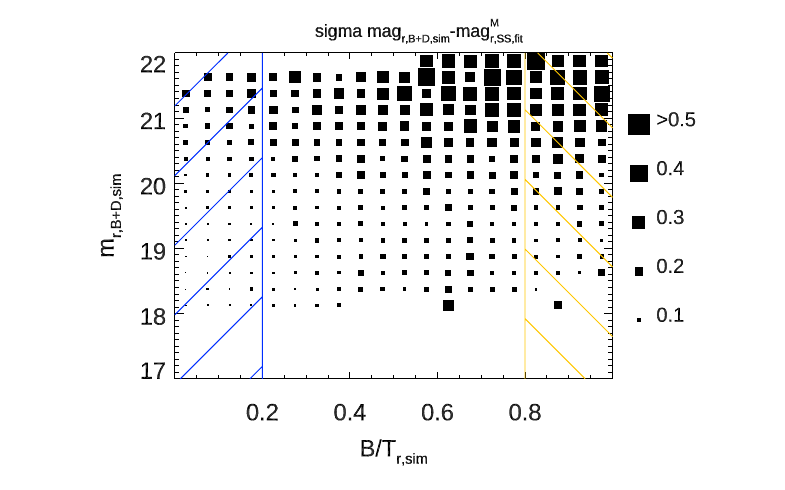}\includegraphics[width=0.48\textwidth,trim=23 5 20 0, clip]{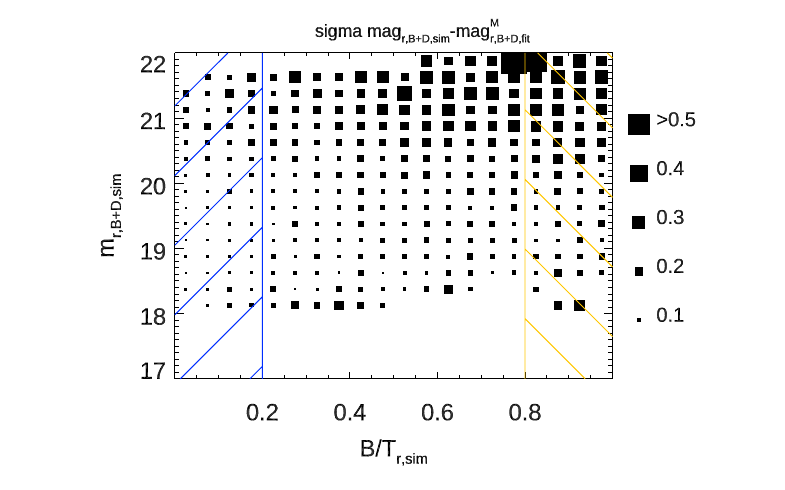}
\caption{Top left: Average offsets of magnitudes of single-\sersic fits, $m_{\rm r,B+D,sim}-m_{\rm r,SS,fit}$ in multi-band fits as a function of $B/T$ and $m_{\rm r,B+D,sim}$ (ideal fits produce small symbols). Bottom left: standard deviation of the same value (small symbols represent good fits and small scatter). Right column: Same plots, but from $B/D$ fits, for the combined magnitude $m_{\rm r,B+D,sim}=m_{\rm r,D,sim}+m_{\rm r,B,sim}$. In general, the total magnitudes of the objects can be well recovered by both single-\sersic and $B/D$ fits, however the scatter increases towards fainter objects and objects with higher $B/T$ values, which are known to be harder to fit than disk-like systems.
Only bins with more than 20 objects are shown, to assure reasonable number statistics in each bin.
}
\label{fig_sim_mendel_mag_ss}
\end{center}
\end{figure*}

However, in order to get a better insight into the systematic effects at work, we can look at the recovery of the total magnitude as a function of $B/T$ ratio and galaxy magnitude, as shown in Fig. \ref{fig_sim_mendel_mag_ss}.
This plot includes galaxies at $m_{\rm r,sim}>19.8$, i.e. our entire dataset.
Both fits, single-\sersic (left) and $B/D$ (right), recover the total galaxy magnitudes well at all magnitudes and $B/T$.
While the scatter/uncertainty (bottom panels) clearly increases for both fainter and more \lq bulge-dominated\rq\ objects in both fits, as one would naively expect, most values for systematic offsets (top panels) are well below 0.1 magnitudes.

From Fig. \ref{fig_sim_mendel_mag_ss}, it is somewhat unclear whether faint galaxies are indeed better fit with single-\sersic models, as was suggested by Fig. \ref{fig_sim_magbd_comparison}.
It seems that the situation is somewhat more complicated than that, and $B/D$ fits in fact give -- on average -- a more accurate total magnitude (less scatter, smaller boxes in bottom panels), depending in detail on the galaxies investigated.
This seems to be specifically true for galaxies fainter than $m_r\sim19.5$ for which all fits produce bad results much more frequently.

At this point, a brief comparison with \citet[][Appendix B]{Mendel2014}, where they show equivalent plots for their catalogues, highlights the strength of the \galapagostwo/\galfitm approach. 
On \sdss data, they reported typical errors on the order of \textasciitilde0.1 mag on recovering the total magnitude of an object for both single-\sersic models and $B/D$ fits at $m_{\rm r}\sim17.5$mag.
In this work, we use the same \sdss data in combination with additional \ukidss data, and we derive total magnitudes with better accuracy at much fainter magnitudes, e.g. \textasciitilde0.05 mag at around $m_{\rm r}\sim21$mag in either method.

For bulge and disk magnitudes, \citet{Mendel2014} see similar effects to our work in that the fainter component is systematically fit badly once it becomes too faint.
They report typical disk uncertainties of >1 magnitudes for all galaxies at $B/T>0.9$.
However, it should be pointed out that for the other extreme, i.e. the brighter component, their method seems to do a better job than what we see in this work, with barely any systematic offset.

Unfortunately, \citet{Mendel2014} only analyse objects at $m_{\rm r}<17.7$mag, resulting in only minimal overlap with this work.
For the objects analysed by both works, we attempt a more direct comparison in \S \ref{sec_real_simard}.
From the brief comparison here and the accuracies cited above, it seems that with our analysis we can indeed push $B/D$ fits several magnitudes fainter at the same accuracy and scatter.

\begin{figure*}
\begin{center}
\includegraphics[width=0.48\textwidth,trim=23 30 8 15, clip]{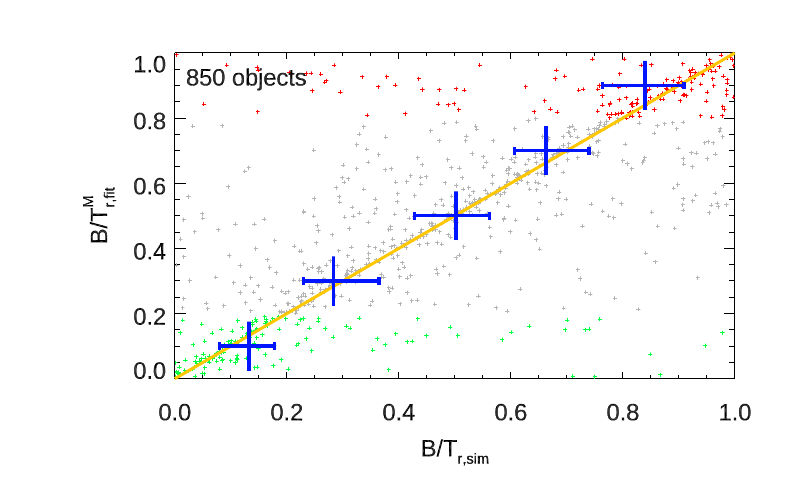}
\includegraphics[width=0.48\textwidth,trim=23 30 8 15, clip]{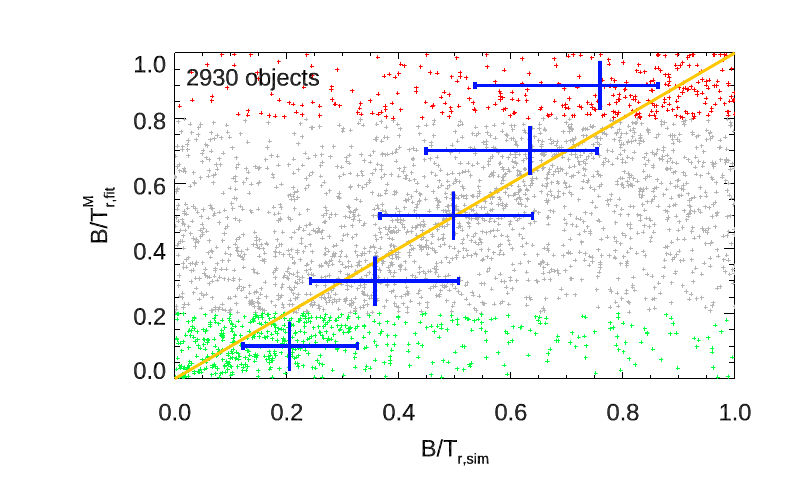}
\includegraphics[width=0.48\textwidth,trim=23 5 8 15, clip]{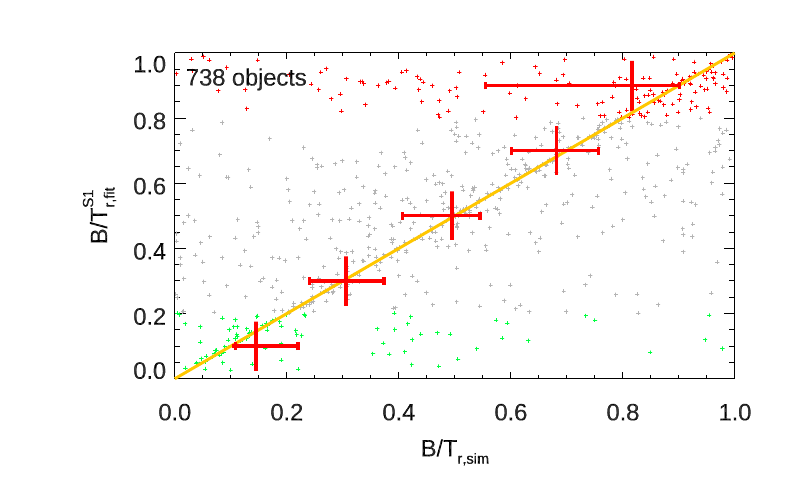}
\includegraphics[width=0.48\textwidth,trim=23 5 8 15, clip]{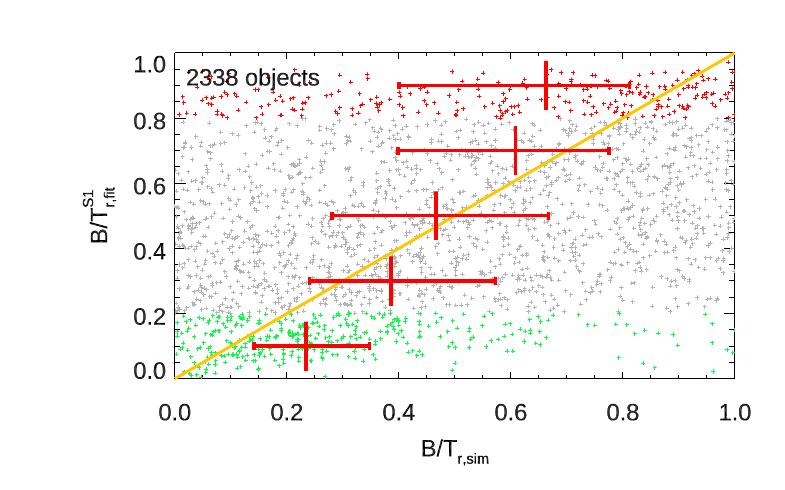}
\caption{Comparison of input (x-axis) and output (y-axis)$B/T$ ratios for multi-band fits (top) and single-band fits (bottom), for bright galaxies $m_{\rm r,B+D,sim}<17.5mag$ (left) and faint galaxies $18.5mag<m_{\rm r,B+D,sim}<19mag$ (right). 
A 1-to-1 line has been over-plotted to guide the eye. The errorbars shown represent median values and 30/70 percentiles in each bin of 0.2 width in $(B/T)_{\rm r,fit}$.
}
\label{fig_sim_BT_comparison}
\end{center}
\end{figure*}

\subsubsection{Bulge/Total ratios}
Finally, there are 2 more important parameters in context of measuring galaxy fluxes/magnitudes other than the component magnitudes themselves: Colours and $B/T$ ratios, both of which are widely used in the community to \lq classify\rq\ galaxies.
$B/T$ ratios are probably the value that is most desired when running $B/D$ decompositions of galaxies.
Unfortunately, defined as a ratio, they are very sensitive to the fluxes in the 2 components and hence hard to measure accurately.
For example, in a $B/T=0.5$ galaxy, distributing the flux between the components wrongly by even 0.2 magnitudes \footnote{a conservative estimate even for bright galaxies, as shown in the left columns of Figure \ref{fig_sim_mag_comp}. Individual galaxies can have mis-measured fluxes well above this limit} changes the $B/T$ ratio by 0.1.
We illustrate this in Figure \ref{fig_sim_BT_comparison}, where we show the input $(B/T)_{\rm r,sim}$ and output $(B/T)_{\rm r,fit}$ ratios of the simulated galaxies.
Objects at $(B/T)_{\rm r,fit}<0.2$, and $(B/T)_{\rm r,fit}>0.8$ are plotted in green and red, respectively, other objects at intermediate $B/T$ in grey.
For bright objects (left panels), both multi-band (top) and single-band fitting (bottom) do a similar job at recovering $B/T$ ratios, although the 1-to-1 correlation is more pronounced in the multi-band fits which also return a result more often, leading to a larger samples size, as indicated in the top left of the panels.
For each bin in $(B/T)_{\rm r,fit}$ of 0.2 width, as one would derive them on real data, we calculate the median $(B/T)_{\rm r,sim}$ and the 30/70 percentiles as an indication of the width of the distribution for both multi-band fits in blue and single-band fits in red, respectively.
While both methods show large scatter, neither method shows any strong systematic offsets.
However, this is different when looking at fainter galaxies (right panels, $18.5mag<m_{\rm r,B+D,sim}<19mag$), where separation of 2 components is naturally harder.
The scatter in both single-band and multi-band fits increases dramatically, measured $B/T$ ratios become very uncertain.
While multi-band mean values are not very accurate on a individual basis, however, their average values are less biased than the values derived from  single-band fitting and actually recover the 1-to-1 line reasonably well.
The average values of the single-band fits are closer to the centre of the plot, indicating a random distribution of $B/T$ values.
For fainter galaxies, the recovery of the $B/T$ ratio becomes even more challenging and the values less reliable.

\begin{figure}
\begin{center}
\includegraphics[width=0.48\textwidth,trim=23 5 20 0,clip]{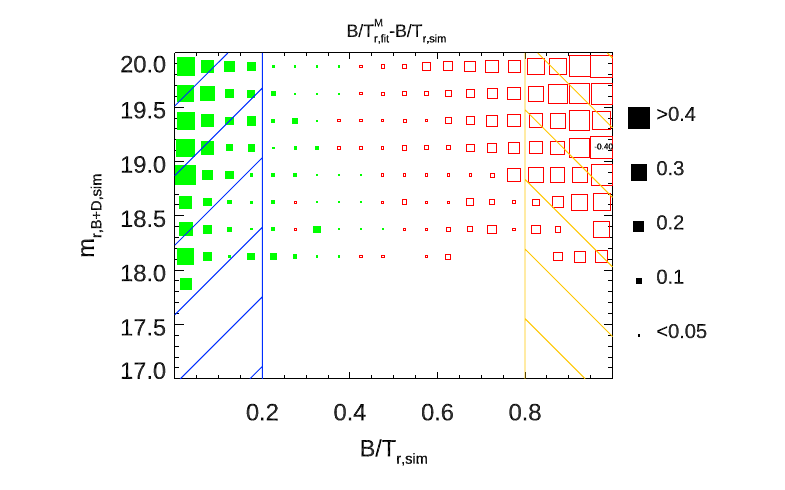}
\caption{Average difference of recovered and true Bulge/Total ratios $(B/T)_{\rm r,fit}-(B/T)_{\rm r,sim}$ measured in multi-band fits as a function of $(B/T)_{\rm sim}$ and $m_{\rm r,B+D,sim}$. Sizes of the symbol represent the average offsets between measured and true $B/T$ measured for a given sample, i.e. in an ideal fit would show no offset (small boxes) over the entire plot range. In the range $0.2<(B/T)_{\rm sim}<0.8$, the measured values agree relatively well with the simulated values. At $(B/T)_{\rm sim}<0.2$ and $(B/T)_{\rm sim}>0.8$, however, the measured values are systematically biased towards more intermediate values.
Only bins with more than 20 objects are shown, to assure reasonable number statistics in each bin.
}
\label{fig_sim_mendel_mag_BT}
\end{center}
\end{figure}

We look at the recovered $B/T$ as a function of $(B/T)_{\rm r,sim}$ and $m_{\rm r,B+D,sim}$ in Figure \ref{fig_sim_mendel_mag_BT}, where we plot the difference $(B/T)_{\rm r,fit}-(B/T)_{\rm r,sim}$.
Green, filled boxes indicate positive values ($B/T$ is overestimated), red, empty boxes indicate negative values, the symbol size represents the amount for which the values are biased. 
A clear trend is visible, in that  objects with low $(B/T)_{\rm r,sim}$ get overestimated in their $(B/T)_{\rm r,fit}$, objects with high $(B/T)_{\rm r,sim}$ get underestimated. 
This trend is, of course, by design, as it's impossible to overestimate the $(B/T)$ of a galaxy with $(B/T)_{\rm r,sim}\sim1$.
However, as we can see at intermediate $(B/T)_{\rm r,sim}$ values, the recovered $B/T$ ratios are in general measured within 0.2 of the true values over a wide range of magnitudes and $0.2<B/T<0.8$.
Measured $B/T$ values in more extreme galaxies are indeed less well recovered, which we will try to explain below.
Please note that the lack of objects at $m_{\rm r,B+D,sim}<18$ (while Fig \ref{fig_sim_BT_comparison} specifically uses objects at $m_{\rm r,B+D,sim}<17.5$) is artificial as we only plot bins with more than 20 objects and entirely ignore objects with $m_{\rm r,B+D,sim}<17$.

In Figures 3\&4 in V14b, we have already shown how well galaxy parameters can be measured in real galaxies that are artificially redshifted to larger redshifts, i.e. becoming smaller and fainter, and show a small shift in observed bands, which plays a minor role in this context).
We have found that multi-band fits are able to recover $B/T$ ratios more consistently over the redshift range tested.
In Figure 9 of the same paper we have further shown that the measured $B/T$ ratios of different galaxy types are harder and harder to measure with increasing (artificial) redshift, as expected.
This effect was visible in both multi-band and single-band fits, but was more pronounced and less smooth (with redshift) in single-band fits.
The results presented in this work underline these findings, but provide a more statistical analysis as the galaxy samples used are significantly larger and span a wider range of parameters. 
Further, as simulated galaxies have been used here, we can compare input and output values directly, rather than relying on the \lq smoothness\rq\ of trends on fitting and artificially redshifting real galaxies.

\subsubsection{Colours and SEDs}
The second important parameter -- and the most critical for many science cases -- that users might want to derive from $B/D$ fitting are colours -- or SEDs -- of the individual components.
When trying to measure e.g. stellar populations of bulges and disks, recovering the SEDs of the components -- at least on average -- becomes vital.

\begin{figure}
\begin{center}
\includegraphics[width=0.48\textwidth,trim=23 30 8 15, clip]{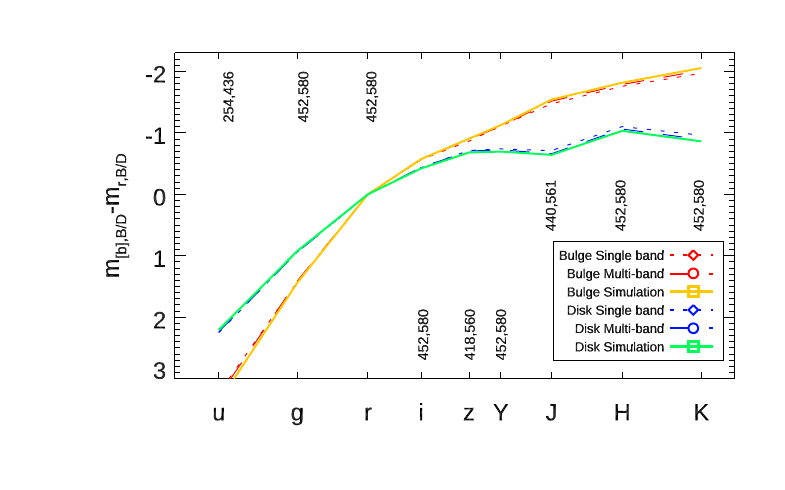}
\includegraphics[width=0.48\textwidth,trim=23 30 8 15, clip]{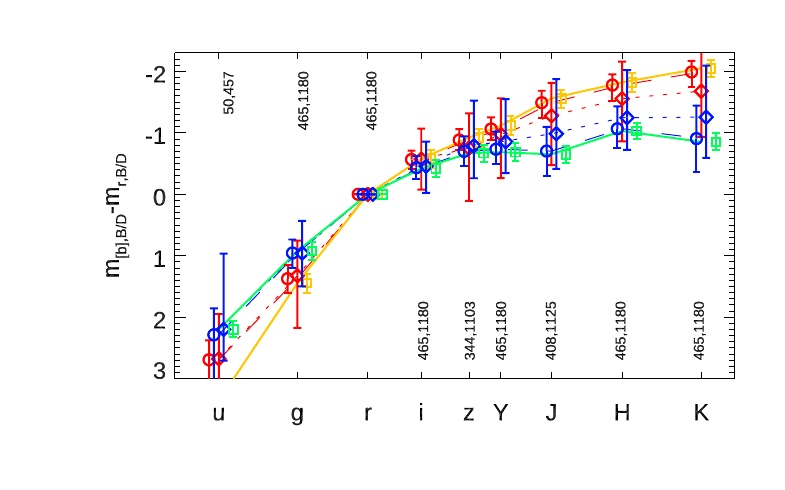}
\includegraphics[width=0.48\textwidth,trim=23 15 8 15, clip]{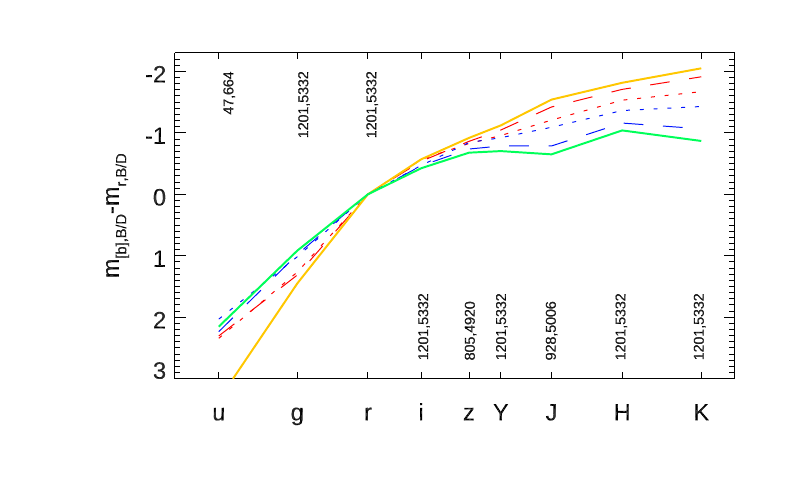}
\caption{Comparison of colours of individual galaxy components shown as an SED for galaxies of different brightness (normalized in the $r$-band). Top to bottom show the SEDs recovered for galaxies at $m_{\rm r,B+D,sim}<17.5$mag (top), $17.5<m_{\rm r,B+D,sim}<18.5$mag (middle) and $18.5<m_{\rm r,B+D,sim}<19.5$mag (bottom panel). Numbers at each wavelength indicate the number of objects used in the analysis at that wavelength for bulges and disks, respectively. In the middle panel, errorbars for all values are indicated, artificially offset to each other along the x-axis. Thick red (bulge) and blue (disk) error bars indicate multi-band fitting, thinner errorbars single-band fitting. Orange and green errorbars indicate the scatter in the simulated values for bulges and disks, respectively, for comparison. Objects used in this analysis are the biggest possible sample for each datapoint.
}
\label{fig_sim_seds}
\end{center}
\end{figure}

The result of such a test can be seen in Figure \ref{fig_sim_seds}, where we show the SEDs recovered by the different fitting methods for galaxy samples of different brightness.
For this test, it is important to highlight once more that all bulges and all disks -- apart from small variations on the simulated values --  show the same SED, respectively, in the simulated data, making this test feasible.
These input SEDs are shown in these plots in solid colours, green for disks, orange for bulges.
The \textbf{average} SEDs of each component recovered by single-band fits are shown as dotted lines (blue for disks, red for bulges), the average SEDs recovered by multi-band fits are shown as dashed lines.
For reasons of readability, we do not show the scatter in the distributions in all panels, but in the second panel only.

For bright galaxies, both fitting methods recover the actual component SEDs very well (top panel, $m_{\rm r,B+D,sim}<17.5$) on average, even in the $u$-band.
An investigation into the scatter, however, shows that multi-band fitting recovers the input values accurately for each object, while single-band fits show significantly larger scatter.
Depending on the band and component, the scatter in the single-band results is 1.5 to 2 times larger.

However, in galaxies with $17.5<m_{\rm r,B+D,sim}<18.5 $mag (middle panel) average single-band SEDs already deviate significantly from the input SEDs, wrongly suggesting that disks and bulges have similar SEDs.
In contrast, multi-band fits recover the \textbf{average} component SEDs much more accurately.
The scatter in the SEDs recovered is significantly worse in the case of single-band fitting, as indicated with the thin error bars in the middle panel, thick error bars show the scatter in multi-band fitting.
Especially in $i$, $z$ and $Y$ band, the scatter in single-band fitting is up to 3 times worse than in multi-band fitting.
Single-band fits make the bulge and disk SEDs statistically indistinguishable at all wavelengths, while multi-band fits statistically allow the SEDs of disks and bulges to be separated much more cleanly.
The exception in all cases is the $u$-band data, in which especially the bulges of these objects are so faint, that their $u$-band magnitudes can not be recovered well in either of the codes and the fluxes recovered basically serve as an upper limit.

At $18.5<m_{\rm r,B+D,sim}<19.5 $mag, small deviations from the input SEDs can be seen for multi-band fits, however, on average, the bulge and disk SEDs can still be recovered relatively accurately and are clearly different.
Single-band SEDs on the other hand, are mostly identical, with very large scatter.

Figure \ref{fig_sim_seds}, above all others, demonstrates the importance and the significant improvements (besides object numbers) achieved by using multi-band fitting, compared to single-band fitting, as it allows -- at least on average -- the analysis of component SEDs.
In the framework of \galapagostwo, this allows the understanding of stellar populations of large numbers of galaxies and their components, present in typical data derived by present-day surveys and inaccessible by single-band fitting, at least in the brightness regime examined in this work.
It is this capability that allows to push the limits of an investigation and science case once multi-band fitting is used.

\subsubsection{Galaxy sizes}
\begin{figure*}
\begin{center}
\includegraphics[width=0.48\textwidth,trim=20 30 8 15, clip]{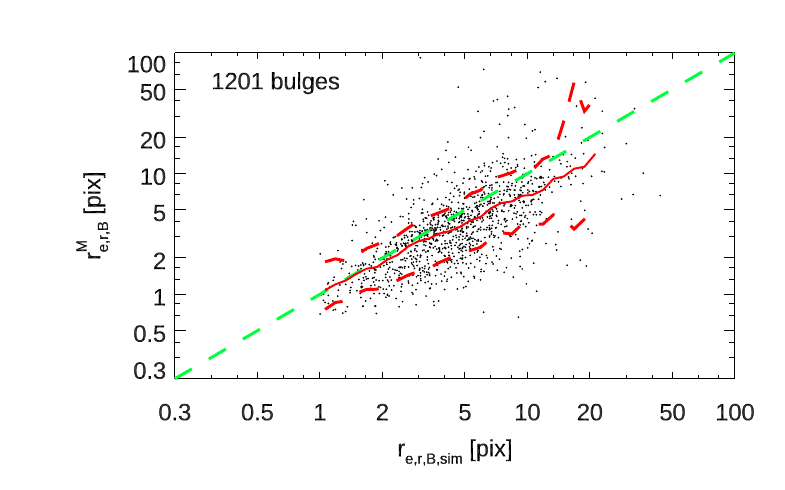}
\includegraphics[width=0.48\textwidth,trim=20 30 8 15, clip]{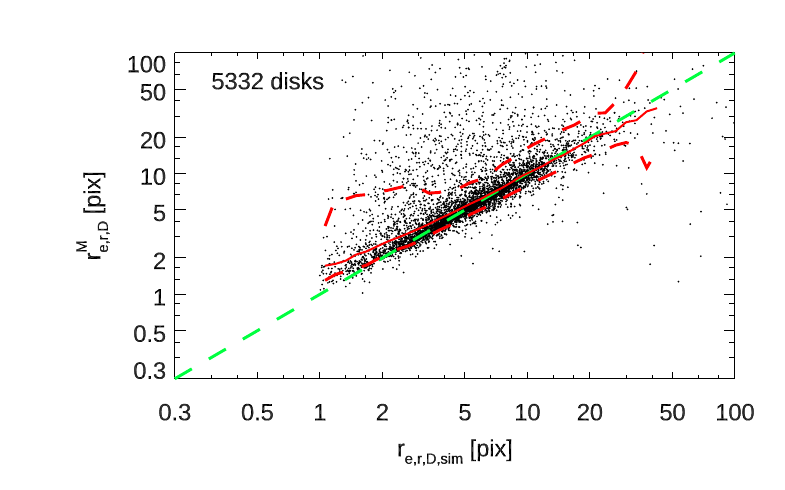}
\includegraphics[width=0.48\textwidth,trim=20 5 8 15,clip]{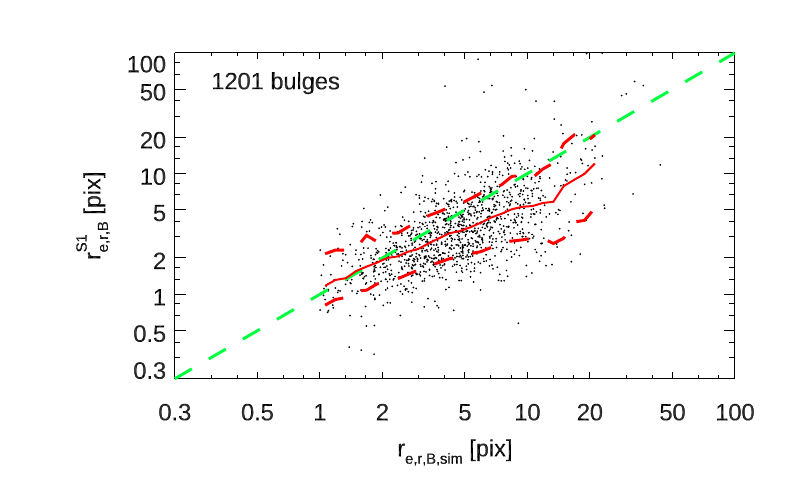}
\includegraphics[width=0.48\textwidth,trim=20 5 8 15,clip]{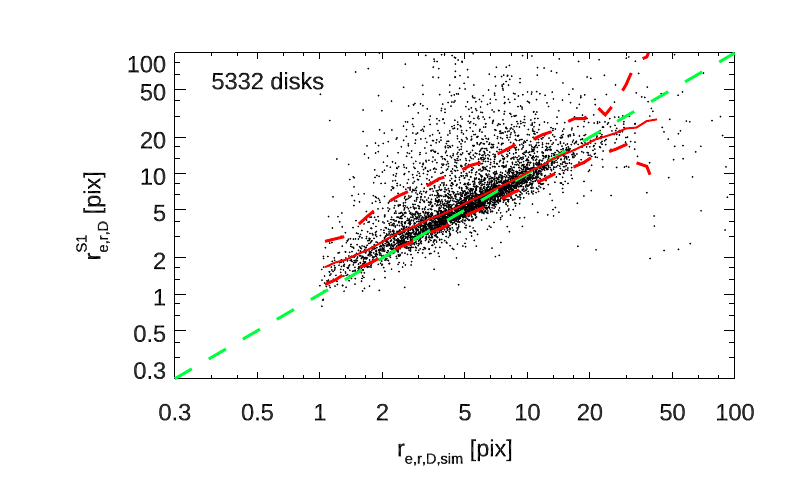}
\caption{Left column : Comparison of input (x-axis) and output (y-axis) bulge sizes  $r_{\rm e,B}$ values for galaxies at $m_{\rm r,B+D,sim}<19.5$mag from multi-band fits (top) and single-band fits (bottom). All measurements given in pixels. Red lines indicate the rolling median along the distribution.
Right column: Same plots, but for disk sizes.}
\label{fig_sim_re_in_out}
\end{center}
\end{figure*}

\begin{figure*}
\begin{center}
\includegraphics[width=0.48\textwidth,trim=23 30 8 15, clip]{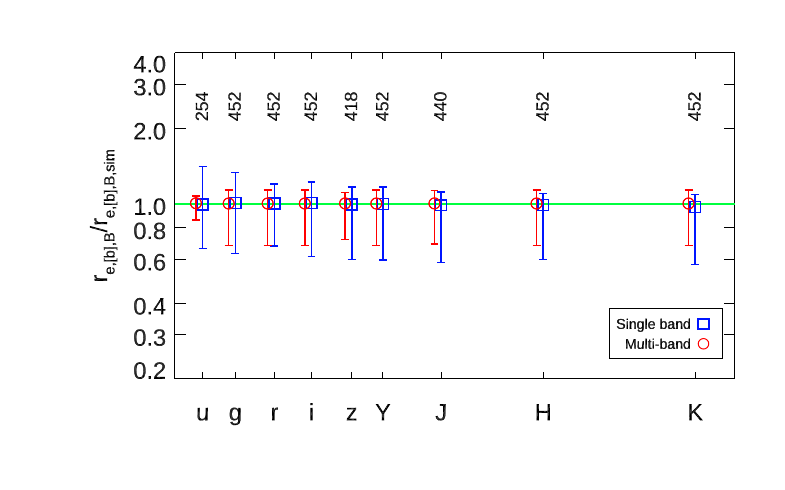}
\includegraphics[width=0.48\textwidth,trim=23 30 8 15, clip]{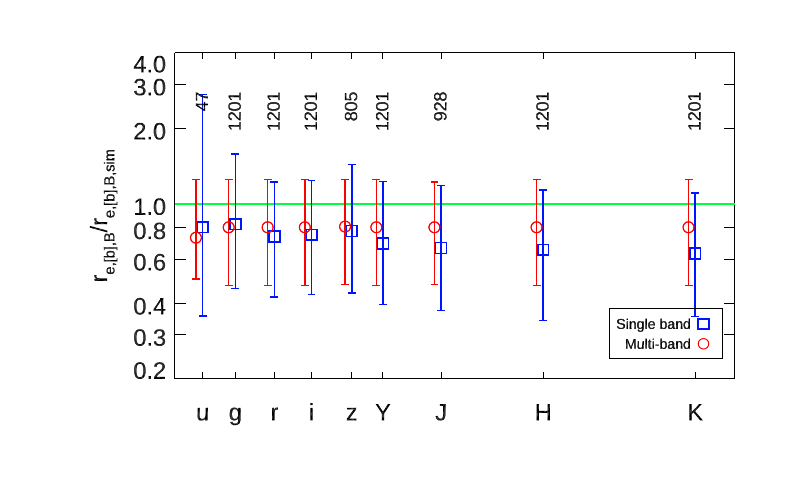}
\includegraphics[width=0.48\textwidth,trim=23 5 8 15,clip]{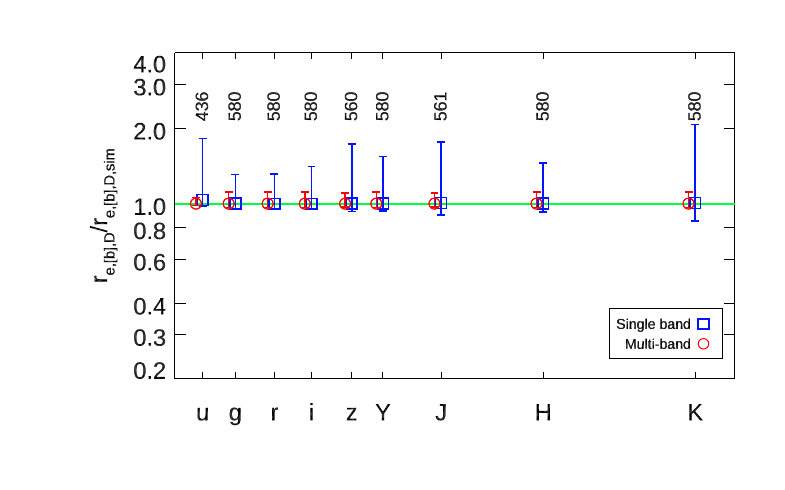}
\includegraphics[width=0.48\textwidth,trim=23 5 8 15,clip]{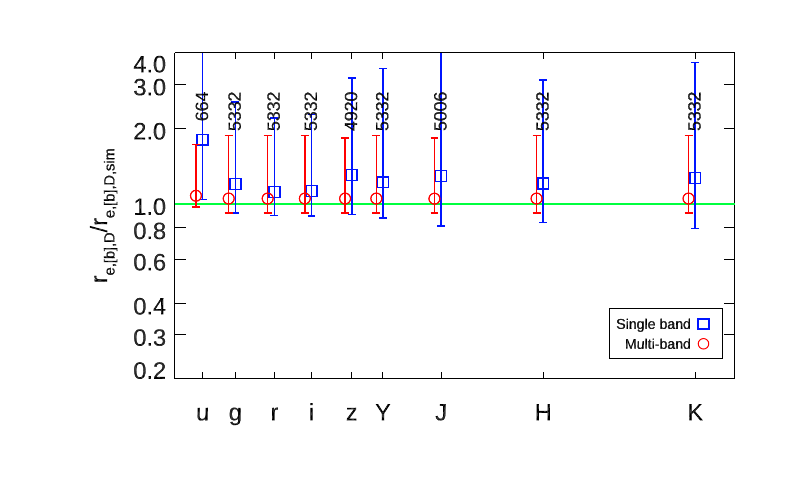}
\caption{Comparison of input and output $r_{\rm e}$ values for bright galaxies $m_{\rm r,B+D,sim}<17.5$mag (left) and $18.5<m_{\rm r,B+D,sim}<19.5$mag (right). Bulges (top), disks (bottom). The numbers above each point show the number of fits included for that band. As in all plots, symbols show median values, errorbars indicate 16 and 84 percentiles.
}
\label{fig_sim_dre}
\end{center}
\end{figure*}

\begin{figure*}
\begin{center}
\includegraphics[width=0.48\textwidth,trim=23 30 20 0,clip]{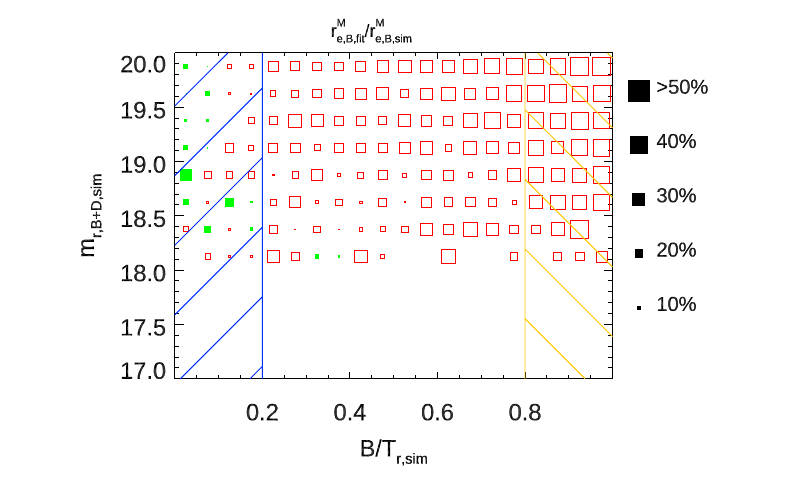}
\includegraphics[width=0.48\textwidth,trim=23 30 20 0,clip]{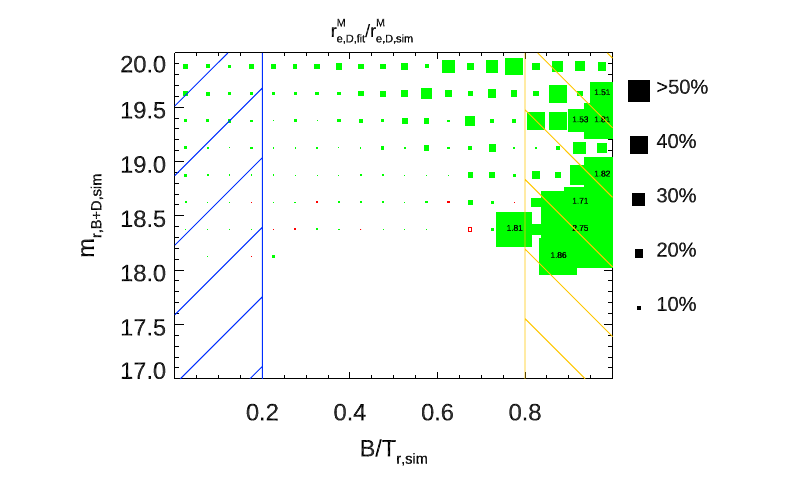}
\includegraphics[width=0.48\textwidth,trim=23 30 20 0, clip]{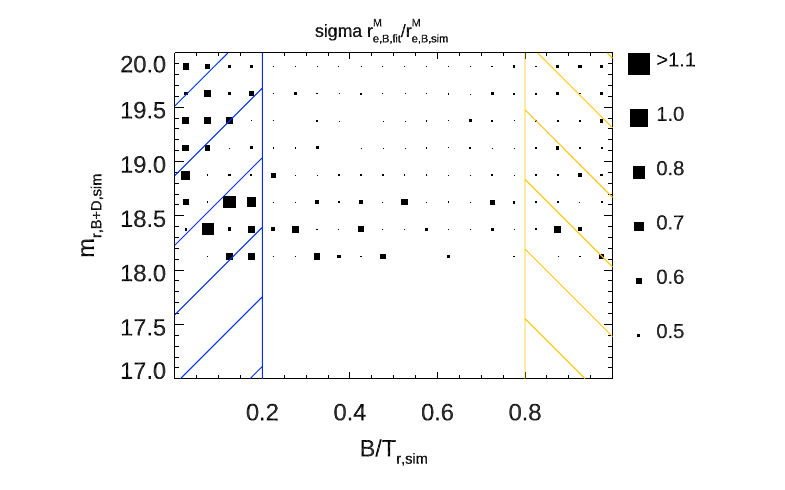}
\includegraphics[width=0.48\textwidth,trim=23 30 20 0, clip]{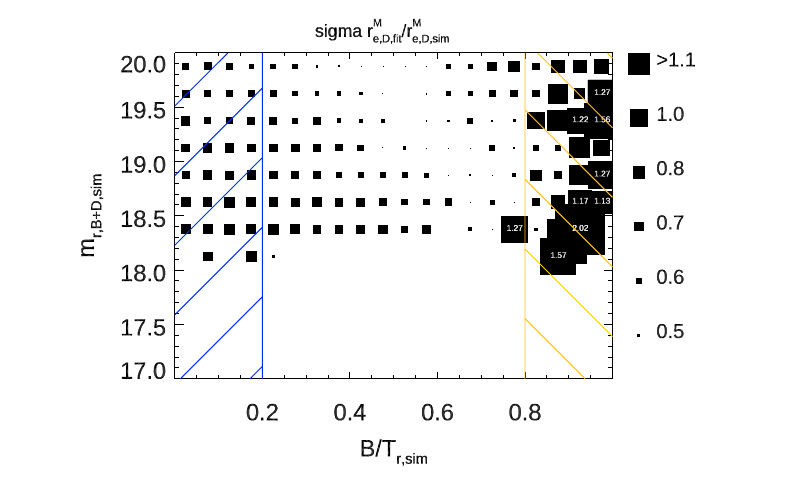}
\includegraphics[width=0.48\textwidth,trim=23 5 20 0,clip]{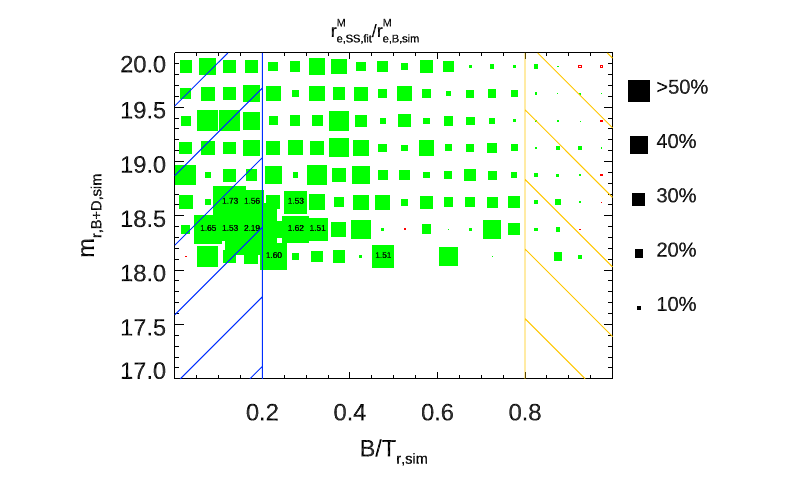}
\includegraphics[width=0.48\textwidth,trim=23 5 20 0,clip]{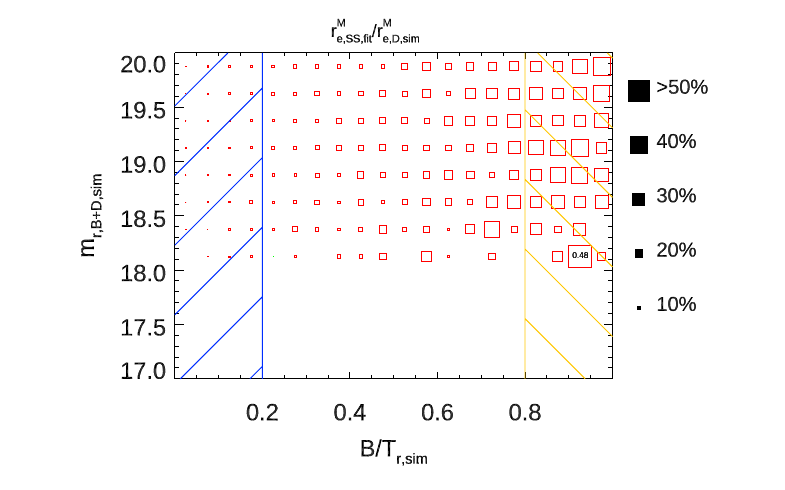}
\caption{Top left: Bulge sizes $r_{\rm r,B,fit}$ from $B/D$ fits, average offsets in multi-band fits as a function of $B/T$ and $m_{\rm r,B+D,sim}$. Symbol sizes indicate the deviation from an ideal value, i.e. 1 (ideal fits produce small symbols). Middle: standard deviation of the same value (small symbols represent good fits/small scatter). Bottom left: Comparison of single-\sersic sizes $r_{\rm r,SS,fit}$ to simulated size $r_{\rm r,B,sim}$. Where these values agree, the symbols should be small. At $(B/T)_{\rm sim}>0.8$, single-\sersic fits fit the bulge sizes well.
Right column: Same plots, but for disk sizes. At $(B/T)_{\rm sim}<0.2$, single-\sersic size agrees well with the simulated disk size.
As bulge and disk sizes are measured constant with wavelength, this figure looks identical in all bands.
Only bins with more than 20 objects are shown, to assure reasonable number statistics in each bin.
}
\label{fig_sim_mendel_re}
\end{center}
\end{figure*}

\begin{figure}
\begin{center}
\includegraphics[width=0.48\textwidth,trim=23 30 20 0,clip]{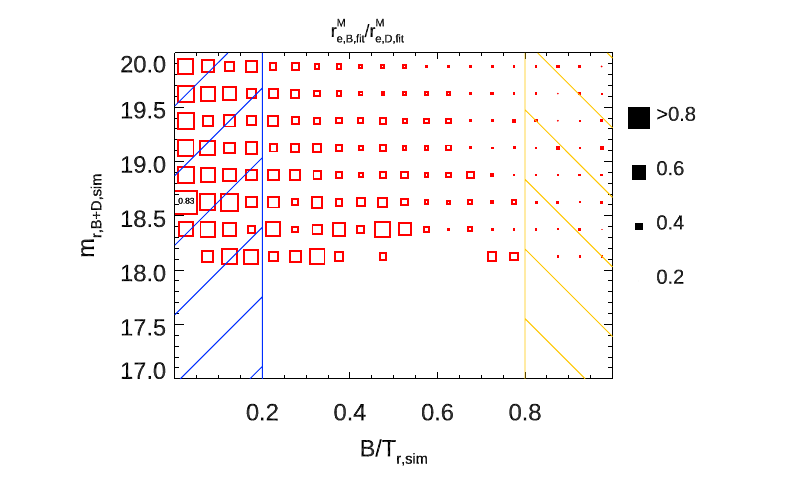}
\includegraphics[width=0.48\textwidth,trim=23 5 20 0, clip]{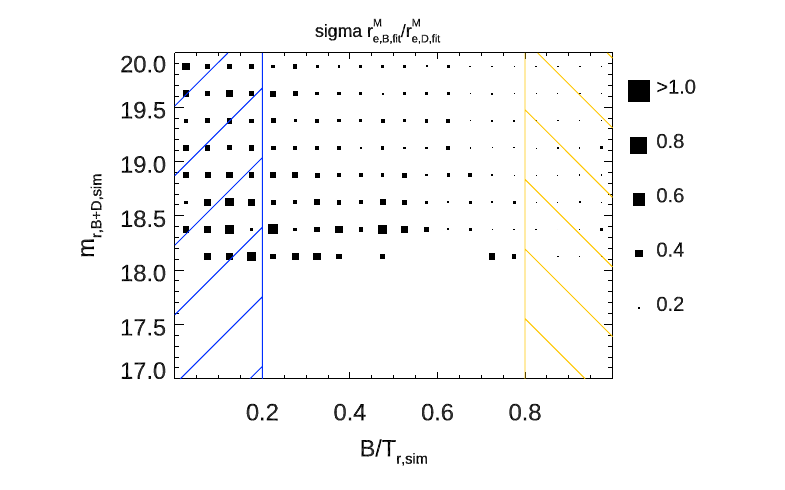}
\caption{Top: Recovered Bulge/Disk size ratio $r_{\rm e,B,fit}/r_{\rm e,D,fit}$ from $B/D$ fits in multi-band fits as a function of $B/T$ and $m_{\rm r,B+D,sim}$. The plot indicates that in galaxies with low $B/T$ values (faint bulges), bulges and disks show very similar sizes, indicating issues with the fitting procedure in this regime, see text for discussion. Bottom: standard deviation of the same value, indicating that this issue is systematic.
Only bins with more than 20 objects are shown, to assure reasonable number statistics in each bin.
}
\label{fig_sim_mendel_re_bd}
\end{center}
\end{figure}

In Figs. \ref{fig_sim_re_in_out} to \ref{fig_sim_mendel_re_bd} we show how well bulge and disk sizes can be recovered by comparing sizes measured to the simulated sizes of the components.
Similar to the previous section, we plot the objects with \lq good\rq\ fits in all $griYHK$-bands in both single- and multi-band fits, to allow easier comparison between bands, and the largest possible subsets of those samples in the $uzJ$-bands, respectively.
Although this is not important in most places as we show ratios of sizes, galaxy sizes are given in units of pixels\footnote{The \gama imaging data shows a pixel size of 0.331\arcsec/pix} throughout this paper, unless indicated otherwise specifically.

In Fig. \ref{fig_sim_re_in_out}, we show a direct comparison of simulated and fitted bulge sizes (left) for multi-band fits (top) and single-band fits (bottom panel) in the $r$-band, before looking at this parameter in a more statistical approach.
This plot shows all objects with fit results in all $griYHK$-bands and multi-band and at $m_{\rm r,B+D,sim}<19.5$mag.
As can be seen, and has been expected, the recovery of the bulge sizes is very noisy, especially when using such a faint limit.
However, a correlation between simulated and fitted values is apparent, albeit the scatter is large.
The single-band results additionally show a more significant systematic offset at sizes at $r_{\rm e,r,sim}>5 pix$, in that the fits on average return smaller sizes, as can be seen from the rolling median lines that are over-plotted in each panel.
We will confirm these findings in Fig. \ref{fig_sim_dre}.

It should be pointed out here that -- as the sizes of each component are held constant with wavelength during both the multi-band fit and in the simulations -- the multi-band plot would look identical when showing the results of any other band.
This fact also explains the identical average and errorbars of the multi-band sizes at all wavelengths in several of the successive plots.
Where differences between bands can be seen, they are a result of the sample selection in this particular band (i.e. a subset of the other bands).

In the right column of Fig. \ref{fig_sim_re_in_out}, we show the equivalent plot for disk sizes.
Both single- and multi-band fitting show a much narrower relation than for bulges, and outliers favour larger fit sizes, although with only small effects on the median values.
The relation of the multi-band results, however, is significantly tighter.
No systematic deviations from a 1-to-1 line can be found, however.

In Figure \ref{fig_sim_dre}, we show the ratio of input and output values for samples of bright and faint galaxies.
Values $\rm >1$ indicate that the fit recovers the galaxy component too large.
For bright galaxies (left column), it can be seen that both single- and multi-band fits recover the component size without systematic offset, but the scatter in the multi-band fits is significantly smaller.
There is also a interesting trend in the single-band fits that in the redder bands, the distribution of bulge values is asymmetrical towards smaller sizes, while in blue bands, it is much more symmetrical, whereas disk sizes (bottom panel) generally seem to be over-estimated more often than under-estimated.
Both effects can not be seen in multi-band fits, as by design the sizes of both bulge and disk are respectively constant with wavelength.
At first glance, it is interesting that the multi-band fits show smaller errorbars in the $u$-band than in the other bands, given that in many previous figures, the $u$-band  showed larger scatter.
This, however, is by design of the multi-band fits. 
Firstly, other than magnitudes, the size measurements discussed here are not independent from the other bands, as they are constrained with Chebyshev polynomials to follow the trend in the other bands smoothly. 
As such, they are less dependent on the image depth of the $u$-band itself. 
Secondly, the $u$-band (as well as $z$ and $Y$) show only a subset of the objects, presumably the \lq brighter\rq\ ones that are \lq easier\rq\ to fit. 
In fact, given that the errorbars returned by \galfitm are, by design, identical in all bands in the case of multi-band fits, these smaller errorbars in the $u$-band is precisely an effect of the sample selection. 

For faint galaxies (right column), both fits behave similarly in that they slightly underestimate the bulges sizes and overestimate the disk sizes.
However, the systematic offsets are smaller in the case of multi-band fits (<5\% for disks, \textasciitilde-20\% for bulge sizes), and the offsets in single-band fits are significantly larger (notice the logarithmic scale on the y-axis).
In faint galaxies, single-band fitting cannot reliably measure component sizes, with bulge sizes being badly underestimated and disk sizes badly overestimated, by a factor of up to a few in individual galaxies.
Multi-band fits, while showing increased scatter compared to brighter objects, return much more reliable and consistent results.
Please note again that the different offset of the multi-band values in the $u$-band is a result of using a different sample and reduced sample size in this band, the numbers of fits for each band are shown above the data points. 
By design sizes in the $u$-band are the same as in all the other bands in the fits.
As $u$-band single-band fits return \lq good\rq\ fits results in only few galaxies, a significantly smaller sample can be analysed.

In the top two panels of Fig. \ref{fig_sim_mendel_re} we look at the ratio of the measured and simulated sizes of the bulges and discs, respectively, using multi-band fits. 
The plots follow the same layout as Fig.~\ref{fig_sim_mendel_mag}, displaying the ratio (offset to unity) of the sizes (top) and the standard deviation (middle) of the measurements by the size of the squares, plotted as a function of $B/T$ ratio and total galaxy magnitudes.
These plots confirm that, in general, the bulge sizes are underestimated by \textasciitilde 10-30\% in faint objects, disk sizes are typically overestimated by 5-15\%, and that some magnitude trends are visible in that deviations are in general larger in fainter galaxies, as one would expect.
The scatter in these distributions are especially high for the fainter component in \lq pure\rq\ galaxies, i.e. bulges in galaxies with $B/T<0.2$ and disks in galaxies with $B/T>0.8$.
In particular, the faint disks embedded in galaxies with high $B/T$ show very large scatter, as one would expect.

As was the case for magnitudes, it is also apparent that \textit{both} components are hard to fit in objects with $B/T<0.2$ and $B/T>0.8$, and the largest systematic trends for both components exist in those areas.
In section \ref{sec_res_sim_mag}, we indicated that this is an effect of the fainter component fitting some of the flux of the brighter component, and indeed, we can see similar effects in these plots as well.
Bulge sizes are overestimated (green, filled squares) more frequently in low $B/T$ objects, which is an effect of galaxy bulges being typically smaller than galaxy disks, although no such restriction was technically imposed on our simulated values.
This result fits the scenario in which the fainter component fits some of the flux -- and the shape -- of the brighter component.

In fact, when we plot the ratio between the bulge and disk halflight radii $r_{\rm r,B,fit}/r_{\rm r,D,fit}$ in a similar plot (see Figure \ref{fig_sim_mendel_re_bd}), we can see that this ratio is between 0.3 and 0.5 on average over most of the $B/T$ and $m_{\rm r,sim}$ range covered, reflecting the distributions we used in the simulated data.
However, in galaxies with $B/T<0.3$ this ratio changes dramatically, reaching values $>0.8$, at small variation, with the bulge in these galaxies fitting the disk profile instead. 
The small scatter despite apparent offsets, suggests that this offset is indeed systematic.
This indicates that the bulge profile in the fit accounts for some part of the disk profile, and we will see the same effect again in \sersic indices in the next section.
Interestingly, in galaxies with $B/T>0.7$ the opposite can be seen, with the disk becoming very large in comparison to the bulge.
We attribute this to the fixed \sersic index of the disk, which can not mimic the $n_b=4$ shape of the bulge profile, and instead fits a background or neighbouring structure.

In the bottom panels of Fig. \ref{fig_sim_mendel_re}, we show the relation of the single-\sersic derived size $r_{\rm e,r,SS,fit}$ to the input values for bulges and disks respectively.
For the areas with $B/T<0.2$ and $B/T>0.8$, we can see much smaller offsets for disk and bulges, respectively, than in the $B/D$ fits.
This suggests that in these galaxies, the single-\sersic fits provide a better fit to the profile of the brighter component.
In this case, one would want to fall back onto using the single-\sersic fits values instead of the results from $B/D$  fits and truly count these objects as \lq pure\rq\ systems.
We had already seen previously that the total magnitudes of these objects are recovered well by the single-\sersic fits. 
Given the low flux in the fainter component, this total magnitude will largely reflect the flux in the brighter component.

\subsubsection{\sersic indices}
\label{sec_sim_results_n}
As the last of the 3 important profile parameters, we need to examine how well we can recover the bulge \sersic index from our simulated data.
In general (e.g. H13 and others), \sersic indices are known to be the parameter hardest to recover, especially for high-$n$ objects. 
This should be especially true given that the \sersic profile of the bulge is now overlaid with an exponential profile of the galaxy disk which will both hide the profile itself and further confuse the \galfitm fits, making $n_{\rm B}$ even harder to recover.
The reader should be reminded at this stage that the disks in our simulated objects are both created and fit using a $n_{\rm D} == 1$ profile, so we do not need to examine how well those can be recovered.

\begin{figure}
\begin{center}
\includegraphics[width=0.48\textwidth,trim=23 30 8 15, clip]{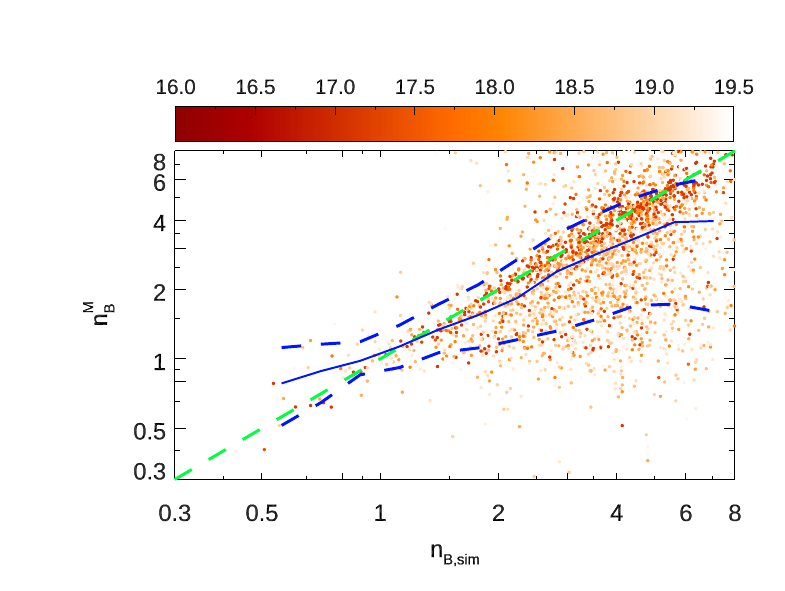}
\includegraphics[width=0.48\textwidth,trim=23 5 8 15,clip]{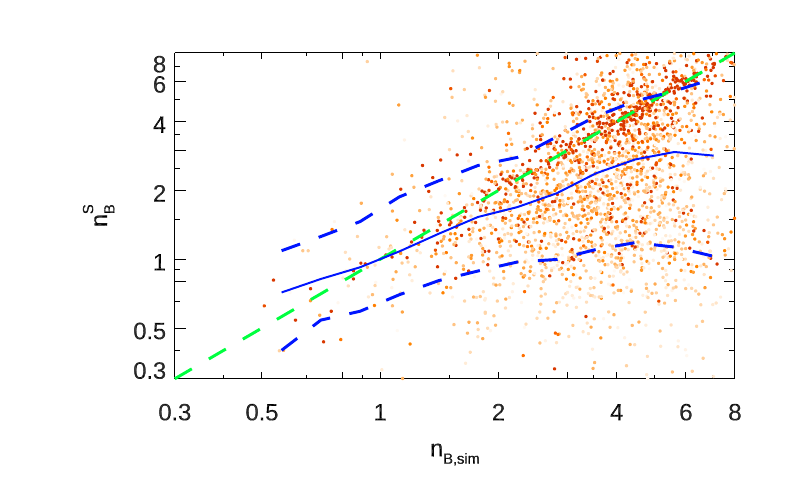}
\caption{Comparison of input (x-axis) and output (y-axis) \sersic indices $n_{\rm B}$ values for galaxies at $m_{\rm r,B+D,sim}<19.5$mag from multi-band fits (top) and single-band fits (bottom). Colour coding indicates brighter objects with darker colours with $16<m_{\rm r,B+D,sim}<19.5$mag. Blue lines indicate median and 16/84 percentiles, the green line represent a 1-to-1 line that indicates perfect fits.
}
\label{fig_sim_nb_in_out}
\end{center}
\end{figure}

\begin{figure}
\begin{center}
\includegraphics[width=0.48\textwidth,trim=23 30 8 15, clip]{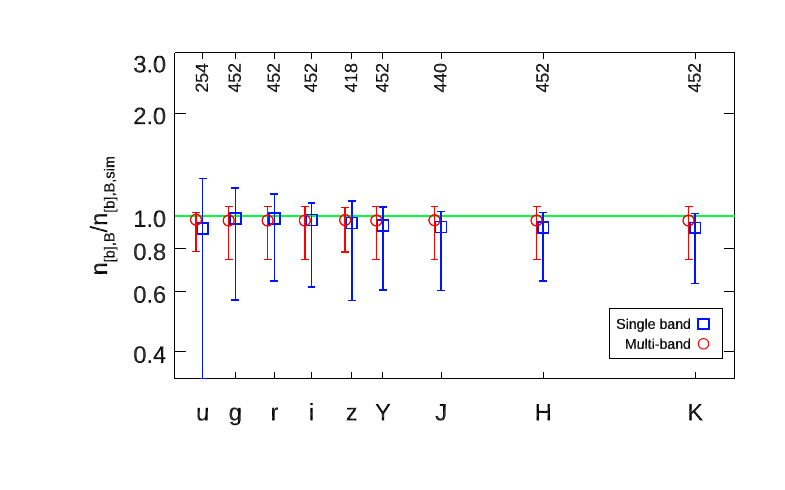}
\includegraphics[width=0.48\textwidth,trim=23 5 8 15, clip]{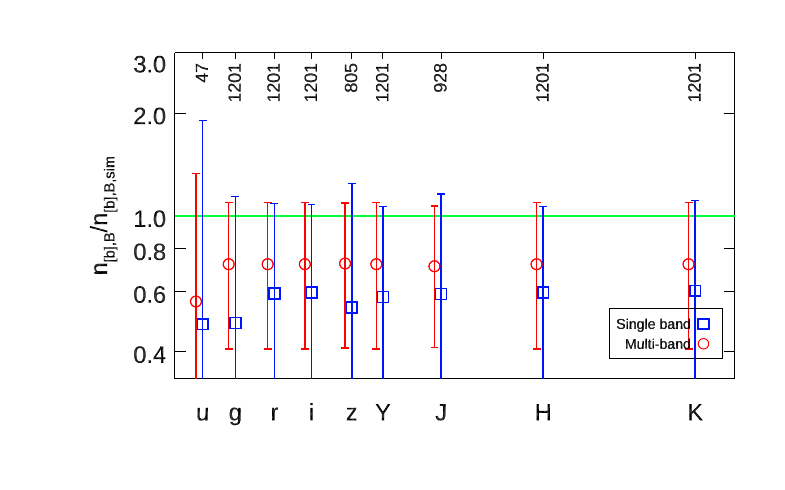}
\caption{Comparison of the difference in the bulge \sersic index $n_{\rm B}$ input and output  values for each band. Top panel: bright galaxies at $\rm <17.5$mag. Bottom panel: faint galaxies with $18.5<m_{\rm r,B+D,sim}<19.5$mag.
}
\label{fig_sim_dn}
\end{center}
\end{figure}

Instead, we show the direct comparison of input and output bulge \sersic indices $n_{\rm B}$ values in Fig. \ref{fig_sim_nb_in_out}, colour coded by simulated object $r$-band magnitude. 
It is clear that $n_{\rm B}$ is indeed the parameter hardest to fit, with both multi-band and single-band fits showing significant scatter, the correlation for bright objects being more visible for multi-band fits.
Additionally, single-band fits show a significantly larger fraction at high $n_{\rm B,sim}$ values that are fitted with lower $n_{\rm B,fit}$ values, also visible in the overplotted lines that indicate rolling median and 16/84 percentiles.
From the individual point in the background, it is visible that in both multi- and single-band fits the scatter increases for fainter objects (lighter colour), as one would expect.
For galaxies at $m_{\rm r,B+D,sim}<17.5$mag, this correlation becomes relatively tight, with a scatter of $\pm1$ as can be confirmed in the top panel of Fig. \ref{fig_sim_dn}.
However, the sample becomes small, which is why we include fainter galaxies in Fig. \ref{fig_sim_nb_in_out}.
As one would expect, the magnitude of the host galaxy is not the only important parameter in this context. 
Colour-coding the plots instead by $B/T$ (not shown) shows the expected trends in that \sersic index values $n_{\rm B}$ are more easily recovered in prominent bulges.
However, while multi-band fits show a very tight correlation for objects with $B/T>0.8$, the single-band fits recover these values much less accurately, and a trend with decreasing $B/T$, while visible, is quite weak.

In Fig. \ref{fig_sim_dn}, we examine the recovery of the bulge \sersic index $n_{\rm B}$ more statistically for bright and faint galaxies.
Similar to what we have seen for magnitudes and sizes, both single-band and multi-band fits on average recover the \sersic index well for bright objects, systematic offsets are small.
However, the deviation in single-band fits is already on the level of 40\% for individual objects, in multi-band fits on a somewhat smaller level.
For faint galaxies, both fitting methods show considerable offset and scatter, both codes underestimate $n_{\rm B}$ significantly.
Interestingly, this is contrary to what we found in H13 where we reported that $n$ is generally somewhat overestimated in single-profile galaxies.
In all bands, the single-band fits recover values considerably worse than the multi-band fits with offsets of up to \textasciitilde 50\% even when excluding the noisy $u$-band.
This systematic underestimation of $n_{\rm B}$ is likely an effect of mixing the bulge profile with an underlying $n_{\rm D}==1$ disk profile, which pulls the fit towards lower $n_{\rm B}$ values.
The scatter of the recovered values is large in both cases, and significantly increased in the single-band fits, as should be expected at this point.
The large offsets and scatter make it impossible in single-band fits to distinguish between a classical \deVa bulge with $n_{\rm B}== 4$ and a pseudo bulge, which generally show lower \sersic index values of $n_{\rm B}\sim1.5$.

\begin{figure}
\begin{center}
\includegraphics[width=0.48\textwidth,trim=23 30 20 0,clip]{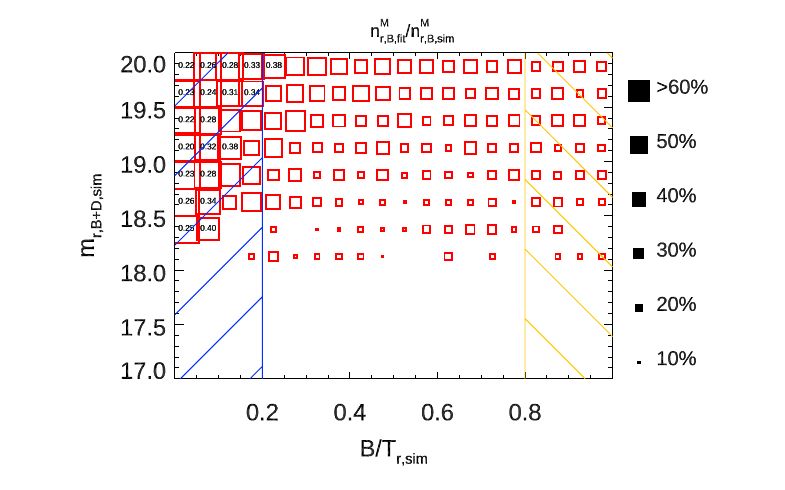}
\includegraphics[width=0.48\textwidth,trim=23 5 20 0, clip]{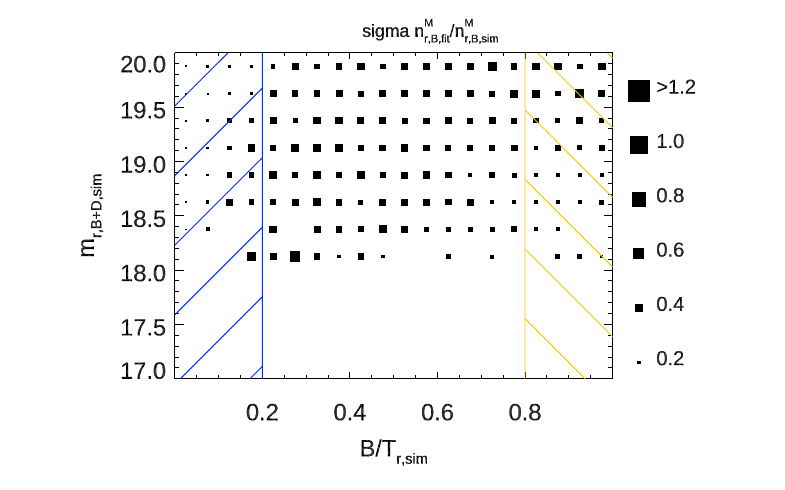}
\caption{Recovery of bulge \sersic indices $n_{\rm B}$ as a function of $B/T$ and $m_{\rm r,B+D,sim}$, showing the ratio in the fitted versus simulated values (top panel) and the standard deviation (bottom panel). Symbol sizes indicate the deviation from an ideal value, i.e. 1. Open red symbols indicate average values of $<1$. Ideal fits would produce small symbols in both panels (small offset, small scatter). $n_{\rm B}$ values can not be well recovered in disk-dominated galaxies.
Only bins with more than 20 objects are shown, to assure reasonable number statistics in each bin.
}
\label{fig_sim_mendel_n_b_rec}
\end{center}
\end{figure}

\begin{figure}
\begin{center}
\includegraphics[width=0.48\textwidth,trim=23 5 20 0,clip]{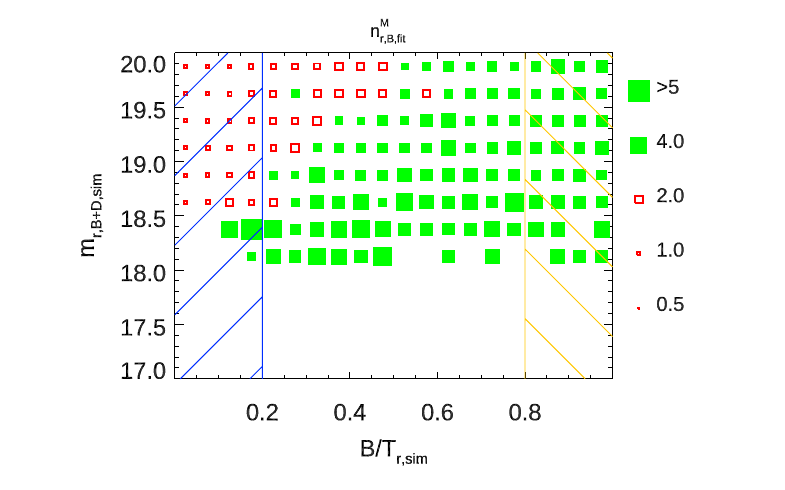}
\caption{Average bulge \sersic indices $n_{\rm B}$ as a function of $B/T$ and $m_{\rm r,B+T,sim}$. Open red symbols indicate average value of $n_{\rm B}<2$, to distinguish bulges with low (red, open) from bulges with high (green, filled) \sersic indices, i.e. \lq pseudo\rq\ from \lq classical\rq\ bulges. As bulges have a simulated distribution of $n_{\rm B}$ around a value of 4, ideal fits would produce equally sizes green boxes at all positions in this plot. In disk dominated galaxies, a median value around $n_{\rm B}\sim1$ is recovered, confirming that in this regime bulges more likely recover light from the disk.
Only bins with more than 20 objects are shown, to assure reasonable number statistics in each bin.
}
\label{fig_sim_mendel_n_b_v}
\end{center}
\end{figure}

In Figs. \ref{fig_sim_mendel_n_b_rec} and \ref{fig_sim_mendel_n_b_v} we again take a closer look at the \sersic indices recovered by the multi-band fitting.
In Fig. \ref{fig_sim_mendel_n_b_rec}, we show how well $n_{\rm B}$ can be recovered by comparing the fit value to the simulated values.
As noted above, the bulge \sersic indices are underestimated by 10\%, even in multi-band fits of bright objects.
The average offset is a function of the object brightness such that $n_{\rm B}$ is underestimated more in fainter galaxies.
However, in faint and disk-dominated galaxies, systematic offsets become very large.
In these galaxies, the average $n_{\rm B}$ is badly underestimated, with very small scatter, indicating that this is a systematic effect.

In fact, when we look at the recovered $n_{\rm B}$ values directly in Fig. \ref{fig_sim_mendel_n_b_v}, we see that the average $n_{\rm B}$ recovered by \galfitm is indeed \textasciitilde3-4 for a large fraction of the galaxies, 4 being the average input value.
However, low $n_{\rm B}$ values are recovered when trying to fit a bulge in faint galaxies (top) or within a bright disk (left).
The $n_{\rm B}$ in these galaxies converges to a value of \textasciitilde 1, which -- together with magnitudes and sizes -- again shows that the bulge profile of the $B/D$  fits in these objects tends to fit part of the disk light.
Magnitudes, sizes and \sersic indices all tend towards those values simulated in the disk profile of these objects.

Open red symbols in this plot indicate average value of $n_{\rm B}<2$.
A value of 2 is often used in the literature to separate classical bulges from pseudo bulges.
In our analysis here, we show that these measurements such a distinction is only reliable in bulge-dominated objects as bulge \sersic indices are hard to recover in other galaxies.

\begin{figure}
\begin{center}
\includegraphics[width=0.48\textwidth,trim=23 30 20 0,clip]{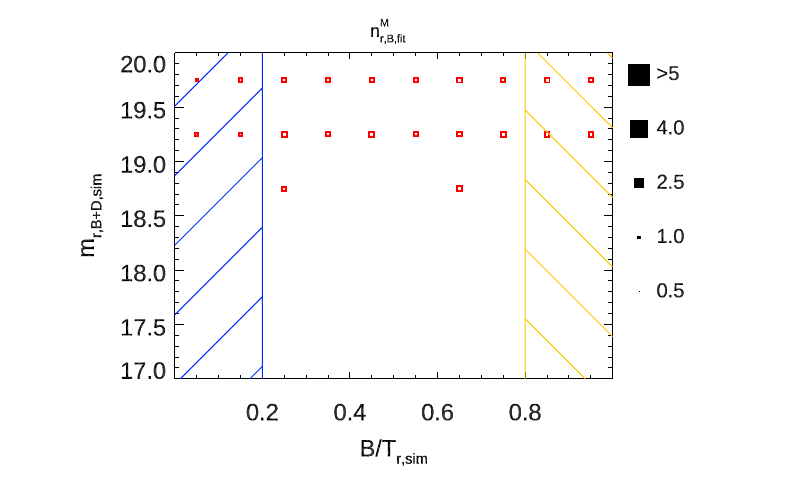}
\includegraphics[width=0.48\textwidth,trim=23 30 20 0,clip]{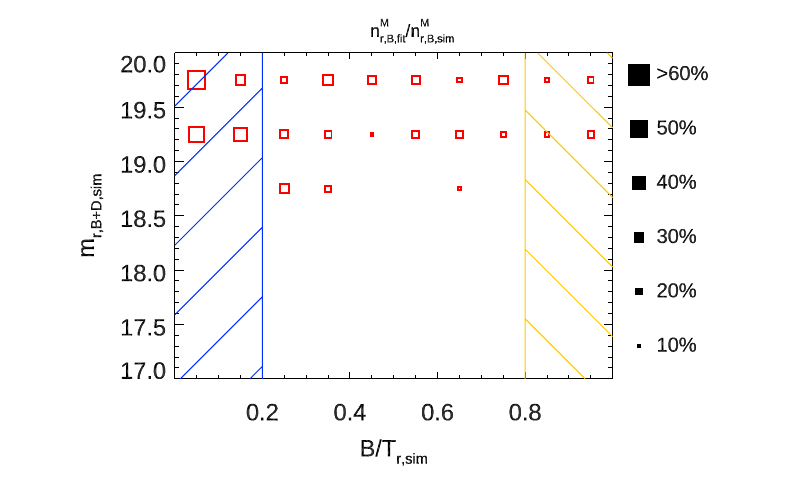}
\includegraphics[width=0.48\textwidth,trim=23 5 20 0,clip]{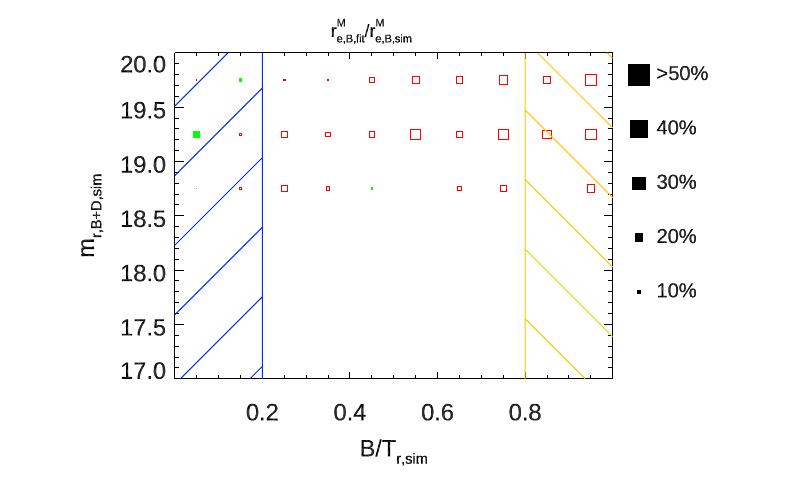}
\caption{Recovery of pseudo bulges: Average $n_{\rm B,fit}$ of galaxies with $1<n_{\rm B,sim}<2$ (top panel, ideal fits would show uniform values of $\sim 1.5$ throughout this panel), comparison with simulated values $n_{\rm B,fit}/n_{\rm B,sim}$ (middle panel), and size comparison $re_{\rm B,fit}/re_{\rm B,sim}$ (bottom panel) for the same sample.
Symbol sizes in the lower 2 panels indicate the deviation from an ideal value, i.e. 1 , ideal fits are hence indicated by small symbols.
Only bins with more than 20 objects are shown, to assure reasonable number statistics in each bin.
}
\label{fig_sim_mendel_pseudo_bulges}
\end{center}
\end{figure}

We have a specific look at simulated pseudo-bulges with $1<n_{\rm B,sim}<2$ in Fig. \ref{fig_sim_mendel_pseudo_bulges}.
In order to avoid low number of objects, we increased the bin size in this plot.
Equivalently to Fig. \ref{fig_sim_mendel_n_b_v} we show the average recovered \sersic index in the top panel. 
On average, \sersic indices are recovered with values of $1<n_{\rm B,fit}<2$, as this was how they were selected in the simulated values.
However, as this might be an effect of \sersic indices being drawn closer to 1 due to the disk, as seen when looking at all bulges, we try to have a closer look at the recovered values.
In the middle panel, we show how well the $n_{\rm B}$ values can be recovered compared to the input values.
Systematic deviations are relatively small.
Only for galaxies at $B/T<0.2$, larger effects can be seen, which can be easily understood as in these objects the bulge would be too faint to be fit reliably, as discussed before.
In these galaxies, the fit values are indeed drawn closer to 1.
Comparison with Fig. \ref{fig_sim_mendel_n_b_rec} reveals similar behaviour, at least qualitatively.
While on average the values can be recovered, we find scatter values of $\sigma n_{\rm B,fit}/n_{\rm B,sim}\sim0.3-0.4$ throughout the entire parameter space (not shown).
Finally, in the bottom panel, we show how well we can recover $r_{\rm e,B}$ for these pseudo-bulges, in comparison with Fig. \ref{fig_sim_mendel_re}.
Again, we find similar behaviour for pseudo- as for classical bulges.

From this result we conclude that the recovery of pseudo-bulges in these data is similar to the recovery of classical bulges.
However, we would like to stress that, while on average the values can be recovered, it is dangerous on an individual galaxy to draw the conclusion whether the bulge is a classical bulge at $n_{\rm B}=4$ or a pseudo-bulge with $n_{\rm B}=1$, the uncertainties on an individual galaxy are too large to allow this conclusion (see Fig. \ref{fig_sim_dn}).
\galapagostwo offers the option to fix $n_{\rm B,fit}==4$ during the fit, to allow only for classical bulges.
However, we have not done so during this analysis and this test is hence beyond the scope of this work. 
Nedkova (in prep), found those fits to generally behave more stable in \candels\ data.
We would also like to stress that these results only apply to the faint and \lq distant\rq\ objects analysed in \textbf{this} dataset.
In nearby galaxies, even an automated approach might warrant some good results, as results could be more reliable due to higher spatial resolution and generally higher S/N.
Certainly, for well resolved galaxies, more care can be taken on individual objects.

%%% LITERATURE MISCONCEPTIONS %%%
\subsection{Danger of using \sersic index to classify galaxies}
\label{sec_bad_habits}

\begin{figure}
\begin{center}
\includegraphics[width=0.48\textwidth,trim=23 5 10 15 15,clip]{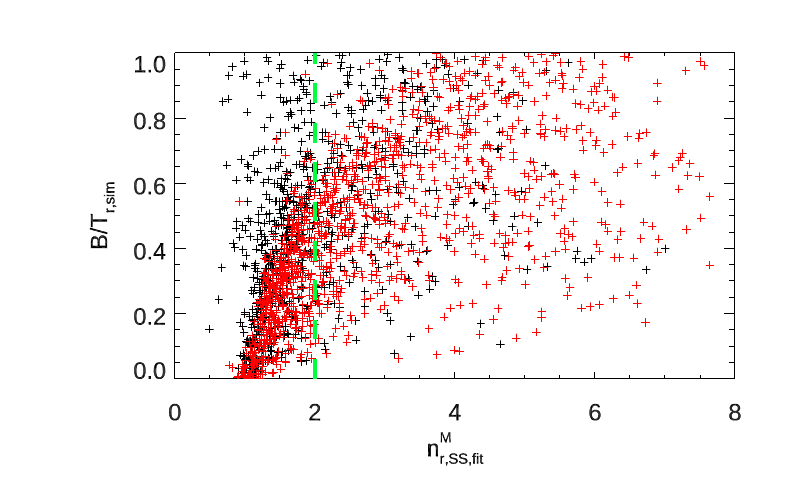}
\caption{A comparison of the ${B/T}_{\rm r,sim}$ values of the simulated galaxies against the \sersic indices $n_{\rm SS}$ derived from single-\sersic fits, for galaxies with $m_{\rm r}<18.5$, using multi-band fits. Red and black points represent galaxies with $n_{\rm B,sim}>3.5$ and $n_{\rm B,sim}<3.5$ respectively. The vertical green lines indicates $n_{\rm SS}=2$, which is used in the literature to separate disk- from bulge-dominated galaxies.
Only a weak correlation of single-\sersic value $n$ and $B/T$ flux ratio is visible, with very wide scatter.}
\label{fig_ss_B/T}
\end{center}
\end{figure}
At this stage, it is important to highlight the danger of using a \sersic-index cut to \lq classify\rq\ galaxies or select specific galaxy samples, as is often done in the literature \cite[e.g.][and many others]{Ravindranath,shen03} and which has been reported by other authors \citep[e.g.][]{GrahamPASA}.
Figure \ref{fig_ss_B/T} shows the \textbf{simulated} $(B/T)_{\rm r,sim}$ values as a function of the measured $n_{\rm SS}$ values as derived from our single-\sersic multi-band fits.
In order to present a cleaner plot, we restrict ourselves to bright galaxies with $m_{\rm r,sim}<18.5$.
Naively, in such a plot, one would expect a relatively tight correlation between \sersic index $n_{\rm SS}$ and $B/T$ ratio, as a more prominent bulge would increase the measured \sersic index of the galaxy as a whole, as their light peak in the centre of the galaxy and their generally smaller sizes drive the $n_{\rm SS}$ to higher values.
And indeed, the \sersic index is often used in this way in the literature, using values $1.5<n_{\rm SS}<2.5$.
While some correlation can be seen, the trend is very weak and not able to select \lq bulge-dominated\rq\ galaxies from a sample, at least without further knowledge.
In order to check whether this is an effect of simulating galaxy bulges at a range of \sersic indices instead of using classical bulges only, we highlight the galaxies with \lq classical\rq\ bulges ($n_{\rm B,sim}>3.5$) in red. 
The correlation is, in fact, very similar, and even for these galaxies, using a cut at n=2 to separate bulge- from disk-dominated objects should be considered unreliable.
Of the 1005 \lq spheroid-dominated\rq\ galaxies with $n_{\rm r,SS}>2$ in the $r$-band, which would be classified as bulge-dominated in such a selection, 350 ($\rm \sim35\%$) were in fact simulated with $(B/T)_{\rm r,sim}<0.5$, of the 944 galaxies with $n_{\rm r,SS}<2$, 175 ($\rm \sim18.5\%$) have $(B/T)_{\rm r,sim}>0.5$, so a somewhat cleaner sample can be selected.

This effect is visible in all bands, although it is somewhat less pronounced in the $H$-band and in fact the opposite for \lq disk-dominated\rq\ objects.
The $H$-band results still show a contamination of $\rm \sim 15\%$ (of 1469 galaxies with $n_{\rm H,SS}>2$  225 have $(B/T)_{\rm H,sim}<0.5$) for \lq spheroid-dominated\rq\ objects, and $\rm \sim 48\%$ (of 480 galaxies with $n_{\rm H,SS}<2$  234 have $(B/T)_{\rm H,sim}>0.5$) for \lq disk-dominated\rq\ objects, respectively.
For completeness, it should be stated that we created similar plots for other wavelengths and mixed bands (e.g. checking for correlation of the $n_{H,SS}$ with $(B/T)_r$), but no strong correlation can be found; in fact trends look even weaker when mixing wavelengths.
From this analysis, we discourage anyone from using such a simple classification approach.

We have also checked for other correlations that could potentially allow such an automatic identification of such a class of objects, but we were unable to identify a different, more reliable method.

%__________________________________________________________________
%%% APPLICATION TO REAL DATA %%%
\section{Application to real imaging}
\label{sec_real}

In \S \ref{sec_sims}, we have presented a detailed analysis of code performance on simulated data and have discussed the advantages of such an approach.
However, using such simulated light profiles assumes that real galaxies actually \textbf{do} precisely follow \sersic profiles -- although it is already known that they do not \cite[e.g. ][and numerous others]{Graham2003,Trujillo2004c}. 
Any deviation from \sersic profiles, e.g. by additional components, galaxy-galaxy-interactions, dust content \citep[especially dust lanes in edge-on galaxies, e.g.][]{Pastrav2013a,Pastrav2013b}, and/or simply substructure like spiral arms within the disk, will not be included the simulations used in this work, and hence their influence on the test results can not be tested.
We discuss the effect of dust specifically in \S \ref{sec_dust}.

In this section we look at comparisons of single- to multi-band fits on real data from the \gama survey, i.e. the very data which we tried to replicate/simulate in the previous sections.
The obvious advantage of this is that all effects apparent in real galaxy surveys can be tested.
However, as the true parameter values of objects are unknown, a detailed analysis as in the previous section is impossible.
We can merely run sanity checks and see whether already known effects can be recovered better with one method than the other.
We can further compare our fit results to results from other works and with fits on alternative/deeper data, which we will briefly attempt in \S \ref{sec_real_simard}, \ref{sec_real_deep-shallow} and \ref{sec_real_profit}.
Again, however, these checks can only serve as sanity checks, a decision on which code is \lq better\rq\ is not possible as true parameter values are unknown. 

\subsection{Data}
\label{sec_data}
The data used in this analysis has been well described in \citet{GAMA}, H13 and others, so we refer the reader to those publications.
In order to be consistent with H13, we carry out the analysis between single- and multi-band fits on part of the \gama-09 survey region using the same data as H13, although newer/deeper data do exist.
Additionally, the simulated data used above have been made to these specifications, so a comparison of results to those data is easier.
The data used here have been provided by \gama and use large mosaics, made from re-\swarp-ing \citep{terapix} data from \sdss and \ukidss LAS archival data \citep{ukidss} onto a common pixel-grid.
Instead, deeper data now uses deeper input data from \kids \citep[Kilo-Degree Survey,][]{KiDS1,KiDS2} and \viking \citep[VISTA Kilo-Degree Infrared Galaxy Survey,][]{Viking}. 
We will use these deeper data in \S \ref{sec_real_deep-shallow} \ref{sec_real_profit}, and Appendix \ref{sec_cat_release}.

\subsection{Setup}
\label{sec_real_setup}
The code version and setup used in this section is largely identical to the one used on simulated data.
Especially, the same versions of \galfitm and \galapagostwo are employed.
All DOF used for the individual parameters are the same, with the difference that it is -- contrary to the simulated data -- unknown whether these choices really do fit the real data used here.

The only difference in this setup is that we made use of a newer feature in \galapagostwo, which allows to target specific galaxies, in order to save CPU time.
Whereas in the simulated images, it was a sensible approach to fit and analyse all objects, we restrict our analysis here to the galaxies which have redshifts provided by the \gama survey, by providing a list of RA\&DEC coordinates to \galapagostwo.
However, as \galapagostwo works through objects in order of brightness, starting with the brightest objects, this approach potentially impacts on the fitting performance -- both in speed and accuracy -- if bright neighbouring galaxies are not included in the target list.
Fitting bright objects several times as \lq neighbours\rq\ costs more CPU time than fitting them once and \lq subtracting\rq\ them from consecutive fits, as \galapagostwo does.
Instead of targeting \textbf{only} the \gama galaxies themselves, we hence additionally target \lq nearby\rq\ galaxies/objects (within \textasciitilde 60\arcsec, i.e. 200 pixels at 0.331\arcsec/pix) that are brighter than said \gama target.
In tests on smaller datasets, it has been established that this minimises the impact of targeting specific galaxies on the fitting parameters (see H07 and H13).

\subsection{Samples}
\label{sec_real_samples}
The selection of good fits in real galaxies presented in this analysis is somewhat different to the one used in \S \ref{sec_sims}.
There, we were able to use simulation values to select galaxies, e.g. by brightness, size, or $B/T$ ratio, in order to get as clean a sample selection as possible.
This is obviously not possible in real data, so a selection based on fit parameters has to be used instead.
Additionally, the simulated data did not include stars, which in real data have to be taken care of and filtered out from any analysed galaxy sample.

The sample used in this section is identical to the sample used in H13, and we therefore refer the reader interested in the details to that work.
In fact, as \galapagostwo requires a first step of fitting single-\sersic profiles to all galaxies, the sample discussed in H13 was directly used as the first step of the fitting process on which these subsequent $B/D$ fits are based.
As neighbouring objects are treated as single-\sersic objects, with parameters kept fixed at their best-fit values in $B/D$ fits in \galapagostwo, the fits of the \gama objects are independent of any $B/D$ fit on neighbouring galaxies.
This means it was sufficient to only target the \gama galaxies themselves in this step, saving further CPU time.
While we have updated the redshifts and masses known for these objects where newer/better data has become available since, this plays only a minor role in this section as fit results and apparent magnitudes/sizes are compared directly.
However, given that we use a redshift selection to separate galaxies from stars, this potentially slightly changes the sample, as these classifications might have changed since.
No update of the \gama object sample selection, however, has been carried out subsequently, we use the same sample used previously.

Equivalently to the simulated data, we excluded any object -- or component -- where the fit parameters were on -- or very close to -- one of the constraint values used during the fit, using the same selection presented in \S \ref{sec_sample}.
This removes bad fits from our analysis in the same way as in H13.

\subsection{Results}
\label{sec_real_results}
\begin{figure}
\begin{center}
\includegraphics[width=0.48\textwidth,trim=25 30 8 15,clip]{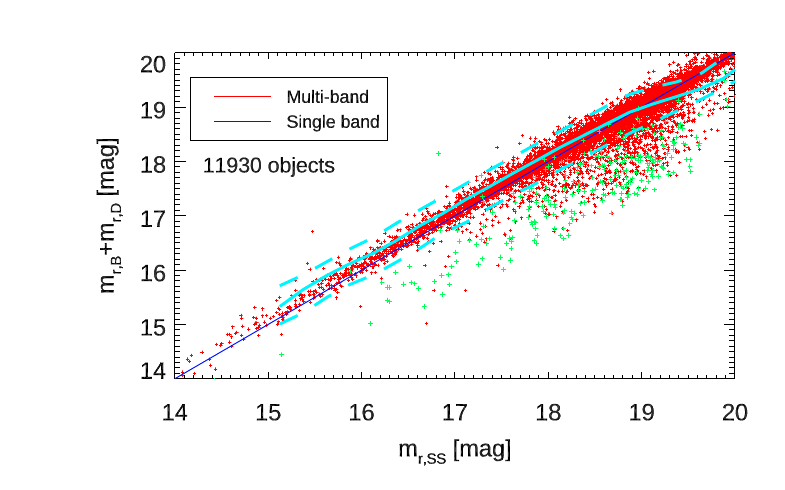}
\includegraphics[width=0.48\textwidth,trim=25 5 8 15,clip]{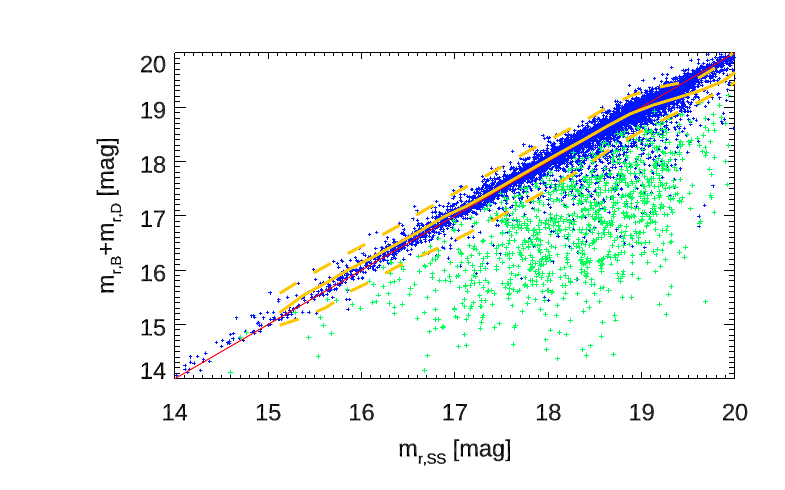}
\caption{\gama: Comparison of total (bulge+disk) magnitude and single-\sersic magnitude in both multi-band (top) and single-band (bottom) fitting. Lines represent rolling median and 16/84 percentiles. Points in green highlight objects where one of the components is fit with $r_e>200$[pix].
}
\label{fig_magbd-vs-mag}
\end{center}
\end{figure}

The most basic possible test on $B/D$  fits is to compare the total magnitude of objects between different fits, 1- and 2-component.
Figure \ref{fig_magbd-vs-mag} shows the comparison between the total $r$-band magnitudes derived from $B/D$  fits and magnitudes from single-\sersic fits, both for multi-band and single-band fits, using the same object sample.
For bright objects, both multi-band fits (top panel, all point) and single-band fits (bottom panel, all points) agree well between the 2 fits.
However, for fainter objects it is clear by visual inspection that single-band fitting produces catastrophic outliers - in which single-\sersic and total 2-component magnitudes do not agree - far more frequently, and much more severely.
While in neither case we find any significant systematic offsets (also compare Fig. \ref{fig_sim_magbd_comparison}),  the number of outliers seen in the multi-band fits is significantly smaller.
However, it should be noted that, given the large number of objects in this sample, this effect is small and does not have a significant impact on the median or 16/84 percentiles shown as dashed lines, nicely following the 1-to-1 line.

The fact that the total magnitude of a $B/D$  fit agrees well with a single-\sersic fit, however, does not prove that the $B/D$  decomposition itself is sensible and physically meaningful; this could simply be an effect of the minimisation process which tries to take all flux in the image into account.
As the neighbouring galaxies are taken into account in identical fashion to the single-\sersic fits, the total magnitudes recovered on the primary target should be expected to be relatively similar and simply reflects the \lq remaining\rq\ flux in the image, with single-band fits showing increased scatter as they use less data.
The larger amounts of extreme outliers in the single-band plot, however, is both interesting and worrying.
As \galapagostwo constrains the position of the 2 components in the $B/D$  fits to be the same, and this position to be within $0.5*re_{\rm SS}$, this can not be an effect of one or both components trying to fit a different object entirely or part thereof by \lq wandering off\rq.
Similarly, objects with extreme axis ratio, as would be the case if one component tries to include any residual flux of a neighbouring objects, should be ignored in the analysis, and in visual inspection are not found to happen very often.
Any extreme such cases would be automatically classified as \lq bad fits\rq\ due to our sample selection which flags and excludes such extreme objects.

However, by visually checking extreme outliers, we have found that many of these objects are indeed galaxies in which one component shows an extreme size.
We have highlighted a number of objects in green in which one of the galaxy components (disk or bulge) shows a size of $r_e>200$ pixels, i.e. possibly unrealistically large, but not filtered out by our sample selection which selects at 395 pixels.
This value has been randomly/empirically derived to be \lq large\rq\ compared to single-\sersic sizes, and only serves the purpose to select these extreme galaxies in this figure.
Especially users of \galapagostwo should take care of this limit and carefully select a value for their own dataset, if they wished to take this into consideration in the sample selection.

As can be seen, these galaxies are in general rare in multi-band fits, but in single-band fits, these objects account for the vast majority of outliers.
Visual inspection revealed that the fitting of the residuals of one or more neighbouring objects or other residual flux in the images is the main reason for these extreme component sizes, but not with such an extreme axis ratio that they would be filtered out by our sample selection above.
Such a large, relatively \lq face-on\rq\ component, despite being quite surface brightness faint, can include a lot flux due to its large size, leading for the brightness of an object to be overestimated badly.
The sensitivity of the single-band fits to such effects causes the large number of outliers in this plot, while the robustness of the multi-band fits to avoid fitting such structures prevents this from happening very often.

\begin{figure}
\begin{center}
\includegraphics[width=0.48\textwidth,trim=25 30 8 15,clip]{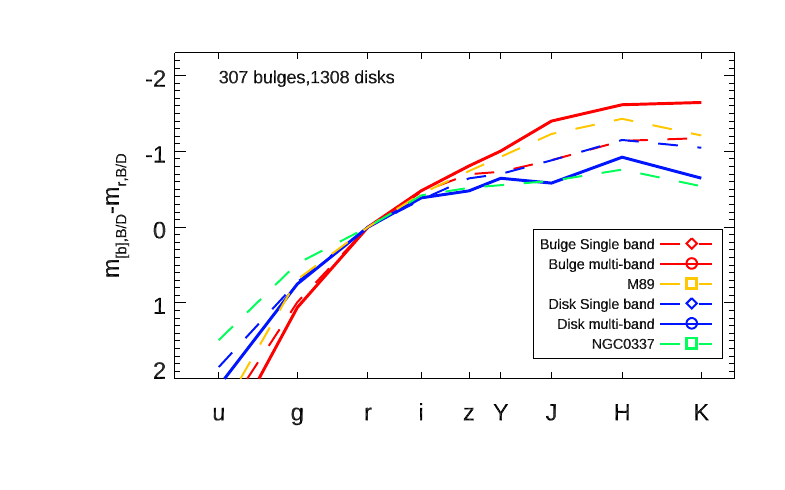}
\includegraphics[width=0.48\textwidth,trim=25 30 8 15,clip]{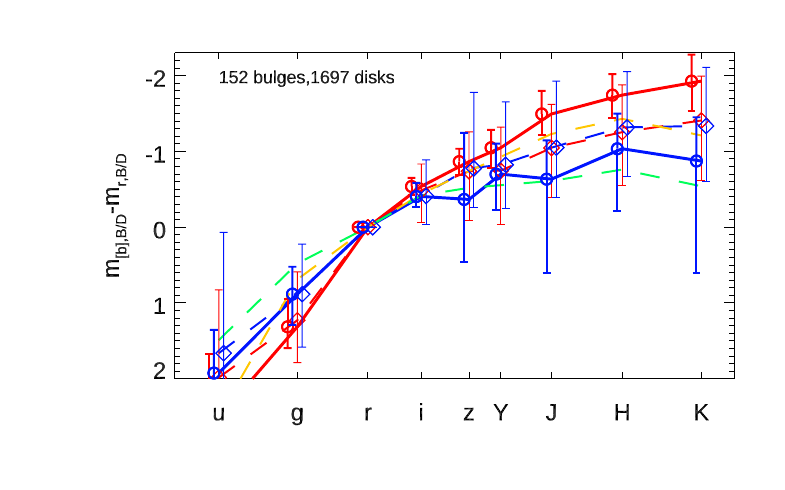}
\includegraphics[width=0.48\textwidth,trim=25 15 8 15,clip]{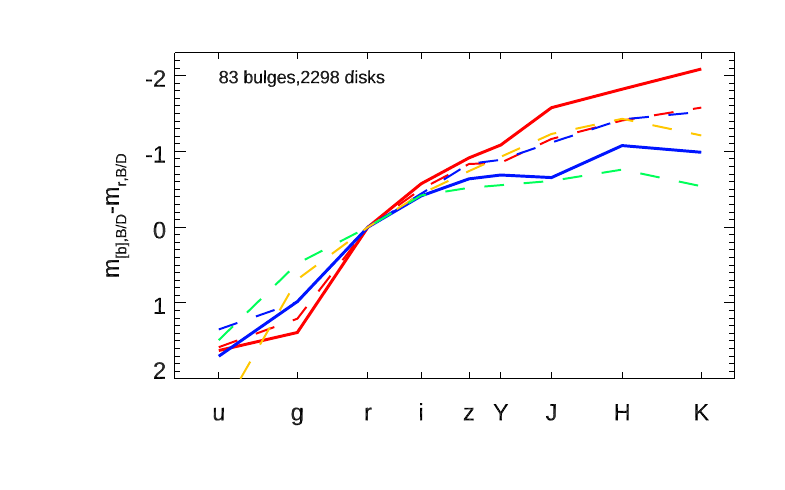}
\caption{Average SEDs of bulges and disks in real galaxies for the largest possible sample, i.e. all single-band and multi-band fits (excluding $u$ and $z$-band fits) produced \lq good\rq\ fit results. The SEDs recovered using single-band fitting are shown as blue and red dashed lines, the values recovered from multi-band fitting are shown in solid lines, respectively. Errorbars in the middle panel indicate the scatter of the distributions, bold/left errorbars for multi-band fits for bulge (red) and disks (blue), respectively, thin/right lines for single-band fits. The brightest galaxies at $m_{\rm r}<17.5$ (harbouring 307 bulges, 1308 disks) are shown in the top panel, $17.5<m_{\rm r}<18.5$ in the middle panel, $18.5<m_{\rm r}<19.5$ in the bottom panel. For comparison, the orange dashed line shows the SED of a typical quiescent galaxy (M89), the green dashed line shows the SED of a typical star-forming, bulge-less galaxy (NGC0337).}
\label{fig_bd_seds}
\end{center}
\end{figure}

In order to evaluate how well the $B/D$  fit truly separates the bulge and disk in a galaxy, we need to look not at the total magnitude, but their fractions at each wavelength, or -- better yet -- the SEDs of the individual components.
This is difficult on an individual basis, especially when the true values are not known and each component/galaxy shows very different stellar populations.
Fig. \ref{fig_bd_seds} shows the average SEDs/colours of bulges and disks-- normalized to the $r$-band -- as derived from the fits for three different magnitude ($r$-band) ranges and their scatter (16 and 84 percentiles) of the values in the second panel.
For this plot, we use the largest possible common sample for each component, i.e. all multi- and single-band fitting produces a \lq good\rq\ result.
However, we exclude the $u$ and $z$ single-band fits from this definition, as they would decrease the sample size too drastically, i.e. we require $griYJHK$ single-band fits to return a \lq good\rq\ result.
For these 2 bands only, we additionally limit the sample size in that band, i.e. $ugriYJHK$ and $grizYJHK$, respectively.
While this selection limits the sample size in this plot and in each band, it makes the average SEDs directly comparable, as the same sample is being used at most wavelengths.
In all panels, more disks than bulges are used, as disks can generally be recovered better, as we have seen in \S \ref{sec_sim_results}, although the difference in numbers is somewhat more extreme here.

It becomes clear from this plot that, at least statistically, single-band fitting can not separate the bulge and the disk in galaxies (blue and red dashed lines, respectively), as both components on average are recovered with the same SEDs, even in the brightest galaxies.
As such, also any $B/T$ ratios derived using single-band fitting in each band would be highly inaccurate, as we have also shown in Fig. \ref{fig_sim_BT_comparison}.
In the middle panel, we show the width of the recovered colour distributions as error bars, thin lines indicating single-band results.
It becomes clear that the colours recovered not only are very similar on average, but also their distributions are nearly identical, indicating that the single-band fits are unable to separate the flux of the two components.
In contrast, the multi-band fits (blue and red solid lines) successfully recover very different SEDs for both components, and there seems little effect of the galaxy brightness on these SEDs.
The error bars, thick lines showing multi-band results, also show much less significant overlap in the distributions.
The exceptions here are the $u$ and $z$-bands in which the reduced sample shows significantly wider spread, especially in the galaxy disks.
We also notice a dip in the $J$-band disk SED (blue solid line), which also seems present in many of the objects at all brightnesses.
We are unsure what causes this dip, but as it is not present in the average bulge SED (red solid line), we conclude that this is not a data (e.g. zeropoint) issue.

For comparison, we also overplot the SEDs (using values from NED\footnote{\href{https://ned.ipac.caltech.edu/byname?objname=M89}{https://ned.ipac.caltech.edu/byname?objname=M89} and\\ \href{https://ned.ipac.caltech.edu/byname?objname=NGC0337}{https://ned.ipac.caltech.edu/byname?objname=NGC0337}}
) of a typical elliptical galaxy (M89, in the Virgo cluster, orange dashed line) and a known galaxy without a prominent bulge (NGC0337, classification SB(s)d, green dashed line).
Especially in case of the NGC0337, multi-band fitting can recover this typical SED quite well as the average SED of galaxy disks. 
The average bulge SED also agrees well with the SED of a quiescent galaxies, but note the offset in the redder bands.
As M89 is a relatively massive galaxy and in a cluster, this difference might be at least partially attributed to differences in metallicity or in the age of the stellar populations. 
Indeed, the difference is found to be smallest for the brightest sample, where the most massive galaxies are more likely to be.

\begin{figure}
\begin{center}
\includegraphics[width=0.48\textwidth,trim=35 30 8 15,clip]{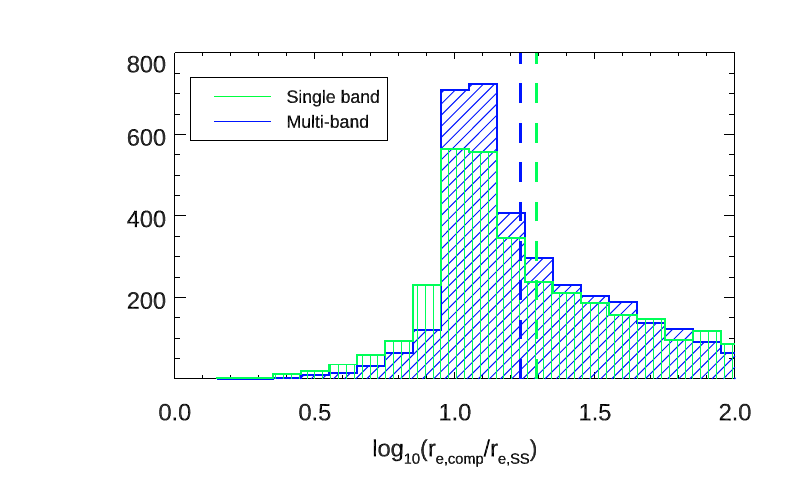}
\includegraphics[width=0.48\textwidth,trim=35 5 8 15,clip]{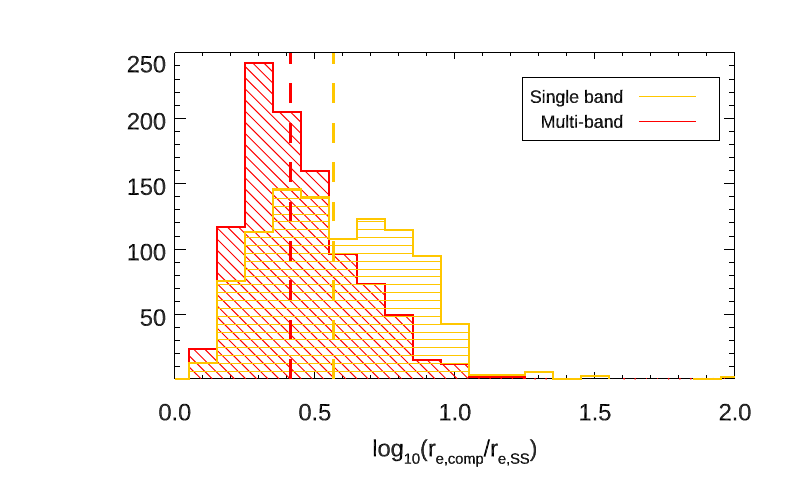}
\caption{Sizes of disks (top) and bulges (bottom, notice the different scale on the y-axis) in comparison to single-\sersic sizes. Blue and red histograms show the values from multi-band fitting, green and orange histograms show the values from single-band fitting. Dashed vertical lines indicate the median values for each histogram.
}
\label{fig_bd_sizes}
\end{center}
\end{figure}

As a further sanity check, we have a look at the component $r$-band sizes in Fig. \ref{fig_bd_sizes} in comparison to the single-\sersic sizes for all galaxies with $m_{\rm r}<18.5$ (samples using different magnitude cuts look similar).
The samples shown for disks and bulges, respectively, are the same in both single- and multi-band fits, i.e. both fits produce a \lq good\rq\ fit result for the respective components, so the histograms for disks (or bulges) should be directly comparable.
Again, one can easily see from the scale on the y-axis that \galfitm produces \lq good\rq\ fits for disks more often, the disk sample being significantly larger.
Multi-band results for disks are given in blue (red for bulges), while green histograms (orange for bulges) show the single-band results for that component.
Vertical dashed lines show the median values for each histogram.
It is well known in the literature that bulges are in general smaller than the disks they reside in \citep[][and many others]{Lange16} and one would expect this to be visible in such a plot.
And indeed, in both cases -- single- and multi-band fits -- the bulges of these galaxies are much smaller than the measured single-\sersic sizes, and the disks are significantly larger on average.
For these real objects, it is impossible to tell which of these methods produces the better fit, but it ties in well with our finding in \S \ref{sec_sim_results} that the 2 methods produce different results, although some trends seem to be reversed here.
For the simulated galaxies, single-band fits were found to provide larger sizes for disks and smaller sizes for bulges (see Fig. \ref{fig_sim_dre}).
In the real galaxies presented here, the behaviour is in fact the opposite with single-band fits producing larger bulges than multi-band fits.
It is somewhat unclear where this difference originates.
However, we can speculate that it might be connected to the fact that in the simulated data both disk and bulge were simulated (and fit) with a constant size across all wavelength, while this might not true in real galaxies \citep[e.g.][]{Xilouris}.
However, real galaxies are still being fit with a constant size across all wavelengths.
These fits can hence be influenced by the colour gradients present within the individual components and by the effect of dust, both of which are not present in the simulated data.

At this point, we would like to remind the reader that the starting values for bulge and disk sizes in these fits are defined as $r_{\rm e,D}=1.2*r_{\rm e,SS}$ and $r_{\rm e,B}=0.3*r_{\rm e,SS}$  (see \S \ref{sec_setup_starts}).
Hence, it seems suspicious that especially the bulge sizes on average are found to somewhat centre around this value.
While in principle, this can be induced by the starting values, we have verified that this is not the case by refitting a large number of objects by using different starting values, e.g $r_{\rm e,B}=1.0*r_{\rm e,SS}$,  $r_{\rm e,B}=0.5*r_{\rm e,SS}$ and $r_{\rm e,D}=1.0*r_{\rm e,SS}$.
The resulting histograms show very similar results, at least for bright galaxies, and we will briefly discuss the different starting values in Appendix \ref{sec_start_params}.
The values found in these tests were in fact one of the reasons why we chose the starting values the way we did.

%__________________________________________________________________
\section{Comparison to other work}
\label{sec_comparison}
In the following sections, we present several comparisons to fits carried out with other light profile fitting codes and/or data, and discuss the shortcomings and uncertainties of some of these codes, including our own.
These sections can serve as a warning against taking the results from any of these codes and methods as the unbiased truth, especially in certain parameters.

\subsection{Comparison to \sdss from Simard, 2011}
\label{sec_real_simard}
\begin{figure}
\includegraphics[width=0.48\textwidth,trim=0 10 8 10,clip]{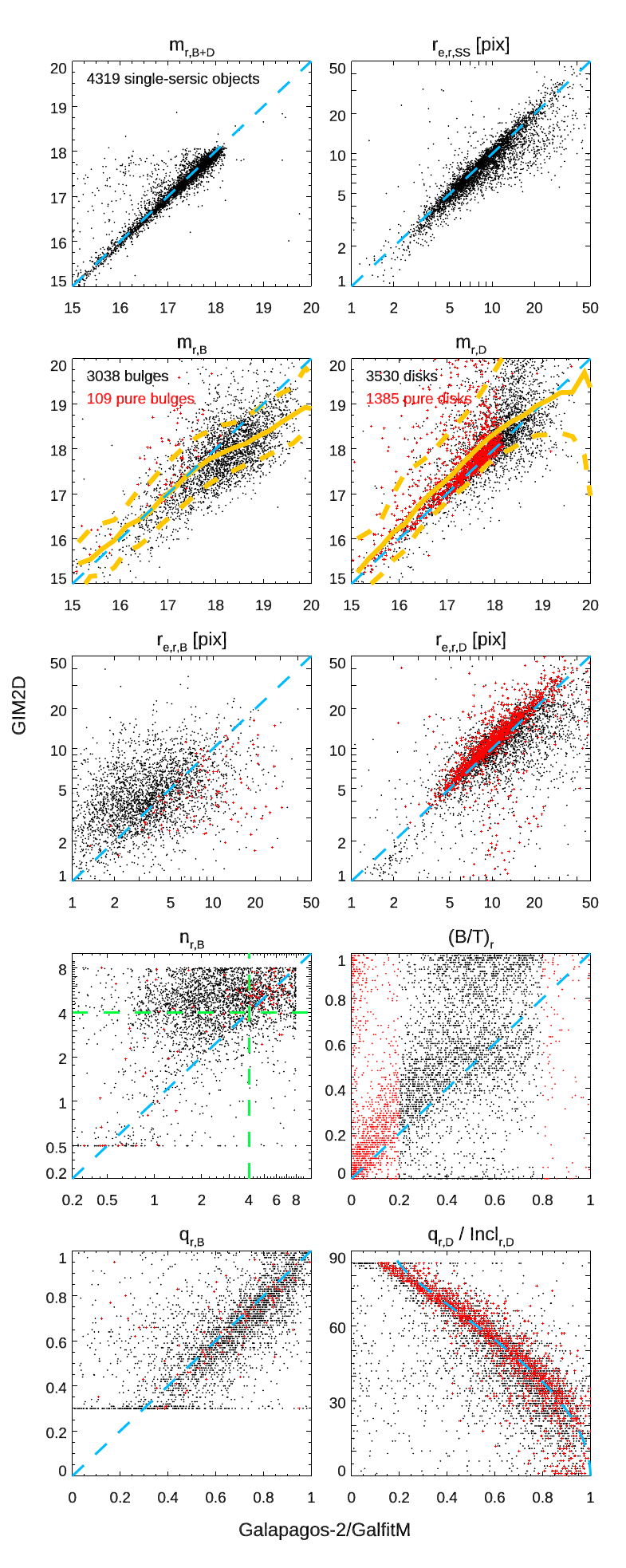}
\caption{Comparison of fitting parameters with \gimtwod from S11. Red points in each panel indicate \lq pure bulges\rq\ and \lq pure disks\rq, respectively, for which we plot the values from the single-\sersic fits. A line showing \lq perfect agreement\rq\ is overplotted as a dashed light blue line on all panels. For further details, please see the text. 
}
\label{fig_comparison_simard}
\end{figure}

The most prominent publicly available catalogue of a large sample bulges and disks has been presented by S11.
In this work, they use \gimtwod to fit 1.2 million galaxies in \sdss data at $m_{\rm petro,r,corr} \leq 17.77$.
By comparison with Fig \ref{fig_magbd-vs-mag}, it is clear that these galaxies compare to the brighter end of our galaxy sample, so supposedly the \lq easiest\rq\ galaxies to fit.
In the previous sections of this paper, we showed that we generally reach objects \textasciitilde2 magnitude fainter than this limit.
Since the optical bands of the \gama data that we use throughout this work were created by resampling the same \sdss data used in S11, i.e. the data used by us and by S11 are nearly identical, these catalogues are well set up for direct comparison.
However, due to the magnitude limit imposed in S11, and the much smaller survey area of \gama compared to \sdss, a direct overlap of the sample is small, so we analyse here \textasciitilde 5000 galaxies present in both datasets.

Some caution has to be taken with the results in this section, however, as these galaxies are also the largest and most detail-rich galaxies in our sample. 
Experience with \galapagostwo shows that these galaxies are somewhat harder to fit using a fully automated method.
Generally speaking, smoother profiles of more distant galaxies are somewhat easier to fit, as no spiral arms or other internal features influence the fit as heavily as in some of these cases. 
Nearby objects are also more likely to be split up into several detections when using a fully automated approach, which we try to minimise by our \lq hot\rq\ and \lq cold\rq\ setup approach for \sex (please see H13 for details).

In Fig. \ref{fig_comparison_simard}, we compare the main parameters of the profile fits in the $r$-band, as this is the deepest image in the dataset and has been used as the \lq main\rq\ band in both these works.
Each panel shows our values from multi-band fitting along the x-axis, and the values from S11 on the y-axis, each plot is labeled with the parameter it shows.
In this figure, we use the largest possible sample for bulges and disks, respectively (3038 bulges, 3530 disks).
As we have seen before, disks are generally fit well more often, which is why the sample in the panels showing disk parameters is somewhat larger, although in these bright objects, this is less of a problem than in \S \ref{sec_real}.
Where both parameters are required (e.g. for $B/T$), the sample is restricted to those objects for which both components returns good results.
The red points in each panel indicate the additional objects with $B/T>0.8$ (109 objects) in case of bulge parameters and $B/T<0.2$ (1385 objects) in case of disk parameters.
For these objects, we have shown in \S \ref{sec_sim_results} that the parameters of \textbf{both} components are not reliably recovered, as the fainter component tends to fit part of the brighter component.
Instead, we plot the results from the single-\sersic fit which we have shown reflects the parameters of the brighter component better in the respective panel.

The top 2 panels show the agreement between the total galaxy magnitude on the left (parameter \verb|RG2D| for \gimtwod, $m_{\rm D}+m_{\rm B}$ for \galfitm), and the single-\sersic half-light radius (parameter \verb|RHLR| for \gimtwod, converted to pixel values) on the right.
To guide the eye, the green dashed line marks the 1-to-1 line, indicating perfect agreement.
For these global parameters, both codes agree very well, although some outliers (e.g. through mis-identified objects, or galaxies split up during the detection process) are visible.

The next 2 panels compare the bulge magnitudes on the left, and the disk magnitude on the right.
The \gimtwod values in these panels are derived by using the total magnitude \verb|RG2D| and the bulge-to-total ratio \verb|__B_T_R| as given by S11, while the values for \galfitm are a direct output of the code.
In general, the agreement between these values is good, but the scatter, even on these bright objects, is substantial. 
Differences of 1 magnitude between the 2 codes are common, although only a systematic deviation for the average bulge magnitudes of fainter objects is visible.
There also is a small offset in that \gimtwod systematically fits disks a little fainter, but this is a small effect.
The red points, indicating the single-\sersic results of \lq pure\rq\ bulges and disks align well with the findings of the other objects, indicating that our decision to use single-\sersic results in those cases helps to recover the true component parameters.

The central panels show the size of bulges (left, parameter \verb|RE| in S11) and disks (right, parameter \verb|RD|), respectively.
As these values are given in $kpc$ in the S11 catalogue, we transferred them back into arcseconds using the redshifts provided by S11, and further into \gama pixels, as this is the parameter we worked with throughout most of this paper. 
We further take into account a factor of $r_{\rm e,D}=1.67*scalelength$ for disks, as S11 provides those values instead of half-light radii.
Disk sizes are generally very consistent between our fits and the ones presented in S11, although some scatter is visible for large galaxies where \galapagostwo seems to measure somewhat larger sizes.
This consistency is not true for the bulge sizes, the scatter between the measurements is very large, and the correlation between the sizes measured is quite weak.
Interestingly, the \lq pure bulges\rq\ in our sample are consistently measured larger in our analysis than in the S11 measurements.
We are unsure why the agreement between the codes is so weak, but it is worth noting that after analysing the same plot for bright bulges only (no plot shown), this effect is not dominated by faint bulges, but seems to be apparent over all bulge magnitudes.
There seems to be no trend with bulge brightness beyond the general effect that fainter bulges seem to be fit smaller by both codes, which naively makes sense in the context of fainter galaxies likely being further away.
One could speculate here that this is again an effect of object detection and might be dominated by split up galaxies, but we cannot show this from the work done here.

A similarly weak correlation between the parameters recovered by the two different codes can be seen in the panel showing the bulge \sersic-index $n_{\rm B}$ (parameter \verb|NB| in S11), 4\textsuperscript{th} row, left panel.
As is increasingly common in the literature, we show the \sersic index on a logarithmic axis.
While both codes generally seem to recover relatively high values of $n_{\rm B}$, the values from \gimtwod seem to be somewhat higher than one would expect for classical bulges (green dashed lines).
Indeed, we have seen in Fig. \ref{fig_sim_dn} that faint bulges seem to be underestimated by \galapagostwo in their \sersic index to some extent.
Upon inspection (no plot shown), we can indeed confirm that this effect can be seen somewhat in this plot, galaxies with low $n_{\rm B}$ values being more likely to be faint objects.
We note that S11 seems to impose a lower limit of $n_{\rm B}\ge0.5$ during their fits, while our values are limited to $n_{\rm B}\ge0.2$.
As these small values are only found in few objects, either choice seems somewhat arbitrary.
Both codes use an upper limit of $n_{\rm B}\le 8$.

As both S11 and ourselves fit galaxy disks with $n_{\rm D}=1$, it makes no sense to show a comparison of this value. 
Instead, in the right panel, we show the comparison of the $r$-band $B/T$ ratio as derived by the codes, where S11 provides this value directly (parameter \verb|__B_T_R|).
Whereas both bulge and disk magnitudes showed clear correlations, this correlation is washed out in this panel, as $B/T$ is defined as a fraction of parameters, and hence more sensitive to small changes.
Although \gimtwod systematically returns higher $B/T$ values (worse to at higher $B/T$ galaxies), a very noisy correlation can clearly be seen.
This difficulty to recover the same $B/T$ values is of course connected to the discrepancies in the bulge sizes and $n_b$ values that are recovered, as these 3 parameters are known to correlate in \sersic fits.
Upon inspection (no plot shown), this correlation is somewhat tighter for brighter galaxies, as one would expect.
Especially catastrophic outliers appear to happen less regularly in objects at $m_{\rm r}\le17$, in agreement with our findings shown in Fig. \ref{fig_sim_BT_comparison}.
The objects with $B/T<0.2$ (left) and $B/T>0.8$ (right) which we call \lq pure disks\rq\ and \lq pure bulges\rq\ can be seen in this plot directly. 
In fact, this is how they are selected.
For the \lq pure disks\rq, the recovered $B/T$ values by S11 are indeed predominantly low, while most \lq pure bulges\rq\ are recovered at high $B/T$ values by S11.

In the last row, we show the axis ratio $q_{\rm B}$ (parameter \verb|E| in S11 of the bulge on the left, and the relation between the axis ratio $q_{\rm D}$ for \galfitm and the disk inclination (parameter \verb|I| in S11) for \gimtwod on the right.
In the case of the disk axis ratio, we have assumed an intrinsic thickness of the galaxy disk\footnote{defined as $h/r$, where h is the scale height of a disk and r is its scale length.} of $0.18$, following \citet{Pizagno} to transfer the axis ratio measured by \galfitm into an expected inclination angle to define a 1-to-1 comparison line (light blue dashed line).
As one can see, both parameters agree well between the 2 codes, despite individual outliers.

In conclusion, \galapagostwo/\galfitm and \gimtwod agree well for the disk parameters in the common sample analysed here.
However, the bulge parameters, especially $n_{\rm B}$ and $r_{\rm e,B}$ differ significantly between the two codes.
As has been mentioned above, this does not necessarily mean that one code performs better than the other.
However in \S \ref{sec_sim_results} we analysed the accuracy of the \galfitm fits themselves, and find better agreement with the input values than the one shown here.

\subsection{Comparison to \kids/\viking fits}
\label{sec_real_deep-shallow}

So far, we have used the same data -- based on \sdss and \ukidss data -- used in H13 throughout this paper to enable direct comparison.
This was also useful for our comparison to S11 in \S \ref{sec_real_simard}, as the data they used was very similar.
However, as has been mentioned in \S \ref{sec_data}, newer data has since become available in \gama, which is substantially deeper, as its uses \kids and \viking data.
As such, this data allows a comparison on how the code performs on these different -- and crucially: entirely independent -- datasets.

In Fig. \ref{fig_deep-shallow} we show this comparison for the galaxies fit in both datasets.
The 2 leftmost columns of plots show the entire common sample, the right 2 columns restrict the sample to $m_{\rm r,SEx}\le18$, using the \sex measurements in the deeper \kids/\viking data as independent reference for comparison and discussion.
While in \S \ref{sec_sim_results}, we have shown that such a simple magnitude cut-off does not fully ensure a clean sample as the fit quality is more likely a function of a mixture of morphological parameters, including $B/T$, it serves here as a simple cut to restrict the analysis to the objects \lq more likely\rq\ to produce accurate fit results.

The panels shown in this plot are analogous to the ones shown in Fig. \ref{fig_comparison_simard}, with the difference that we can compare the axis ratios of disks directly, rather than comparing an axis ratio to an inclination angle, assuming some intrinsic disk thickness, as had to be done before.
Please note that in this plot, we show the results from the deeper \kids/\viking data on the x-axis, and the results from the \sdss/\ukidss data on the y-axis.
In both datasets, the selection of \lq good\rq\ sources was carried out using the same definitions as given above. 
An additional selection of extended objects, using the $\verb|CLASS_STAR|<0.8$ values provided by \sex on the shallow data, has also been applied to remove likely stars within the field. 
We will comment on the effect of this additional selection below.

In the first panels of Fig. \ref{fig_deep-shallow}, we again compare the single-\sersic fits.
Both fits recover very similar magnitudes, visible as the good agreement between the measurements, with a small offset of 0.07 mag over all objects with brighter objects recovered on \kids/\viking data.
With \textasciitilde 6000 objects shown in this panel, there are only few outliers visible.
However, for galaxy sizes, it becomes clear that the shallower \sdss/\ukidss data fits somewhat larger sizes to most objects, on a level of 10-15\% (median: 12.7\%).
This is even true for the \textasciitilde 1700 brightest galaxies, shown in the 2 rightmost panels, where a similar offset is visible. The difference between the recovered magnitudes is reduced to 0.05 mag in this bright sample.
From our data, we can not trace the exact reason for these offsets.
However, it seems likely that the reason for the size offset is connected to one of 2 issues.
First, the native pixel scales of the images that went into the \swarp-ed images used in the 2 datasets are different.
\sdss has a native pixel scale of 0.396\arcsec/pix and \ukidss has a pixel scale of 0.4\arcsec/pix (using \wfcam), while the data used for the newer/deeper images shows a native pixel scale of 0.21\arcsec/pix for \kids (\omegacam) and 0.339\arcsec/pix for \viking (\vircam).
These smaller initial pixels should in principle allow resolving smaller objects, even if the data is resampled to a common 0.339\arcsec/pix scale in all cases.
Second, the typical seeing reported in these surveys is different, with a reported median PSF full-width-half-max (FWHM) of \textasciitilde1.4\arcsec\ in the r-band for \sdss \citep{StoughtonSDSS} and <1.2\arcsec\ for \ukidss \citep{ukidss}, compared to <0.7\arcsec\ for \kids \citep[e.g.][]{Venemans2015} and <1.0\arcsec\ for \viking.
While \galfitm in principle takes the PSF into account and should be able to measure galaxy sizes significantly smaller than the PSF FWHM if the PSF is well known (see e.g. H07 and H13), it is not unfeasible to assume that this works better in data with a smaller PSF.
This could \textit{especially} be true if the data additionally shows a smaller pixel scale (leading to better spatial sampling of the PSF) and if the data is significantly deeper, as is the case here.
Furthermore, the quality of the PSF itself will be different between the 2 datasets, even though the same software, \psfex \citep{PSFEx}, has been used with slightly different setups.
Unfortunately, neither of these are effects that we can easily identify in the simulated data, as the input PSF was by design identical to the one used by \galfitm.
Both effects should become visible more strongly for small objects, which indeed seems to be the case, as we will discuss below.

Bulge and disk magnitudes (second row of panels) seem to be recovered equally well as in the comparison between \gimtwod and \galfitm presented in Fig. \ref{fig_comparison_simard}, especially noting the different magnitude ranges shown. Galaxies in this deep-shallow comparison are up to 1-1.5 magnitudes fainter than the ones examined in the previous section.
Over all, no systematic offsets but significant scatter can be seen.
Visible trends are very similar when looking at the bright objects only.
These plots also confirm our decision to call objects with $B/T>0.8$ \lq pure bulges\rq\ (305 objects) and objects with $B/T<0.2$ \lq pure disks\rq\ (3402 objects) and use the single-\sersic results instead, as shown in red.
The distribution of these points generally seems to follow the overall distribution of other objects -- 1644 bulges and 2569 disks, respectively -- but seems to be asymmetrical in the magnitude comparisons.
This offset to fainter magnitudes in the shallower \sdss/\ukidss fits of the \lq pure\rq\ objects is a result of the way we define these \lq pure\rq\ objects in the deep data, but base the decision which parameter to plot ($B/D$ or single-\sersic) on each dataset individually, in order to be consistent with a scientific analysis if only one dataset existed.
It is hence possible that a \lq deep\rq\ single-\sersic parameter is plotted against a \lq shallow\rq\ $B/D$/component parameter, which would, by design, favour fainter magnitudes for the \sdss/\ukidss data.
Instead of over-plotting those objects for which \textbf{both} the shallow and the deep fits measure $B/T>0.8$ or $B/T<0.2$, i.e. both fits show single-\sersic parameters, we refer to the top left panel, where we can see the much better agreement between the single-\sersic magnitudes from both fits.

Bulge sizes agree relatively well within the scatter between the two datasets, with the exception of very small galaxies at $r_{\rm e,B}<3 [pix]$, which are fit larger in the shallower \sdss/\ukidss data. 
As one would expect, small bulges, embedded in a galaxy disk are the most challenging components to fit, and uncertainties are expected to be large.
Unfortunately, in our simulated dataset very few bulges showed $r_{\rm e,B,sim}<2 [pix]$ as those simulations were created to mimic the results from the shallower \sdss/\ukidss data, where those small objects are missing, so we are unable to conclude whether the fits from the \sdss/\ukidss or the \kids/\viking data are more accurate in this size range.
Figure \ref{fig_sim_re_in_out} revealed no such systematic offset between input and output sizes in the few galaxies present, which seems to suggest that the fits on \sdss/\ukidss data indeed did behave as expected on simulated galaxies.
However, uncertainties in the PSF shape present in real data, but absent in the simulated dataset, further prevents us from determining which of the datasets and fits reveal the true value.
The arguments discussed on image resolution and additional image depth would/should suggest that fits on \kids/\viking data are better.

No such size dependent effect can be seen in the panel showing disk sizes $r_{\rm e,D}$.
Apart from the general offset discussed above, no systematic trends are seen and the measured disk sizes generally agree well.
This might be an effect of no small disks being present in either dataset, so the issue at small radii is simply masked out.
Indeed, there is a hint of a similar flattening of the relation just before the cutoff at small values.
It is, however, interesting that all \lq pure disks\rq\ (as defined by the deeper \kids/\viking data) seem to recover larger sizes in \sdss/\ukidss data.

The next panel, as in Fig. \ref{fig_comparison_simard}, shows the agreement between the codes in the bulge \sersic index $n_{\rm B}$.
There seems to be a trend in that the fits on \sdss/\ukidss data recover somewhat smaller $n_{\rm B}$ values especially at high $n_{\rm B}$ values, something that we did not see in Fig. \ref{fig_sim_nb_in_out}.
Again, this trend is still visible when only taking the bright objects into account.
However, profiles with high $n_{\rm B}$ values rely particularly strongly on the innermost pixels for a good fit.
The PSF and resolution arguments made above hence would also apply here and could explain the trend seen.

The right side panel once again compares the $B/T$ values recovered.
As before, this plot shows only weak agreement as ratios of noisy magnitude measurements are inherently sensitive to small changes.
The red points indicate galaxies with $B/T_{\rm deep}<0.2$ and $B/T_{\rm deep}>0.8$ as measured on the deep data.
It is interesting to see that, while low $B/T$ values seem to agree relatively well between the 2 fits, high $B/T$ values are not very consistent.
On the contrary, objects with $B/T_{\rm deep}>0.8$ show a pretty uniform distribution in $B/T_{\rm shallow}$, although number statistics are small.
This effect is smaller when only bright objects are considered, but they still do not agree perfectly.
This is possibly an indirect effect that the different data quality and depth has on the fitting parameters.

Finally, in the bottom two panels, we show a comparison of the axis ratios of the recovered galaxy components.
There is a effect that $q$ values are higher -- objects are rounder -- when derived on \sdss/\ukidss data.
This effect is much more pronounced in bulges, disks seem to agree relatively well apart from a small and constant (\textasciitilde 0.05) offset, which decreases towards larger values of $q$.
Once again, this behaviour could be explained as an effect of data resolution and/or PSF uncertainties.
Especially in small objects, the axis ratio measurements become very sensitive to data resolution and the accurate knowledge of the PSF becomes critical.
Especially if the PSF model used is somewhat too elongated for any reason, measuring an elongated underlying objects becomes very challenging and the code tries to overcompensate for the effects in the PSF.

Most of the effects present in faint galaxies are also visible in the bright objects, presented in the right 2 columns, which is why we do not discuss those plots in detail.
However, we can use them to show that all these features are less pronounced in bright objects, as one would expect.
Especially $B/T$ values agree much better and a clearer correlation becomes visible.

However, there are a few features visible in these plots which we have ignored so far.
We have already discussed the offset in $r_{\rm e,B}$ at small sizes, but there are further prominent features at $n_{\rm B,deep}\sim8$ and $q_{\rm B,deep}\sim0.1$, both visible as vertical features.
These features are more pronounced for faint objects, and are less visible in the right panels that show bright objects only.
However these clusters of objects are not exclusively connected, i.e. made up by the same objects.
While many of the objects at low $q_{\rm B, deep}$ values indeed also show large $n_{\rm B,deep}$ values, and while \lq many\rq\ of these objects indeed show small $r_{\rm e,B}$ sizes, none of these features vanishes by excluding objects in the other.
However, the features seem to be connected to stars or stellar like sources, which is why we introduced the cut at $\verb|CLASS_STAR|<0.8$ in this analysis.
When further removing faint objects, these features become even weaker.
This \verb|CLASS_STAR| cut removes the vast majority of these objects, but such a hard cutoff values does not work perfectly.
We have tried cutoff at different values in the 2 different datasets, and have found that we can achieve a similar effect by using a lower cutoff $\verb|CLASS_STAR|<0.7$ in the \kids/\viking dataset.

\subsection{Comparison to ProFit}
\label{sec_real_profit}

Finally, we can compare our fits with fits derived using \profit \citep{PROFit}, a new profile fitting code that uses MCMC minimisation to derive the best fit profile, potentially outperforming the Levenberg-Marquardt approach used by \galfit and \galfitm.

\profit results galaxies on \kids/\viking data have been presented by \citet[][and Casura et al. in prep]{Casura} for the lowest redshift \textasciitilde13000 galaxies in the \gama equatorial survey regions, out of which \textasciitilde 3100 unique galaxies fall into the correct \gama-9h survey region analysed in this work.
Importantly, \profit is not run on the \swarp-ed mosaics that we use for \galapagostwo, but the original \kids images (single-band $r$-band only\footnote{Fits in several other bands are available, but are not discussed here. Results are qualitatively the same.}), which provide a significantly better spatial resolution than our images.
\citet{Casura} provide the parameters for 3 models: single-\sersic, \sersic + exponential and a point source + exponential.
In this section we only analyse the first 2 of these models, where appropriate, and select the single \sersic parameters for all galaxies which were flagged as 'good' fits by \citet{Casura}, and the bulge and disk parameters only for those objects where the flag provided by \citet{Casura} indicated that the 2-component model is indeed the best fit model to the galaxy.

This selection leaves a sample of 1903 objects for a single-\sersic comparison (top 2 rows, right columns in Fig. \ref{fig_profit}) and, after selecting only the objects with \lq good\rq\ fits from \galfitm, 427 2-components objects with \lq good\rq\ disks (and $0.2<B/T<0.8$), and an additional 168 \lq pure disk\rq\ objects, as well as and 227 \lq good\rq\ bulges (and $0.2<B/T<0.8$) and an additional 9 \lq pure bulge\rq\ objects (remaining panels), for which we plot the single \sersic parameters in red.
Numbers for the overlapping sample of the \sdss/\ukidss fits can be found in the respective panels.

There are a few important differences between how \profit is run compared to our own fits.
Most importantly, contrary to \galfitm, \citet{Casura} do not provide values integrated to infinite radii (although \profit itself does).
Instead, they provide a \lq segment radius\rq\ \verb|RAD_SEG| within which the model is valid, missing out additional flux at larger radii, and which they suggest is more reliable and should be used. This choice was made to produce better fit results in the postage stamp region and less sensitivity to potential problems in the outskirts of the galaxies, but makes a direct comparison somewhat challenging.
In order to allow a fairer comparison, it was necessary to convert the \galfitm values to match this definition for all profiles independently, single \sersic, bulges and disks.
In order to do this, we integrated our derived profiles out to this radius (assuming semi-major axis, while this is not strictly true as \profit allows segments to have any arbitrary shape), correcting the magnitude values, and then derived a new half-light radius which contains half of this flux.
This correction indeed improves the comparison of \galfitm and \profit values drastically.
As \sersic indices are untouched by this correction, comparing them could be challenging, as often the \sersic index of a profile is dominated by not only the innermost pixels, but the outskirts of a profile as well; profiles with high $n$ values contain more light in their outskirts.

Fig. \ref{fig_profit} shows the comparison of the fit parameters derived by \profit with our \sdss/\ukidss fits on the left, and our \kids/\viking fits on the right.
As \citet{Casura} use deeper \kids data, one would expect the comparison to our \kids/\viking values to be more accurate.
While this seems true in some parameters, e.g. $m_{\rm r,B}$ which shows a smaller number of outliers, it does not seem to be true in all parameters.
Particularly, $r_{\rm e,r,B}$ and $q_{\rm r,B}$ agree significantly better with the values from the \sdss/\ukidss fits.
Specifically, we see the same effects on both parameters that we saw in \S \ref{sec_real_deep-shallow} in that the \kids/\viking fits produce smaller and rounder objects than either \profit or the \sdss/\ukidss fits, while the agreement between those 2 codes seems to be better.
Given the assumed data quality, this seems surprising and it will require additional investigation why this is the case, PSF effects being one of the main suspects for this behaviour.
The agreement in the other parameters is generally good, especially for the disk parameters.
Also, at least for these bright objects, there is a clear correlation in the $B/T$ values, albeit noisy, which we deem encouraging.

\subsection{Discussion}
\label{sec_real_discussion}

As we have seen in this section, $B/D$ decomposition of real galaxies is a challenging task, and different methods/codes are prone to give different answers.
Even the same code on different data -- e.g. \galapagostwo on \sdss/\ukidss and \kids/\viking data -- can seemingly return different results for certain galaxies.
It might hence be worth commenting on some of the issues at hand when fitting galaxies and their components.

It is somewhat speculative what exactly leads to the differences seen, because we can not prove this with the data at hand in this work.
However, apart from the obvious differences in the imaging data used and the advantages of the \kids/\viking data regarding image resolution and depth discussed above, the only major difference between our runs on \sdss/\ukidss and \kids/\viking data, was the handling of the PSFs used for the fits.
To our understanding, many of the effects we see in the comparison between these fits -- and in fact, with the fits carried out using \profit -- can be explained by PSF effects. 
In this section we will briefly discuss the effects as we might expect them, and compare them to the Figures shown above.
While investigations into these PSF effects have not revealed any issues with the PSFs in any of the datasets, we discuss here the theoretical implications, as an attempt to explain the differences we see between the different fits.

In the comparison with other codes and data, 3 different methods for deriving PSFs for the fits were being used.
We do not include the \gimtwod fits in this discussion as the details of deriving the PSFs are unknown to us, although they are briefly discussed in their paper.

\begin{itemize}
\item In our \sdss/\ukidss fits, we used PSFs provided by \citet{Kelvin2012}.
These PSFs were derived on-the-fly, object by object, on a subsection of the images centred on each galaxy, using \psfex, in a mode that derives a PSF for the central galaxy/position only.
\psfex uses basis vectors to model the images of stars, and then interpolates their parameters to an intermediate position in order to create an artificial PSF image at that position.
While \sigmakelvin ensures that at least 10 stars are available for the task, using an area of up to 1501x1501 pixels in size, this method uses the limited number of stars available in this subsection of the image.
On average, \textasciitilde 24 stars have been used to create a PSF of 25x25 pixels in size, see \citet{Kelvin2012} for details.
While it was ensured that no major effects were visible in the resulting PSFs, the limited number of stars used can create some issues, as we will discuss.
The advantage of this method is that the PSF in principle was created for the exact position of the galaxy, especially as in the process it was ensured that only stars from the same \sdss and \ukidss images were being used, avoiding possible issues with galaxies close to regions of overlapping input images.
\item The PSFs used by \profit were created following a similar idea, using \profound \citep{PROFound} to identify up to 8 \lq good\rq\ stars around the object of interest.
These stars were then fit using Moffat functions, and the median value of each parameter was used to create a (perfectly centred) PSF model which is then used for the fit.
As \profit fits were run on the \kids images directly, no issues with overlapping tiles are expected in this dataset.
However, the limited number of stars used by this method can again affect the resulting PSF in the same manner discussed above.
Bad centring of individual stars could have a significant effect on the PSF shape, e.g. when a star falls directly between 2 pixels.
In such a case, the resulting fit could be too elongated and too large along its major axis, impacting on the resulting PSF.
While quality checks on the PSFs used have been carried out, it is unlikely that they work perfectly in all situations.
Specific investigations by \cite{Casura} looking for such effects found no indication of elongated PSFs.
\item For our fits on \kids/\viking data, we instead followed the advice of the \psfex author \footnote{communication on the \astromatic forum} and ran the code on much larger tiles, \textasciitilde 8500x8500 pixels, creating a grid of 10x10 or 20x20 PSFs (depending on the band, i.e. less PSFs in the noisy $u$-band), in order to use the full strength of the \psfex interpolation over large numbers of stars, using 3rd and 4th order polynomials.
The advantage of this approach is that it lead to significantly higher $S/N$ PSFs as they were typically derived from \textasciitilde 2200-2500 stars in all cases.
This method ensured that there was a PSF model within \textasciitilde 300 pixels (\textasciitilde100\arcsec) of the galaxy, but the PSF has not been created for the precise position of the galaxy.
While this would have been possible, it would have been unfeasibly CPU intensive to run this method for each object.
For the fit, we simply chose the \textit{closest} 75x75 pixel PSF to the galaxy of interest. 

As the \lq tiles\rq\ used in \galapagostwo are not identical to the native \kids or \viking tiles, but were arbitrarily sized cuts of the \swarp-ed images provided by \gama, edge effects would be expected in those areas where multiple \kids or \viking tiles overlap.
E.g. in an area where several \kids images/tiles overlap, a sudden jump of the PSF parameters could be expected, which would be difficult to model using the smooth functions used by \psfex.
This could, for example, lead to PSFs that are systematically too round/uniform, as they smooth over areas that show elongated or drop-shaped PSFs (i.e. the corners of \kids images), making them non-ideal for the galaxies in those areas of the images.
However, due to the number of stars used, centring effects or bad fits in individual stars should play much less of a role.
Visual inspection confirmed that the residuals between the derived PSFs and the individual stars used are minimal in the large majority of cases, residual level are typically less than 1.5\% in the central 20x20 pixels of the PSFs (2.5\% in the $u$-band).
In the images provided by \psfex to check the PSF quality, no jumps of parameters can be detected in either the PSF parameters themselves, or -- crucially -- in the strengths of the residuals along the overlap areas of images, indicating that these effects, if present, are not very strong.
\end{itemize}

The potential problems discussed in these cases can lead to different effects on the fitting parameters.
Unfortunately, as has been mentioned before, we can not test our hypothesis directly by using our fits on the simulated data, as by design the PSF used to create the images and the PSF used for the fit were identical.
A full analysis of these PSF effects would require very sophisticated simulated images, that mimic the overlap of different tiles in the \swarp-ing process and is well beyond the scope of this work. 
However, we can find indications in the plots where we compare the fitting results between different codes.

In case that the \kids/\viking PSFs are generally too round and uniform compared to the true PSF shape present in a galaxy image, one would expect the measured galaxy axis ratios to be systematically under-estimated, as some of the \lq roundness\rq\ of a galaxy is already presented in the PSF.
This effect should be pronounced in small objects, as the PSF plays a more important role when fitting those.
Indeed, such an effect is visible when comparing the \kids/\viking fits to \sdss/\ukidss fits in \ref{fig_deep-shallow} and \profit fits in Fig. \ref{fig_profit}, especially in the bulge axis ratios.
Fits on \kids/\viking data show systematically lower axis ratios.
Bulges are generally smaller, so such an effect would be more pronounced.
However, the effect seen in the -- generally larger -- disk galaxies, is much weaker.
While it is possible that this effect plays some role in our fits, no such effect is seen when we compare our fits to the \gimtwod fits by S11, where the axis ratios of even the bulges show no systematic offset and generally agree well.

On the other hand, if the PSF shows some elongation -- especially at its centre -- due to centring issues or low $S/N$ from a small number of stars going into its creation, one could expect the opposite effect.
Fits would be overestimating the axis ratios of small objects, as the PSF already takes care of some of the elongation of a galaxy, at least when the elongations somewhat align.
This, again, matches what we see in the bulge axis ratio $q_B$ in Figs. \ref{fig_deep-shallow} and \ref{fig_profit}.
In order to make up for the residual flux perpendicular to the elongation of the PSF, the resulting size of the object would be over-estimated to minimise the residuals.
Indeed, this is what we see in the measured bulge sizes in both our \sdss/\ukidss and the \profit fits and additionally in the single-\sersic sizes in the \profit fits.
Interestingly, the bulge sizes measured by both \galapagostwo on \sdss/\ukidss data and by \profit on \kids/\viking, seem to show a hard lower limit.
Nearly no objects are fit at sizes $r_{\rm e}<2]$[pix] or \textasciitilde 0.6\arcsec.
There is an argument to be made that PSF and galaxy elongation don't usually align, so this effect should balance out for co- and mis-aligned (specifically at 90 degrees) galaxies. However, whether this is in practice true, and whether this is a linear effect that could in fact balance out on average, can not be solved here without significant follow-up studies.
\cite{Casura} also exclude very small objects from their catalogue during the model selection of 2-component objects. While this could explain the lack of objects with very small $r_e$ in \profit fits, the remaining objects suggesting a systematic upturn in the comparison with \galapagostwo fits on \kids/\viking data, it can not explain the differences between those fits and those on \sdss/\ukidss, where such a cleaning of the sample is done in a consistent manner.

We can not from our data conclusively decide which of the datasets provides the better fit and in this paper we concentrated on comparing multi-band to single-band techniques for $B/D$ decompositions.
However, this section highlights that deriving a good PSF model for the fit is paramount to a good fit, no matter which fitting technique is used, especially for small objects, both in single-\sersic fits and $B/D$  decompositions.
If the PSF is perfectly known, in principle all techniques should be able to separate bulges and disks well, as long as the data has sufficiently high $S/N$.

Besides the PSF effects, there are other issues that should be taken into account when fitting galaxies.
For example, we have seen even in the simulated data, which removes some sources of errors, that the \sersic indices of the bulges are very hard to measure.
Multi-band fitting helped deriving this parameter, but even with this technique and on relatively bright galaxies (see Fig. \ref{fig_sim_nb_in_out}), measuring $n_{\rm B}$ is challenging.
Arguably, the uncertainty is too large to separate classical from pseudo-bulges, at least in the kinds of data used in this work.
While this might be possible in well-resolved, nearby galaxies, where our multi-band approach using \galfitm would certainly improve the fit quality, this does not seem possible in large surveys of distant galaxies.
This could indeed suggest that a fit using a classical $n_{\rm B}==4$ bulge could be considered superior as it removes one large source of error and minimises the confusion between bulge and disk profile when both try to fit the same galaxy component.
However, we have analysed the fits with \galapagostwo of a sub-sample of galaxies with fixed $n_{\rm B}$ values on the \sdss/\ukidss and \kids/\viking data.
While we do not show plots for this test, we can report that we did not see a significant improvement in the comparison of these fits. 
In fact, both $r_{\rm e,b}$ and $q_{\rm B}$ seem to agree less well between the fits, while the other parameters do not seem to change significantly.

Similarly, it is clear from this work, that measurements of $B/T$ flux ratios are very unreliable on a individual basis, at least for \lq faint\rq\ galaxies.
If a science case requires such measurements, care has to be taken in the selection of a bright, reliable and/or sufficiently large galaxy sample such that individual bad measurements do not influence the science results.

Our comparison between codes also highlights the necessity for any user of any light profile fitting code to verify that the galaxy parameters are actually well measured, ideally by comparing different codes on a small sample of objects and/or carrying out tests on simulated images, similar to the ones used in this work.
More and more tools for creating such simulated images become available, and the detailed description in this paper can serve as a guideline on how to create such data.

%__________________________________________________________________
%%% Dust! %%%
\section{Effects of Dust}
\label{sec_dust}
As mentioned previously, we have not incorporated effects of dust attenuation in this work, since these are notoriously difficult to take into account and can only be properly quantified with radiative transfer calculations \citep{Xilouris,Popescu2000,deGeyert14}. 
This is well beyond the scope of this paper. 
Nonetheless, it is well known that the effects of dust attenuation are very strong in the ultraviolet, and, while in the optical bands they become very small for face-on disks, they can still produce a significant impact for non face-on galaxies \citep[e.g.][]{Driver2007}. 
Dust attenuation does not only affect the spatially integrated SEDs \citep{Tuffs2004,Natale22}, but also the surface brightness distribution of the direct stellar light \citep{Moellenhoff,Gadotti2010,Pastrav2013a,Pastrav2013b,Thirwall}.  Because of this, the derived photometric parameters, like exponential scale-length, effective \sersic radius, \sersic index, and total luminosity, would be different from the intrinsic ones as obtained in the absence of interstellar dust \citep{Pastrav2013a}. 
In addition, the bulge-to-disk decomposition itself would suffer an extra effect due to dust, meaning that the decomposed disks and bulges in the presence of dust would be different from those derived if the galaxy would only have an attenuated disk or an attenuated bulge \citep{Pastrav2013b}.
All these effects would depend on wavelength, inclination angle, dust opacity and bulge-to-disk ratio.
 
In order to account for dust effects one would need to use realistic, dust attenuated simulations of attenuated disks and bulges, instead of simple exponential or \sersic functions, since no analytic functions exist to describe the complex modifications induced by dust. 
Nonetheless, such modified surface brightness distributions can be calculated with radiative transfer codes, and indeed, such simulated images already exist in the literature \citep{Tuffs2004,Popescu}. 
So the question then arises of why we would not attempt to use these simulated images to test our \galapagostwo/\galfitm codes. 
The problem is that then, instead of fitting only one or two analytic functions with a few free parameters, as done in this work, we would need to find the best-fit distribution from a large data set of simulations corresponding to all combinations of parameters describing dust effects. 
In this paper we already showed that even simple function fitting is computationally a difficult task when dealing with large samples of galaxies. 
It becomes immediately apparent that complex distribution-fitting, is at present computationally impractical.
 
There is, however, a different solution to this problem, which has already  been addressed and solved outside the development of a photometric code, by \citet{Pastrav2013a,Pastrav2013b}. 
The idea in those works was to provide corrections for dust effects on parametric models. 
\citet{Pastrav2013a,Pastrav2013b} followed as closely as possible the procedures and algorithms used in photometric codes like the present one.
It is just that instead of using observations of galaxies or the simple simulated images used here, they used radiative transfer simulations for which the input parameters describing the distributions of stellar emissivity and dust were known. 
By comparing the input values of the parameters describing the simulations with the values of measured parameters describing simplified distributions (exponential and \sersic functions), like those used in this work, they were able to quantify the degree to which the photometric codes underestimate or overestimate the intrinsic parameters of galaxies. 
These corrections are listed in \citet{Pastrav2013a,Pastrav2013b}. 
They are provided in the form of coefficients of polynomial fits to the corrections as a function of inclination, for different wavelengths and a large range of dust opacity. 
 
We thus recommend using the dust corrections from \citet{Pastrav2013a,Pastrav2013b}\footnote{These dust corrections from \citet{Pastrav2013a,Pastrav2013b} can be downloaded at the following links:\\
\href{http://cdsarc.u-strasbg.fr/viz-bin/qcat?J/A+A/553/A80}{http://cdsarc.u-strasbg.fr/viz-bin/qcat?J/A+A/553/A80}\\
\href{http://cdsarc.u-strasbg.fr/viz-bin/qcat?J/A+A/557/A137}{http://cdsarc.u-strasbg.fr/viz-bin/qcat?J/A+A/557/A137}
}, to adjust the \lq apparent\rq\ (dust attenuated) photometric parameters derived from our code to \lq intrinsic\rq\ (dust de-attenuated) parameters. 
The corrections could be either considered for a typical dust opacity \citep[e.g. the Milky Way opacity from][]{Natale22} or could be derived on an individual basis from other methods (SED fitting, Balmer decrement, etc).

%__________________________________________________________________
%%% Discussion %%%
\section{Summary and conclusions}
\label{sec_summary}
In this paper, we have presented and released \galapagostwo, a multi-band fitting code to be run on large surveys that automatically carries out profile fitting (both single-\sersic and $B/D$ fits) to all objects in the survey.
We have presented a detailed analysis of the capabilities of \galapagostwo and \galfitm, as well as a comparison with other works.
As implications for single-\sersic fitting have already been discussed in H13, we have focused on presenting the advantages of multi-band fitting in the context of bulge and disk parameters, by comparing the code performance to the single-band fitting as usually carried out by using other codes.

Throughout the paper, and using both simulated and real data, we have shown that multi-band fitting presents a significant and important improvement over single-band fits, not only in terms of parameter accuracy itself, but also in terms of returning an actual \lq good\rq\ results.
Multi-band analysis allows a significantly larger sample to be used for scientific studies.
The improvements on measured parameters is also -- and especially -- true for the magnitudes and colours/SEDs of the different components, see Figs. \ref{fig_sim_seds} and \ref{fig_bd_seds}.
While in single-band fitting it is very difficult to derive a reliable SED for the individual components, multi-band fitting allows -- at least on average -- to measure SEDs of bulges and disks down to fainter magnitudes than is possible with other techniques, both on simulated and real data.
Examining parameter accuracy as a function of $B/T$ and galaxy magnitude, we were able to identify which galaxies can be reliably fit with 2 component models.
In this context, we were able to show that in objects with extreme $B/T$ values ($B/T<0.2$ or $B/T>0.8$), \textbf{both} components can not be reliably fit, and instead the single-\sersic fit provides a better match to the profile of a galaxy and the brighter component.
As this requires measuring the $B/T$ itself, this is somewhat of a circular argument, however, and can not seemingly be avoided. 
In such scenarios, conservative values should be used.

We have further compared the outcome of these $B/D$ fits to the results from other codes, where possible, and found that while they generally agree well, there were several specific differences, e.g. bulge sizes recovered by \gimtwod.
A detailed comparison to the literature has already been carried out in \S \ref{sec_real_simard} to \ref{sec_real_profit}, so we will not go into further detail here.

Lastly we have shown that using a \sersic index of a single-\sersic fit as a measure for the prominence of a bulge in a galaxy may not select high $B/T$ galaxies as intended, as the correlation between \sersic index and $B/T$ ratio is weak at most.

We emphasise that this work is based on 1-\citep{MMHaeussler1} and 2-component profiles (this publication), and the inclusion of additional components beyond a simple Bulge/Disk model (e.g. bar, halo, central light cusp, etc) might be advisable depending on the science case, data at hand, spatial resolution, as well as data quality/depth.
However, such profile fits often overfit the data and should hence, for now, only be carried out on an individual basis as they are by nature very hard to automate.
To the best of our knowledge, no systematic attempt has been carried out in the literature, but the \galapagostwo output files might serve as a good starting point for such an approach.
Machine learning techniques by themselves, or in combination with profile fitting, might open a way to such an analysis, but so far have not been proven to solve this issue entirely.

With this paper, we release the code\footnote{\galagithub} to the community for wider use.
In principle, this code is ready to be used on any large dataset.
However, due to a IDL limitation of dealing with maximum 16 galaxies in parallel, it would be advisable to develop a faster wrapper script for application to the large next generation surveys like \euclid, \lsst or \wfirst.
This could be done in a one-to-one translation of the IDL code, or by adapting the same principles, and consecutive testing.
We invite any developer to further improve the IDL version of \galapagostwo and submit a pull request on github such that new features can potentially be implemented in the code. 
We are also happy to provide guidance for such development.

Finally, we publish 2 catalogues of fit results for all \gama objects, using the \kids/\viking imaging data (see appendix \ref{sec_cat_release}). 
These catalogues provide both single-\sersic and $B/D$ fits for 234,239 objects.

%__________________________________________________________________
\begin{acknowledgements}
This publication was made possible by NPRP grant \# 08-643-1-112 from the Qatar National Research Fund (a member of Qatar Foundation). The statements made herein are solely the responsibility of the authors. BH and MV were supported by this NPRP grant.
SPB is supported by an STFC Advanced Fellowship. We thank Carnegie Mellon University in Qatar and The University of Nottingham for their hospitality.
E.J.J. acknowledges support from FONDECYT de Iniciaci\'on en Investigaci\'on Project 11200263.

GAMA is a joint European-Australasian project based around a spectroscopic campaign using the Anglo-Australian Telescope. The GAMA input catalogue is based on data taken from the Sloan Digital Sky Survey and the UKIRT Infrared Deep Sky Survey. Complementary imaging of the GAMA regions is being obtained by a number of independent survey programmes including GALEX MIS, VST KiDS, VISTA VIKING, WISE, Herschel-ATLAS, GMRT and ASKAP providing UV to radio coverage. GAMA is funded by the STFC (UK), the ARC (Australia), the AAO, and the participating institutions. The GAMA website is http://www.gama-survey.org/ . This work is based on observations made with ESO Telescopes at the La Silla Paranal Observatory under programme IDs 179.A-2004 and 177.A-3016.

The authors thank Kalina Nedkova for publicly providing a python script to interpolate the Chebyshev polynomials to arbitrary wavelengths, and/or for restframe correction.
\end{acknowledgements}

%-------------------------------------------------------------------
%\begin{thebibliography}{}
%[...]
%\end{thebibliography}
\bibliographystyle{aa} % style aa.bst
\bibliography{references} % your references Yourfile.bib

%-------------------------------------------------------------------
\begin{appendix}
\section{Introduction}
The following Appendices serve as a resource for additional information regarding the findings above. They also provide a overview of the software and catalogues released with this paper. 
In \S \ref{sec_app_sim_sims} we describe in detail the creation of the simulated data used throughout \S \ref{sec_sims}.
\S \ref{sec_start_params} will briefly discuss the effect of using different starting values for the $B/D$ fits.
\S \ref{App_gala_features} will introduce and explain features of \galapagostwo introduced into more recent versions of the code, some motivated by the results presented above, including a brief section on possible further improvements in \S \ref{sec_prospect} that might be implemented into the code in the future.
\S \ref{sec_cat_release} will describe the content of the catalogue of \kids/\viking fits - both single-\sersic and $B/D$ fits - released with this paper.

\section{Additional figures}
\begin{figure*}
\includegraphics[width=0.46\textwidth,trim=0 10 8 10,clip]{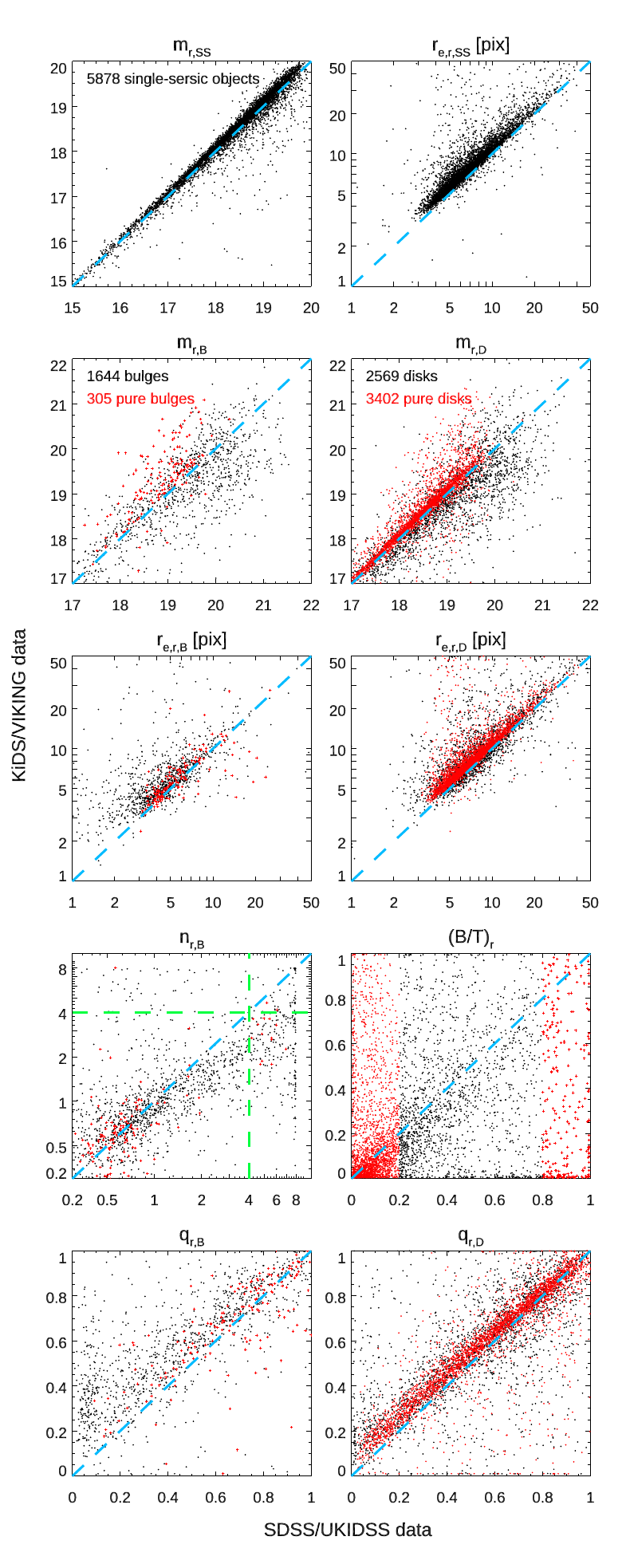}
\includegraphics[width=0.46\textwidth,trim=0 10 8 10,clip]{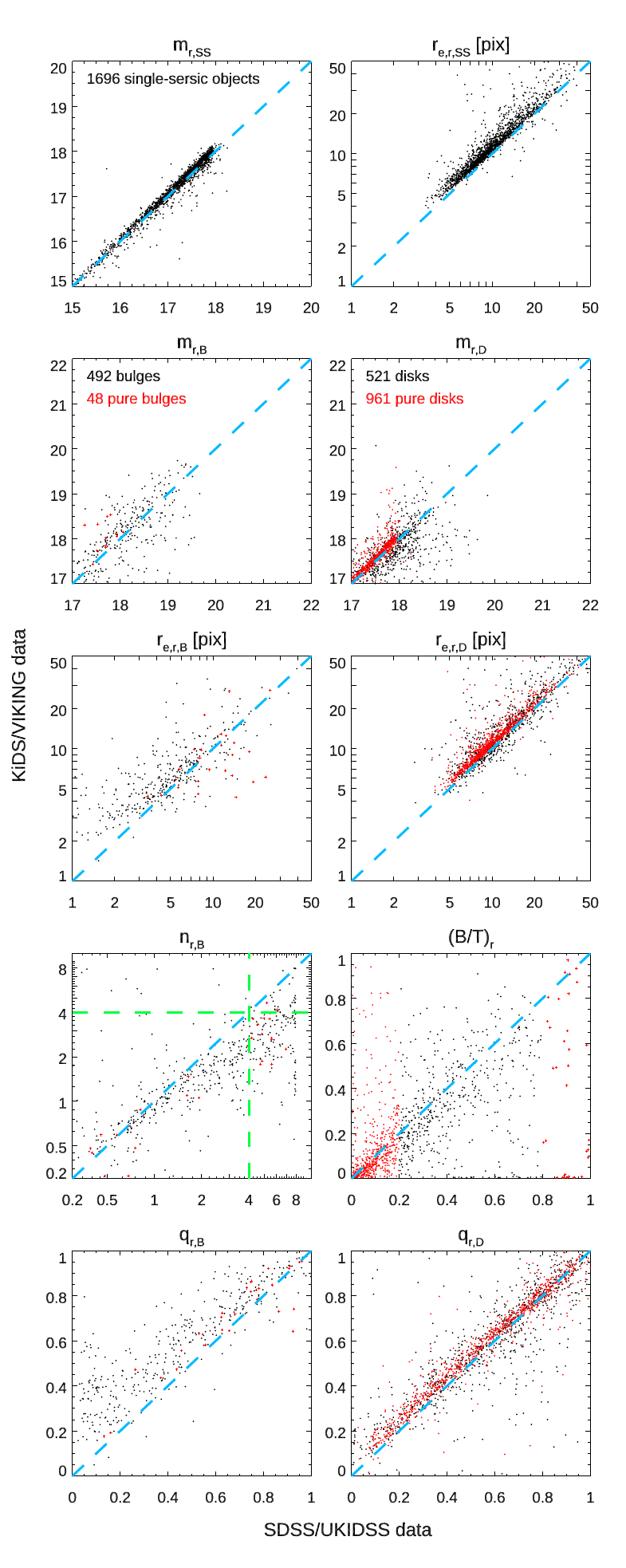}
\caption{Comparison of fitting parameters between \galapagostwo fits on \sdss/\ukidss and \kids/\viking data. Red points indicate objects with $B/T_{\rm deep}<0.2$ and $B/T_{\rm deep}>0.8$. Panels on the right show the same information for bright galaxies with $m_{\rm r,SEx}\le18.$. See text for detailed discussion.}
\label{fig_deep-shallow}
\end{figure*}

\begin{figure*}
\includegraphics[width=0.46\textwidth,trim=0 10 8 10,clip]{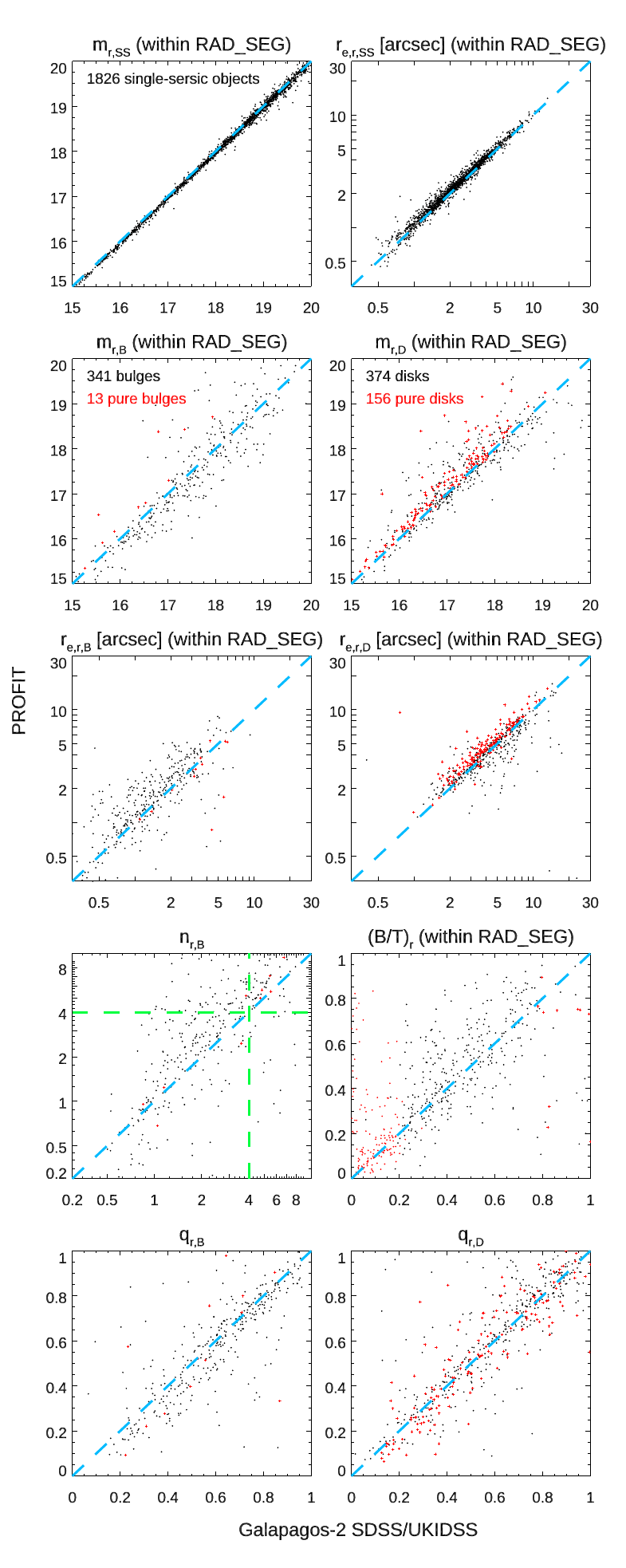}
\includegraphics[width=0.46\textwidth,trim=0 10 8 10,clip]{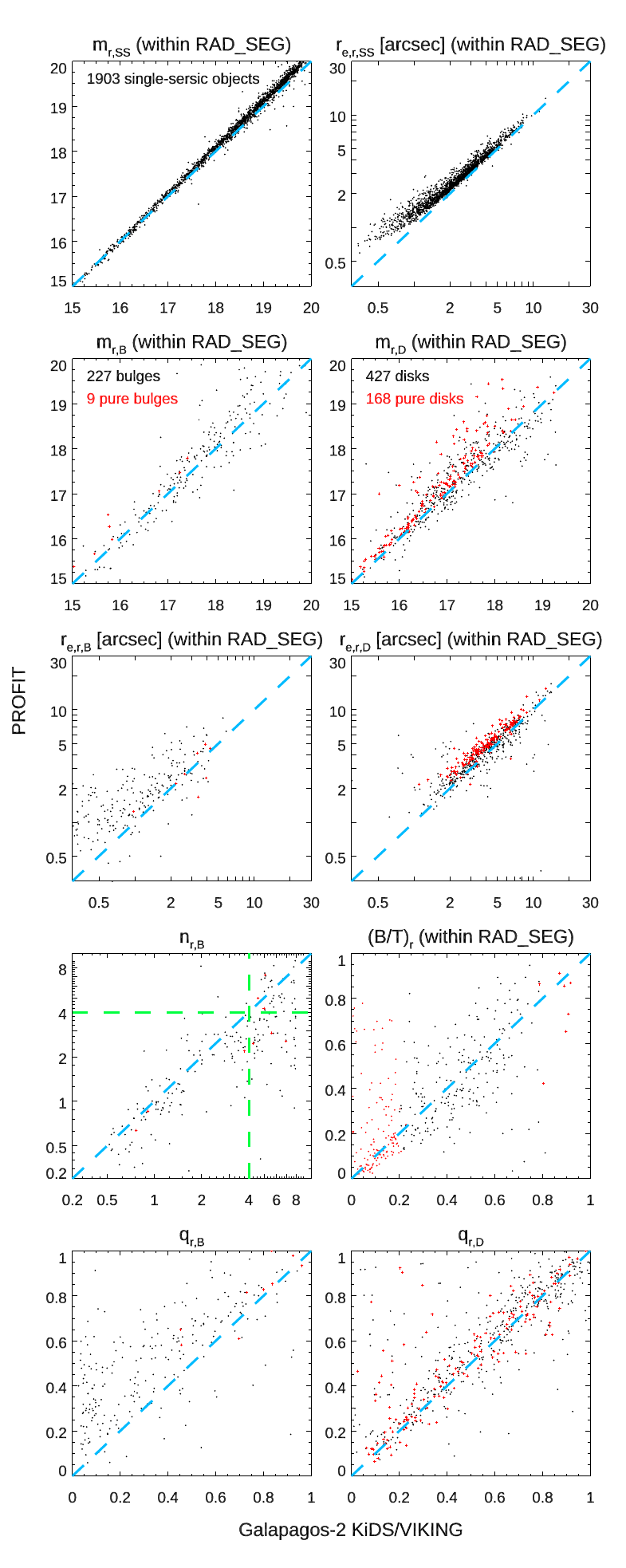}
\caption{Comparison of fitting parameters with \galapagostwo on both \sdss/\ukidss (left) and \kids/\viking (right) data with \profit fits (on \kids/\viking data only).}
\label{fig_profit}
\end{figure*}

\clearpage
\section{Image/Profile simulations}
\label{sec_app_sim_sims}
As mentioned in \S \ref{sec_sim_sims}, the simulated data used in this work have been created using the same scripts described in \S~5.1 of H13 and \S~2 of H07.
Naturally, however, given that 2-component objects are the main target in this work, a different input catalogue for the component parameters had to be used.
In this section, we explain the process of deriving the simulated parameters in detail for those readers interested.

We started from a catalogue of results from fitting profiles to real galaxies (please see \S~\ref{sec_real}), in order to get a realistic distribution of galaxy parameters.
In a first step, we created a catalogue of single-\sersic objects equivalently to the way we did in H13.
This assured that the overall parameters of objects (e.g. the distribution of sizes and total magnitudes) stayed comparable to the analysis carried out in the real data, making sure that our findings on simulated data apply to real data.
However, instead of passing this catalogue on to be simulated directly as was previously done, we introduced a few additional steps in order to transfer these objects into $B/D$  systems to be decomposed and to optimise the simulations to their purpose in this work.

While many of the objects in H13 were too faint to be detected by \galapagostwo/\sex, objects just above the detection limit were required to realistically represent faint neighbouring sources.
As the purpose of these simulations is somewhat less basic, we decided to include fewer very faint objects, in order to minimise the time needed for the simulations and fits.
We have already shown in H13 and H07 that these faint objects do not have a significant effect on the fitting results of the brighter primary sources.
As deriving a detection completeness was not part of this analysis, reducing the number of these objects around the detection threshold should not influence our results significantly.
However, we tried to keep the object density in the images roughly constant, in order to not introduce additional challenges to the code by increased object overlap/confusion or -- vice versa -- make it too easy for the code to fit galaxies as no neighbouring objects are present.

While we ran \galapagostwo on all detected galaxies -- the object magnitude of simulated sources peaks around $m_{\rm r} = 20$ -- objects with a brightness close to the sky brightness are unlikely to be successfully decomposed by any method. 
As such, these faint galaxies were of somewhat less importance in this analysis, as only few of them were likely to be fit well, especially in single-band fits, and indeed only presented results for galaxies at $m_{\rm r,B+D}<19.5$, well above the detection limit of these data.

We found that bright galaxies were very rare in this simulated catalogue, making an analysis difficult with \textasciitilde 10 objects in certain plots/bins (e.g. galaxies with $m_{\rm r}<17.5$ and successful fits in most/all single-band fits).
We hence artificially increased the number of bright objects at the expense of very faint objects, to keep the over all number of galaxies constant, by forcing 1\% of objects to have magnitudes in the range $11.5< mag_r <18$.
As we have already established in H13 that the object density and the number of neighbours is not critical in these data, i.e. the objects don't strongly influence each others fits given a typical object density in \gama, this adaptation is unlikely to change our conclusions, but provides better statistics for the objects we are most interested in.

We also decided to exclude {\bf very} small objects $r_{\rm e} < 1px$ for this work as these objects were not expected to be decomposed into components successfully and the performance of the single-\sersic fitting has already been tested in H13.

Of all simulated galaxies, we kept 20\% as single-\sersic objects using the parameters chosen at this stage, in order to be able to test any single- vs multi-component classifiers from our simulations. 
While this is a very important issue, it has not yet been solved satisfactorily.
Using these simulations, we are in general able to examine this problem at least in the case of idealised, symmetrical, smooth galaxy profiles.
As this is beyond the scope of this paper, we leave this analysis to a later paper and instead only focus on separating the 2 components of the 2-component galaxies themselves.

In an additional step, we transferred the remaining 80\% of objects into $B/D$  systems.
This is done following these rules for each parameter:

\begin{description}
\item[\textbf{Position:}] The $x$ and $y$ centre coordinates of the bulge and the disk are kept to be identical in the simulations, we simply kept the randomly chosen position from step 1.
While in real galaxies, a slight offset between the two components could in principle be seen, either because of an intrinsic offset, or due to dust in the galaxy affecting the different components in different ways, we decided to avoid this additional confusion.
This is hence directly in line with the setup used when fitting these images with \galfitm, where the position of bulge and disk are constrained to be identical.

\item[\textbf{Half-light radius:}] For the purposes of this test and analysis, we have decided to ignore gradients in colour and stellar populations within the galaxy components, mainly for simplicity, but also as in first order (only) this is a good approximation of reality, as has been discussed above.
This means that both $r_{\rm e}$ and \sersic index $n$ are to be kept \textbf{constant} with wavelength in the simulated galaxies in each component.
The sizes of the galaxy components were hence simply calculated from the overall half-light radius as derived in step 1 as the median of the values at the different wavelengths.

The distributions used for this purpose followed the distributions found by fitting $B/D$  profiles to the real galaxies (see \S~\ref{sec_real} for discussion of these fits).
We used the following distributions:
For disks, we used:
\begin{equation}
r_{\rm e,D} = median(r_{\rm e,SS})*(rand_{\rm D}*0.7+1.2)
\end{equation}
where $rand_{\rm D}$ is a random number (gaussian, FWHM of 1, centred on 0). 
This effectively creates most (68\%) of disks within $\rm 0.5*median(r_{\rm e,SS}) < r_{\rm e,D} < 1.9*median(r_{\rm e,SS})$, which is typically found in our real galaxies.

For bulges, we used
\begin{equation}
r_{\rm e,B} = median(r_{\rm e,SS})*(rand_{\rm B}*0.5+0.85)
\end{equation}
with $\rm rand_B$ being a different random number compared to the disk value (i.e. bulge and disk size are determined independently).
This effectively creates most (68\%) of bulges within $0.35*median(r_{\rm e,SS}) < r_{\rm e,D} < 1.35*median(r_{\rm e,SS})$.

Please note that we explicitly do \textbf{not} restrict component sizes to $r_{\rm e,B}<r_{\rm e,D}$ as we wanted to test the code for all cases, even though galaxies with $r_{\rm e,B}>r_{\rm e,D}$ are rarely found in nature.
A similar constraint is not used within \galapagostwo for the fitting process, either.

We do, however, impose a less strict constraint at $r_{\rm e,B}<2.5*r_{\rm e,D}$, in order to avoid extreme values.
We further impose our usual minimum and maximum sizes to $1[px]<r_{\rm e,B/D}<400[px]$.
If any of these limits are violated, we simply derived $r_{\rm e,B}$ and $r_{\rm e,D}$ again from $r_{\rm e,SS}$, using different randomisation.

\item[\textbf{Magnitudes/Colours:}] The magnitudes in the simulated objects were by far the trickiest of the parameters to decide on, as they could potentially have strong effects on both the results from the test and its implications on real galaxies.
\megamorph will perform best if the 2 components have very different SEDs, making a separation easier than if both components showed the same colours, i.e. identical SEDs.
We have created several different sets of simulations, with tens of thousands of simulated galaxies each, before we settled on a simulated dataset to be used in this paper.
Previous sets basically lead to the same conclusions, but with lower statistical significance and in a way that was more complicated to present in simple plots (e.g. due to the lack of bright objects, or each disk and bulge showing an individual colour).

For all simulations, we used the distribution of the total magnitudes $m_{\rm SS}$ from step 1 and used a random$B/T$ ratio (uniformly between 0 and 1) to divide the flux of each galaxy in the $z$-band (in the middle of the covered wavelength range) between the 2 galaxy components.
As this split was carried out in the $z$-band, we gave both components a boost of 0.2 mag to account for the typical magnitude difference between the $r$-band used in step 1 and the $z$-band used here.
As we wanted to concentrate this analysis on generally brighter galaxies, this worked in our favour.

After defining the $z$-band brightness this way, we needed to define a SED for disks and bulges, to derive their respective magnitudes in the other bands.
The result of this approach is that all disks and all bulges in the simulations actually \textbf{do} show the same SEDs, respectively.
This makes a comparison of parameters -- especially the magnitudes -- much cleaner and easier to present.
While leading to similar conclusions in our internal analysis, they show results at higher statistical significance than previous sets of simulations, which is why we present them here.

We settled on the approach to use a median colour of the bulges and disks (i.e. \lq typical\rq\ bulges and disks) from the \galapagostwo fits on our real galaxies (see \S \ref{sec_real}), respectively, restricting the sample to a tight subset of bright, large galaxies with high-$n$ bulges, where a $B/D$  decomposition is most likely to be successful.
In the end, we used for disks (colours compared to the $r$-band): 

\begin{itemize}
    \item $u$-$r$ = 2.195
    \item $g$-$r$ = 0.910
    \item $i$-$r$ = -0.429
    \item $z$-$r$ = -0.683
    \item $Y$-$r$ = -0.705
    \item $J$-$r$ = -0.655
    \item $H$-$r$ = -1.047
    \item $K$-$r$ = -0.871
\end{itemize}
    
and bulges
\begin{itemize}
    \item $u$-$r$ = 3.384
    \item $g$-$r$ = 1.442
    \item $i$-$r$ = -0.577
    \item $z$-$r$ = -0.917
    \item $Y$-$r$ = -1.123
    \item $J$-$r$ = -1.543
    \item $H$-$r$ = -1.820
    \item $K$-$r$ = -2.047.
\end{itemize}
These values are somewhat extreme in their colours, especially in the low signal-to-noise bands, e.g. $u$, but produced images which looked over all more similar to the real \gama images than previously tried datasets.
Finally, the magnitudes of each component are modified by adding Gaussian noise with a 0.1 standard deviation for each component, in order to simulate observational noise and re-introduce some variety into the component colours.

\item[\textbf{\sersic indices:}] 
Although for testing purposes this might have been a valid approach, we wanted to avoid the over-simplified case that all bulges have $n_{\rm B}==4$, making a successful decomposition more (and unrealistically) likely.
Instead, we wanted bulges to show a variety of $n_{\rm B}$ values as this was a more realistic approach and tested the codes for a wider variety of objects.
We wanted to test whether our codes can successfully decompose galaxies even if they contained low-$n$ bulges (i.e. pseudo-bulges), with a \sersic index similar to galaxy disks, as long as the two components showed different colours. 
For simplicity, however, we decided against randomly choosing a \sersic index from \lq real\rq\ bulges, and instead chose to use an artificial/simple distribution of \sersic indices.
\begin{equation}
n_{\rm B} = rand*2+4
\end{equation}
with $\rm rand$ again being a random number (gaussian, FWHM of 1, centred on 0). 
This results in a $n$-distribution centred around 4, with a FHWM of 2.
We restrict the $n_{\rm B}$ value as $0.5<n_{\rm B}<8$ as values outside of this range are rarely found in nature. 
We simply re-calculate the value if this constraint is violated.

Equivalently to the $r_{\rm e}$ values, by deciding to ignore gradients in colour and stellar populations within the galaxy components, \sersic indices have to be constant with wavelength, i.e. $n_{\rm B}$ is the same at all wavelengths.

For disks, we simply use $n_{\rm D}==1$ at all wavelengths, as it is well established that the \sersic index of disk components is very close to this value, while the bulges in real galaxies show a much wider spread.

\item[\textbf{Axis ratio:}]
For the axis ratio (q) of each component, we also use an artificial distribution for the 2 galaxy components.
We choose a random orientation $\Theta$ of the galaxies plane (randomly uniform in $cos(\Theta)$) and assume an intrinsic thickness of $\rm 0.18$ for disks and $\rm 0.5$ for bulges, which we use to calculate the intrinsic axis ratio of each component given $\Theta$.
These values are assumed to be constant with wavelength using
\begin{equation}
q_{\rm D} = \sqrt{0.18^2 + (1-0.18^2)*\cos(\Theta)^2}
\end{equation}
for the disk, and 
\begin{equation}
q_{\rm B} = \sqrt{0.5^2 + (1-0.5^2)*\cos(\Theta)^2}
\end{equation}
for the bulge. 

\item[\textbf{Position angle:}] 
The PA of the galaxy is chosen randomly and assumed to be the same for both bulges and disks.
While there are physical reasons why the PA could be (at least somewhat) different for the different galaxy components, we have chosen to ignore these special cases and instead stick to the much more common case that bulge and disk semi-major axes are aligned with each other.
Technically, this is the more challenging case for the fits as 2 elongated components with very different PAs should be easier to separate.
However, please note that no such constraint is being imposed during the fit.
\end{description}

\noindent The result of this procedure is a catalogue of objects that can be turned into \lq observed\rq\ images using the same scripts as used in H13.
These images contain a mixture of single-\sersic (20\%) and 2-component objects (80\%) which leaves us the possibility to test any 1- vs 2-component classifiers in future work.

Following the scripts used in H13, the resulting images used in this work include Poisson noise, and the galaxies are added into an image made up of empty sky patches from real \gama imaging.
This leads to more realistic noise properties in the image and allows a realistic estimation of the recovery of fitting parameters.

%%% A few words about $B/D$ starting parameters %%%
\section{Varying starting parameters}
\label{sec_start_params}
During the development of \galapagostwo, and specifically during the implementation of $B/D$ fitting, we have experimented with different starting parameters for the $B/D$ fits, especially in $r_{\rm e}$ and $n$, as the other parameters in our setup are usually free in each band (magnitudes) or held constant at all wavelengths, and usually easier to fit ($q$ and $PA$).
These choices should be presented in the simulations, and in the choice of starting parameters for the fit.

Throughout most of this work, we have used a \galapagostwo version that uses \begin{itemize}
    \item $r_{\rm e,D}=1.2*r_{\rm e,SS}$ (lower limit 1 pixel) and 
    \item $r_{\rm e,B}=0.3*r_{\rm e,SS}$ (lower limit 0.5 pixels) 
\end{itemize}
    for component sizes, and
\begin{itemize}
    \item $n_{\rm D}==1$ and
    \item $n_{\rm B}=median(n_{\rm SS})$ (lower limit 1.5)
\end{itemize}
for \sersic indices for disk and bulge respectively.

We have, however, tried different combinations of these and the following starting values:
\begin{itemize}
    \item $r_{\rm e,B}=1.0*r_{\rm e,SS}$
    \item $r_{\rm e,B}=0.5*r_{\rm e,SS}$
    \item $r_{\rm e,D}=1.0*r_{\rm e,SS}$
    \item $n_{\rm D} = 1$ (as starting value, not fixed value)
    \item $n_{\rm B} = 4$ (as starting value, not fixed value)
    \item $n_{\rm B} = 1$ (as starting value, not fixed value)
\end{itemize}

Over all, we found limited impact on the fitting results derived in multi-band fitting derived once extreme values were avoided, at least in bright objects.
In single-band fitting, however, there was indeed a difference between the individual test runs, particularly in the number of objects that return a \lq good\rq\ result at all, as separating bulges and disks becomes increasingly difficult in those data and convergence of $B/D$ fits becomes challenging.
Maximising this number became one of the goals of an ideal solution, although the impact on multi-band data was much smaller.
However, we saw no reason why multi-band fits should not also benefit from such a choice of starting parameters, even if it might only be for the number of iterations required in the fit, hence speeding up the analysis.

In this section, we show examples of the differences between 3 of these runs on the real \gama galaxies.
Throughout, we label the 3 different runs as \lq Fit1\rq, \lq Fit2\rq, and \lq Fit3 as follows:
\begin{itemize}
\item \lq Fit1\rq\ uses as starting parameters of the $B/D$ fit:
    \begin{itemize}
        \item $r_{\rm e,D}=r_{\rm e,SS}$ (lower limit 1 pixel)
        \item $r_{\rm e,B}=r_{\rm e,SS}$ (lower limit 0.5 pixels) 
    \end{itemize}
\item \lq Fit2\rq\ uses:
    \begin{itemize}
        \item $r_{\rm e,D}=r_{\rm e,SS}$ (lower limit 1 pixel)
        \item $r_{\rm e,B}=0.5*r_{\rm e,SS}$ (lower limit 0.5 pixels) 
    \end{itemize}
\item \lq Fit3\rq\ (used throughout most parts of this paper) uses:
    \begin{itemize}
        \item $r_{\rm e,D}=1.2*r_{\rm e,SS}$ (lower limit 1 pixel)
        \item $r_{\rm e,B}=0.3*r_{\rm e,SS}$ (lower limit 0.5 pixels) 
    \end{itemize}
\end{itemize}

\begin{figure}
\begin{center}
\includegraphics[width=0.48\textwidth, trim=0 0 0 10, clip]{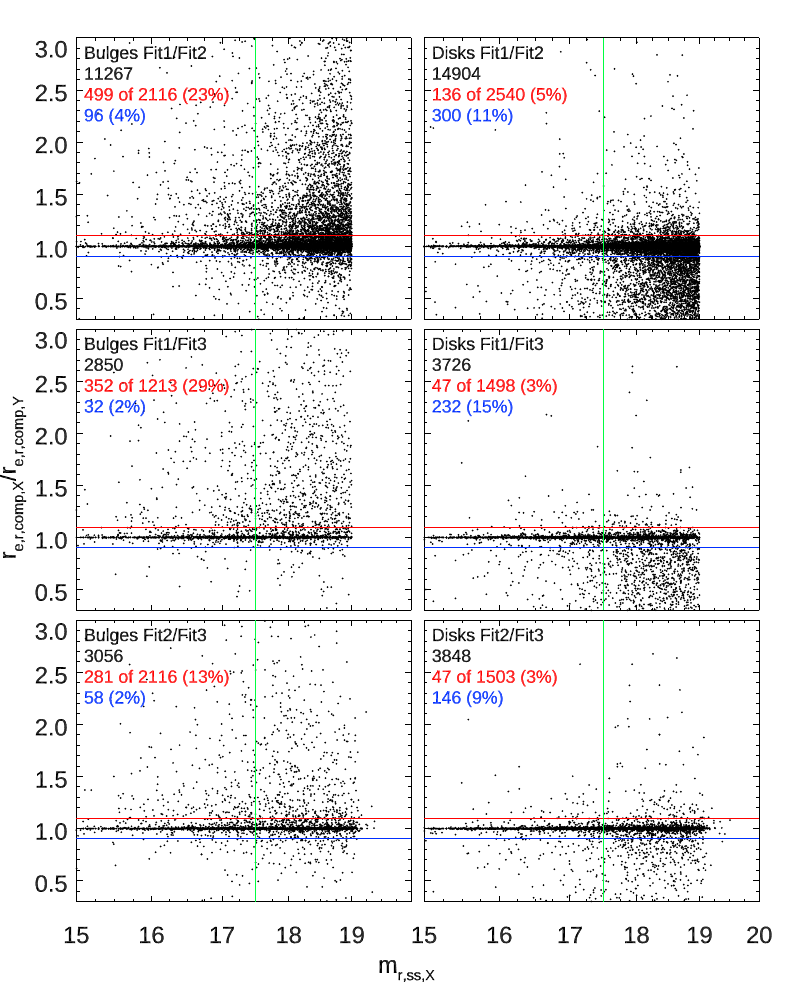}
\caption{Comparison of fitting parameters using different starting values. See text for details.
}
\label{fig_start_comparison}
\end{center}
\end{figure}

In Fig. \ref{fig_start_comparison} we present the comparison of these fits on the example of the bulge size in the $r$-band $r_{\rm e,r,B}$ on the left and the disk size $r_{\rm e,r,D}$ on the right side. 
The labels in each panel indicate which values are compared, and the black number shows the number of objects in each panel (different sample sizes largely reflect the fact that different samples of galaxies were fit in these tests). 
In each panel, we show the ratio of the derived parameters ($cat1/cat2$) as a function of the single-\sersic magnitude in $cat1$. 
The green vertical line indicates a galaxy brightness of $m_{\rm r,SS}=17.5$, red and blue horizontal lines indicate 10\% deviation from perfect agreement between the fits above and below, respectively. 
Finally, the numbers in red and blue indicate how many galaxies at $m_{\rm r,SS}<17.5$ are above and below these 10\% limits, respectively.

From these plots, we can see that Fit2 and Fit3 largely agree on the bulge and disk sizes, relatively few galaxies show sizes that disagree by more than 10\% (15\% of bulges, 12\% of disks) between the different runs (bottom panels).
However, when comparing either of these runs to the fits from \lq Fit1\rq (top and middle panels), it is obvious that the disagreement between the runs is pronounced and systematic. 
Bulges are systematically fit larger in \lq Fit1\rq, disks smaller.
We attribute this behaviour to the starting values for the sizes of bulge and disk being identical in this run, indeed making the separation of the components harder.
In this case, besides the obvious difficulty in separating the 2 components, it more likely happens that bulge and disk \lq flip\rq\ position, which explains the larger bulges and smaller disks.
We find similar and consistent behaviour in other fitting parameters like \sersic index (no plot shown).

However, once the bulge is started at a smaller size than the disk, the exact starting value seem to become less critical, and fitting values become more consistent, although some offsets can still be seen for faint galaxies.

This exercise can serve as a indication for the accuracy of the fitting values, i.e. to what brightness galaxies can be reliably decomposed, and as a warning to not over-interpret the finding of disk and bulge parameters found in faint galaxies.

We further tried using different constraints on the fitting values, including, and most importantly, the position of the object.
This was either allowed to vary over the entire image, or within $0.5*r_{\rm e,SS}$, but this choice had very minor impact on the fitting results.
A larger impact was found when we did not constrain the disk and bulge centres to be identical.
However, these cases usually turned out to be bad fits in which one of the components tried to fit the residual of a neighbouring object, so we found it important to implement the constraints to keep disk and bulge centres identical in \galapagostwo, which prevents the vast majority of these cases.
It also makes these cases easier to detect, as in this setup these objects require extreme axis ratios in order to minimise the residual of neighbouring objects, see Fig. \ref{fig_magbd-vs-mag}.

Additionally, one could individually test the impact of using different degrees of freedom for the polynomials used for each parameter.
As this is a user choice in \galfit and \galapagostwo and far beyond the scope of this work, in which a truly complete analysis of all parameters is impossible, we only present the results for the limited models that we have used in this paper.

%%% NEW FEATURES IN NEWER VERSIONS OF THE CODE %%%
\section{Additional features in the most recent version of Galapagos-2}
\label{App_gala_features}
As this paper serves as a code-release paper for \galapagostwo, we deem it important to summarise further improvements of the code. 
In case of further questions, or in case of requests for new/additional features, please do not hesitate to contact us.
As was previously mentioned, we invite users to adapt \galapagos themselves by cloning the GitHub repository\footnote{\href{https://github.com/MegaMorph/galapagos/}{https://github.com/MegaMorph/galapagos/}} and feeding any new features back into the main code via a pull request.
In this chapter, we present a brief list of the main improvements of the code since version 2.2.7, which was used throughout this work. 

\begin{itemize}
\item A potentially critical bug was fixed in which, depending on \textbf{why} galaxies were fit as neighbours, they were put on a fixed position which was off by 1 pixel in both x and y. This had minor impact on the data that we have seen, but was potentially critical in dense fields with many stars, as the artificial residuals created would impact the fitting parameters of the main target.
\item A new parameter \verb|primary-|$\chi^2$ was introduced, which measures the $\chi^2$ value in a area around the main target only, by masking out all other objects (no actual fit is done in the process). This parameter could potentially be used as a measure of the \lq goodness\rq\ of the fit and decide which of the models -- single-\sersic or $B/D$ -- better fits the data. However, we have not fully followed this approach through in this work. This selection of the better model is a largely unsolved problem in astronomy, for which several approaches have been attempted, including statistical methods \citep[e.g.][]{LacknerGunn,Simard2011}, visual morphologies \citep[e.g.][]{GZ1}, machine learning \citep[e.g.][]{HuertasDL}. None of these methods, however, has yet been found to work in a universal fashion, so it can be easily employed to any dataset. We have briefly looked at cross-validation to measure how good the fit (in some areas of the images) fits the data in other areas of the images. As this method is very CPU intensive, requiring several hundred fits to the same galaxy, it is unfeasible for large samples. However, it is a promising approach for small samples or individual -- well resolved -- galaxies and seems to work well on those.
Our approach here of presenting a \verb|primary-|$\chi^2$ value attempts to give a good starting point for statistical approaches, e.g. BIC and AIC methods, and has been employed by Nedkova et al. (in prep.) to analyse bulges and disks of galaxies in HFF \citep{HFF} and \candels data.
\item Several parts of the code have been significantly sped up.
\item Several bugs have been fixed to avoid crashes of the code in very specific circumstances. E.g. the code can now deal with surveys where the footprint of each image is different. Earlier versions of the code would crash if a postage stamp for an object in one band contains only masked pixels.
\item Several checks have been implemented at the start of the code, e.g. whether all data exists. Previously, it was possible due of a typo (e.g. in the name of the PSF used) to produce a fit without using a PSF, which was hard to notice afterwards. This is no longer possible, as \galapagostwo checks that all PSF files exist.
\item The output of the code into the command line has been changed, so it's easier to monitor the progress.
\item Several parameters for object counts now use \verb|LONG| instead of \verb|INTEGER| in order to deal with individual images with more than 32,767 detected objects.
\item A option \verb|\galfitoutput| was added, such that the \galfit terminal output for each object are saved, in case a user wants to examine them at a later stage.
\item A option \verb|\sex_skip| was added, to avoid re-running \sex on all the individual tiles, e.g. in case the code previously crashed during the stage of combining those catalogues.
\item The lower constraint for the axis ratio of objects was changed from 0.0001 to 0.01, to avoid very extreme cases which, due to sampling issues, can create a line of dots across the image, rather than a smooth profile. As both values are very extreme, no significant impact on the fits has been expected, other than avoiding these specific issues in very isolated cases.
\item While \galapagostwo includes a experimental feature to use super-computing facilities, the limitations of IDL (16 objects in parallel) do not allow this to be used efficiently, and to our experience actually slows the fits down while creating additional work for the user. While this feature has not technically been removed from the code, we stopped any further development of this feature.
\item Several new utility scripts have been created to make using \galapagostwo and its data products more efficient. These are somewhat experimental and we don't guarantee that they work perfectly, but they can be used for a variety of useful tasks. Please check the description \verb|*/src/utilities/Utility_README.md| for details.
\end{itemize}

%%% PROSPECTS/PLANS %%%
\subsection{Prospects, plans and further improvements}
\label{sec_prospect}

It should be stressed, that \galapagostwo, besides the obvious fits carried out, generally also works well as a setup for further analysis by readily creating all necessary data and \galfitm setup files.
Adding another component (point source, bar, etc) and enabling Fourier or Rotational Modes becomes a trivial undertaking for specific objects and science cases.
As implementing and testing such features is a major task and not easily done in such a general, fully automated code, we do not currently have plans to implement these features.
Instead, we encourage users to use the output files from \galapagostwo to create their own add-on routines for further analysis of specific objects.

However, several improvements of \galapagostwo are, of course, possible, some have already been mentioned in \S \ref{sec_setup_improvements}.
In this Appendix, we will discuss some of the more obvious improvements.
We do not currently have plans to implement all/any of these, but if users request them, we are happy to work together with them to improve the code in this direction and implement such new features.

\begin{itemize}
    \item One obvious improvement is possible by using different starting values for the magnitudes. The starting values for magnitudes in different bands are currently defined by using the \sex on the detection image, and applying a typical offset to the other bands. Such an approach is obviously not ideal and instead one could run \sex in dual image mode on all images, in order to derive better starting values for the single-\sersic fits in all bands. It is unclear how this idea could be used for the $B/D$ fits, but it's possible that tweaking the starting values for those fits (currently starting from the same SED for disk and bulge, that was derived in the single-\sersic fit) would improve the robustness and accuracy of the $B/D$ decomposition, i.e. by assuming redder SEDs for bulges and bluer SEDs for disks.
    \item Similarly, better starting values for other parameters are possible, e.g. by using some measure of light concentration to estimate the \sersic index of the single-component fit. Such an attempt would require extensive testing, however, as these parameters have been found to not correlate well in the past, so it is currently not planned.
    \item Depending on the fit result, one could imagine a second fitting step. For example one could re-start the fit of \lq failed\rq\ fit objects using different starting values, even multiple times \cite[e.g. as employed in][]{Lange16}. Alternatively, one could employ a scheme in which one uses the single-\sersic axis ratio to create azimuthally averaged one-dimensional profiles and fit those in order (especially small bulges might be more easily detected in such a scheme) to derive better starting values for another 2-D fit using \galfitm, possibly even with a reduced set of free parameters. Such a system, however, would require intensive development and testing.
    \item A new method for the determination of the sky values in data was presented and discussed by \citet{JiSky}. While we have no current plans to implement this new method as the current method seems to be working well, this is, of course, possible and might further improve the fits.
\end{itemize}

The ultimate improvement of \galapagostwo and any other, similar code would be to combine them with their individual strengths and different features into one code and skipping the development of several, largely duplicate codes altogether as a community.
A new, modular fitting code wrapper -- written in a more flexible (and free) coding language than IDL, e.g. Python, Julia, R, or others -- could be implemented in a modular fashion, in which a user could chose to use a multitude of source detection codes (e.g. \sex or \profound), different fitting codes (\galfit, \galfitm, \profit, \gimtwod, \imfit, or others), different setup versions, etc.
Such a code would be the ultimate tool for galaxy profile fitting, but can only be developed by the community as a whole.

%%% Catalogue Release %%%
\section{Catalogue release, fits of \kids/\viking data}
\label{sec_cat_release}
The work presented in this paper has been carried out with \sdss/\ukidss data, in order to be consistent with the test carried out in H13.
However, more recent \kids/\viking data is much deeper as discussed in \S \ref{sec_real_deep-shallow}, and has recently been fit using \galapagostwo.
With this paper, we also release two catalogues of \galapagostwo fits of these deeper data, as we think this would be highly beneficial for the larger community.
\footnote{The full table D.1 is only available in electronic form at the CDS via anonymous ftp to \href{http://cdsarc.u-strasbg.fr/}{cdsarc.u-strasbg.fr} (130.79.128.5)
or via \href{http://cdsweb.u-strasbg.fr/cgi-bin/qcat?J/A+A/}{http://cdsweb.u-strasbg.fr/cgi-bin/qcat?J/A+A/}}
In this section, we briefly describe the catalogues released. 

\begin{itemize}
    \item The catalogues released contain both single-\sersic and two-component Bulge/Disk model fits to the 2D surface brightness distributions of 234,239 objects in re-\swarp-ed \kids/\viking data -- as provided by \gama as large mosaics -- using \galapagos version 2.3.2. and \galfitm version 1.2.1. This represents the full sample of objects presented in \cite{GAMADR3}, with a few exceptions described below.
    \item Re-\swarp-ed imaging data from the \gama G09, G12, and G15 survey regions are being used. These data include four-band optical ($ugri$) imaging from \kids \citep{KiDS1,KiDS2} plus five-band near-infrared (ZYJHK) imaging from the \viking survey. 
    \item A number of bright stars and nearby areas are masked out of the data by visually identifying them and setting the values in the weight images in an area around those stars to a value of 0, so both \sex and \galfitm ignore those areas during the object detection and fit, respectively. Similarly, areas where the data were saturated have been masked out generously as to not influence the fits of nearby objects. This removes some objects from the sample that would be difficult to be fit due to their proximity to very bright stars. Other than these adaptations, we only cut the large mosaics provided by \gama into images of \textasciitilde 8500x8500 pixels with overlap (termed \lq tile\rq), in order to make the file sizes more manageable for \galapagostwo, which further cuts the images into postage stamps centred on each object.
    \item We use \sex on the $r$-band image to detect sources. This image is deemed the deepest data available amongst the available data. While objects with extremely red colour could potentially be missed, it is sufficient to easily detect any object with \gama spectroscopy. As was found to be ideal for \galapagostwo, this is done in a  2-stage (hot + cold) process process. \lq Cold\rq\ detects the brightest objects without splitting them up, \lq hot\rq\ detects the faint objects. Both catalogues are then combined as outlined in the \galapagos manual \citep{galapagos}.
    \item The PSFs used for these fits are obtained by using \psfex to create a large number of PSFs for each input tile, choosing the closest PSF in RA and DEC to the targeted object for each fit. These PSFs were constructed on a grid of 10x10 ($u$-band) and 20x20 ($grizYJHK$-bands) PSFs per tile of \textasciitilde8500x8500 pixels in size as outlined in \S \ref{sec_real_discussion}. This means that PSFs are available on a regular grid with a density of one PSF every \textasciitilde400 or \textasciitilde800 ($u$-band) pixels. Detailed setup values can be provided upon request.
    \item We fit single-\sersic profiles to all objects that have a redshift in the \gama spectra catalogue \lq SpecAll27\rq\ as provided by the \gama survey, at a search radius of 5\arcsec (any object within this radius is fit). Additionally, we fit all objects within 100 arcseconds of any of these objects, but brighter than said object. This is ideal for \galapagostwo as objects are dealt with in brightness order. By leaving these objects out, the sample selection could potentially change the fitting results. In the fitting process itself, fainter neighbours are taken into account as normal, but these will never become primary objects. Successive two-component fits are only carried out on the \gama objects themselves, as is sufficient, see discussion above.
    \item The degrees of freedom used for the individual parameters are as discussed in \S \ref{sec_setup_choices} and in H13:
    \begin{itemize}
        \item single-\sersic fits use free magnitudes, 2nd order polynomials (3 free parameters) for halflight-radii and \sersic index. All other parameters are constant with wavelength.
        \item Two-component fits use full freedom for magnitudes, and values constant with wavelength for all other parameters (where $n_{\rm D}==1$ for disks).
        \item The second provided catalogue contains identical single-\sersic fits, but additionally uses $n_{\rm B}==4$ for bulge \sersic indices. This catalogue is released, but has not been tested in detail in this paper. Results from Nedkova et al. (in prep.) to  HFF and \candels data suggest that these bulge parameters are actually more reliable. 
    \end{itemize}
    \item We make no attempt to quantify the quality of an individual fit, nor do we indicate whether the single or double component fit is to be preferred for a given galaxy. When using the data from these catalogue, specifically the cleaning of the sample, please follow the guidelines discussed in \S \ref{sec_sample} and in \citet{MMVulcani} and \citet{Kennedy16a}.
\end{itemize}

\begin{figure}
\includegraphics[width=0.48\textwidth,trim=0 0 8 10,clip]{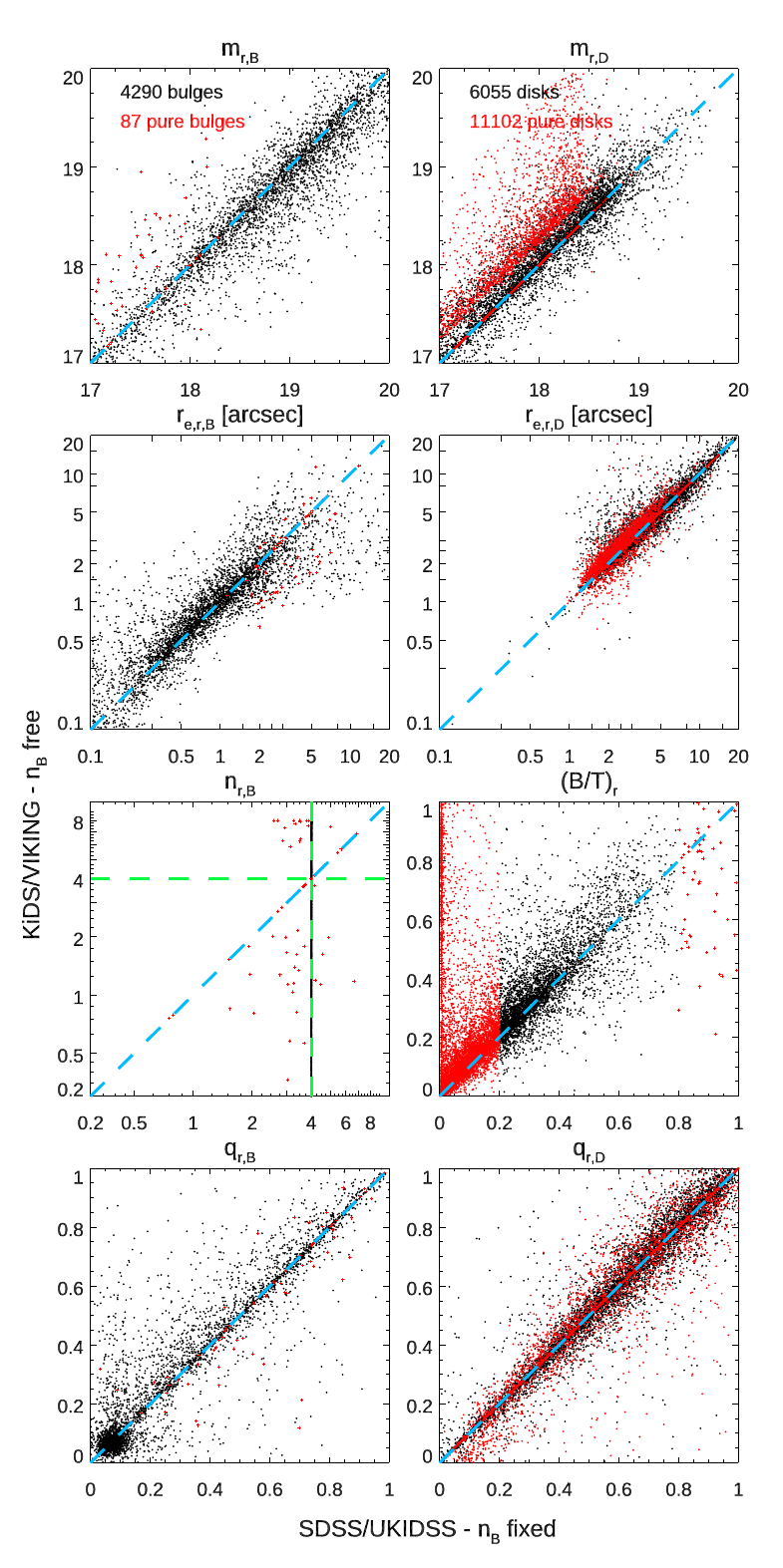}
\caption{Comparison of the 2 $B/D$ catalogues with $n_{\rm B}==4$ (x-axis) and $n_{\rm B}$ free (y-axis) for objects with \sex MAG\_BEST$< 18.5$.
}
\label{fig_comparison_n4}
\end{figure}

In Fig. \ref{fig_comparison_n4}, we show a comparison of the main fit parameters for the 2 catalogues for a bright sample (\sex MAG\_BEST$< 18.5$) and $\verb|CLASS_STAR|<0.8$ .
The plot looks very similar for fainter objects, but there are too many objects, so that features become difficult to see.
As the same single-\sersic fits are used for both fits, we do not show their comparison as we did in previous figures, as they would exactly show a 1-to-1 line.
In general, the agreement of the $B/D$ parameters is good in all parameters, even the $B/T$ ratio shows a very clear and relatively tight correlation.
There is, however, a concentration of objects with $B/T\sim 0$ in the fits with fixed $n_B$ (on the left side of the $B/T$ panel).
While we did not investigate those objects further, we can speculate from the findings above, that these are at least in part objects in which the bulge with free \sersic index fits the disk profile, and the fit with $n_B=4$ correctly finds only a very weak bulge.
In the panel comparing the $n_{\rm B}$ values (although for one of the fits this was fixed at $n_B=4$), it is interesting to see that most of the \lq pure bulges\rq\ plotted in red indeed show \sersic indices in the range of $2.5<n_{\rm SS}<5$ as one would expect for such \lq bulge-dominated\rq\ galaxies.
The user should be reminded that this selection, as defined in \S \ref{sec_comparison}, is done on the fitted $B/T$ ratio and no \sersic index selection is used.

There is also a prominent bulk of objects at low $q_{\rm B}$ values in both fits.
This has already been discussed in \S \ref{sec_real_deep-shallow} and can be further reduced by a stricter cut on \verb|CLASS_STAR|, which also somewhat reduces the scatter in the other panels, indicating that most outliers are indeed very small objects.

In Table \ref{tab_catalogue}, we present a short summary and explanation of the parameters in the released catalogues. 
As a multitude of parameters are very similar, and a full description would be very long, we only give examples.
However, following the nomenclature, other parameters should easily be understood.
Any user of either of these catalogues should be careful to follow the advice on deriving clean object samples given in \S \ref{sec_sample}.
In summary, the catalogues contain
\begin{itemize}
    \item The GAMA \verb|CATAID|
    \item several useful \sex parameters
    \item several additional \galapagostwo parameters
    \item general parameters (single numbers) and \galfitm parameters (9-element arrays). 
    \item single-\sersic parameters indicates as \verb|_galfit|, BD parameters as \verb|_BD|, bulge parameters \verb|_B| and disk parameters \verb|_D|.
    \item parameters (e.g. \verb|MAG_GALFIT_BAND|) and errors as derived by \galfitm (e.g. \verb|MAGERR_GALFIT_BAND|)
    \item Two sets of fit parameters. One containing the fit values at the effective wavelengths of the filters used \verb|_BAND|\footnote{3543\AA, 4770\AA, 6231\AA, 7625\AA, 9134\AA, 10305\AA, 12483\AA, 16313\AA, 22010\AA\ for $ugrizYJHK$-band, respectively.} and the other containing the fit parameters for the Chebyshev polynomials \verb|_CHEB| themselves, where they make sense (i.e. not if values are constant with wavelength). The latter can be used to easily interpolate between values to other wavelengths, but a user should be aware of \S \ref{sec_setup_choices} where interpolation of polynomials with high degrees of freedom are being discussed. Where \verb|_BAND| values are given, they are in the order of $r$,$u$,$g$,$i$,$z$,$Y$,$J$,$H$,$K$-band, due to the $r$-band serving as the main band in the \galapagostwo run\footnote{A python script for interpolating the Chebyshev polynomials, kindly provided by Kalina Nedkova, is available at \href{https://github.com/kalinanedkova/mass\_size}{https://github.com/kalinanedkova/mass\_size}. 
    A similar IDL script can be provided on request.}.
    \item Sizes/half-light radii are given in arcseconds.
\end{itemize}

The catalogues provide fit values even for those objects where the fit crashed. 
These objects can easily be identified by \verb|FLAG_GALFIT| and \verb|FLAG_GALFIT_BD| for single-\sersic and $B/D$ fits, respectively, see details in table.
Their parameter values and errors are set to obvious standard values, i.e. 0 for position, \verb|-999.| for magnitudes, and \verb|-99.| for sizes and \sersic indices, generally \verb|99999.| for error values.
It should be noted that there can be objects with error values of \verb|99999.| despite having fit values provided.
For these few objects, the error estimation within \galfitm has failed to provide a value, while the fit itself did converge on a final solution.
Visually, we have not been able to identify anything special about such objects.

\begin{table*}
\centering
\caption{Summary of the released tables, available on \href{http://cdsweb.u-strasbg.fr/cgi-bin/qcat?J/A+A/}{http://cdsweb.u-strasbg.fr/cgi-bin/qcat?J/A+A/}.}
\scriptsize
\begin{tabular}{@{}lccl@{}}
\hline
Parameter & Format & Example (First Object) & Explanation \\
\hline
\hline
\verb|CATAID| & LONG & 3396309 & \gama \verb|CATA_ID| as given in catalogue \lq SpecAll27\rq\\
\verb|GALA_ID| &  STRING  &  \lq G09\_t14\_8.4640 \rq & unique identifier with \galapagostwo, in the format \verb|FIELD_TILE.SEX_NUMBER|\\ 
\verb|RA|       &       DOUBLE      &     134.7202 & RA as provided by \gama\\
\verb|DEC|      &       DOUBLE      &     1.9992 & DEC as provided by \gama\\
\verb|RA_GALA|       &       DOUBLE      &     134.7202 & \sex \verb|ALPHA_J2000|\\
\verb|DEC_GALA|      &       DOUBLE      &     1.9992566 & \sex \verb|DELTA_J2000|\\
\verb|THETA_IMAGE| & DOUBLE &  2.630 & \sex \verb|THETA_IMAGE|. All images are created so north is up.\\
\verb|ELLIPTICITY| &  DOUBLE    &  0.0680 & \sex \verb|ELLIPTICITY|\\
\verb|FLUX_RADIUS| & DOUBLE &   0.7011 & \sex \verb|FLUX_RADIUS| [arcsec]\\
\verb|KRON_RADIUS| & DOUBLE &   1.1865 & \sex \verb|KRON_RADIUS| [arcsec]\\
\verb|BACKGROUND| &      DOUBLE   &    -0.0064 & \sex \verb|BACKGROUND|\\
\verb|FLUX_BEST| &  DOUBLE & 363618.50 & \sex \verb|FLUX_BEST|\\
\verb|FLUXERR_BEST| &  DOUBLE & 432.8541 & \sex \verb|FLUXERR_BEST|\\
\verb|MAG_BEST| & DOUBLE  &  16.0984 & \sex \verb|MAG_BEST|\\
\verb|MAGERR_BEST| &     DOUBLE &  0.0013 & \sex \verb|MAGERR_BEST| \\
\verb|FWHM_IMAGE| &     DOUBLE  & 1.010 & \sex \verb|FWHM_IMAGE| [arcsec]\\
\verb|FLAGS|    &       INT & 0 & \sex \verb|FLAGS|\\
\verb|CLASS_STAR|  &    DOUBLE      &    0.9780 & \sex \verb|CLASS_STAR|\\
\verb|FLAG_GALFIT|   &  INT & 2 & Flag on single-\sersic fits. 0=not attempted, 1=fits started, but crashed, 2=fit finished \\
& & & In case of crashed fits, \galfitm parameters contain standard values.\\
\verb|NEIGH_GALFIT|  &  INT & 0 & Number of neighbouring profiles/objects fit (or fixed) during the single-\sersic fit\\
\verb|CHISQ_GALFIT|  &  FLOAT & 736530. & \galfitm $\chi^2$ value, single-\sersic fit\\
\verb|CHISQ_GALFIT_PRIME| & FLOAT &  68963.2 & \galfitm $\chi^2$ value within primary ellipse only, see Appendix \ref{App_gala_features}, single-\sersic fit\\
\verb|NDOF_GALFIT|  &   LONG    & 29888 & degree of freedom during the fit, includes \#pixels, used to derive reduced $\chi^2$:\\
& & &  $\chi^2/\nu$, single-\sersic fit\\
\verb|NDOF_GALFIT_PRIME| &  LONG  & 3154 & same for \lq fit\rq\ within primary ellipse, single-\sersic fit\\
\verb|CHI2NU_GALFIT| &  FLOAT  & 24.6430 & \galfitm reduced $\chi^2$: $\chi^2/\nu$, single-\sersic fit\\
\verb|CHI2NU_GALFIT_PRIME| & FLOAT & 21.8653 & \galfitm reduced $\chi^2$: $\chi^2/\nu$, within primary ellipse only, single-\sersic fit\\
\verb|SKY_GALA_BAND_[X]| &  FLOAT(9)  &  (0.621,[...],-1.12100) & \galapagostwo sky values at each band.\\
\verb|SKY_SIG_BAND_[X]|  &  FLOAT(9)    & (0.0350,[...],0.205246) & \galapagostwo uncertainty of the sky estimation in each band. While the sky value is fixed \\
 & & & during the fit, this loses it's meaning, but can be used to detect \lq difficult\rq\ fits.\\
\verb|SKY_FLAG_BAND_[X]| &  INT(9)    & (0,[...],0) & \galapagostwo sky flag as described in \citet{galapagos}\\
\verb|X_GALFIT_DEG| &   INT &  1 & Degree of Freedom (DOF) of x-position of the single-\sersic fit\\
& & & (0=fixed, 1=constant with wavelengths, [...], 9=full freedom)\\
\verb|X_GALFIT_BAND| &  FLOAT & 30.83 & \galfitm x-position of the single-\sersic fit (relative to its postage stamp). \\
& & & Provided for completeness as a uncertain position might give useful insight.\\
\verb|XERR_GALFIT_BAND_[X]| &  FLOAT & 0.9421 & \galfitm x-position uncertainty of the single-\sersic fit\\
\verb|[...]| & & & \\
\verb|MAG_GALFIT_DEG| & INT & 9 & DOF of magnitudes of the single-\sersic fit. Full freedom has been used. \\
\verb|MAG_GALFIT_BAND_[X]| & FLOAT(9) & (17.9330,[...],15.9602) & \galfitm magnitudes in each band, single-\sersic fit\\
\verb|MAGERR_GALFIT_BAND_[X]| & FLOAT(9) & (228.705,[...],148.161)& \galfitm magnitude uncertainties in each band, single-\sersic fit\\
\verb|MAG_GALFIT_CHEB_[C]|& FLOAT(9) &  (5.7690,[...],0.9317) & \galfitm magnitudes, Chebyshev polynomial parameters, single-\sersic fit\\
\verb|MAGERR_GALFIT_CHEB_[C]| & FLOAT(9) & (38.3268,[...],9.28216) & \galfitm magnitude uncertainty on Chebyshev polynomial parameters, single-\sersic fit\\
\verb|RE_GALFIT_DEG|  & INT & 3 & DOF of $r_{\rm e,SS}$ of the single-\sersic fit. Second order polynomials have been used.\\
\verb|RE_GALFIT_BAND_[X]| & FLOAT(9) & (0.1101,[...],0.1402) & \galfitm halflight-radii $r_{\rm e,SS}$ at each wavelength [in arcsec].\\
\verb|RE_GALFIT_CHEB_[C]| & FLOAT(3) & (0.125,0.015,-0.0001) & \galfitm halflight-radii $r_{\rm e,SS}$ Chebyshev parameters [result in arcsec].\\
\verb|[...]| & & & \\
\verb|MAG_GALFIT_BAND_B_[X]| & FLOAT(9) & (18.5300,[...],17.2254) & \galfitm Bulge magnitudes in each band\\
\verb|[...]| & & & \\
\verb|N_GALFIT_DEG_B| &  INT & 1 & DOF of Bulge \sersic index $n_{\rm B}$. \\
& & & Either =1 (free values) or =0 (fixed values ==4), depending on the catalogue\\
\verb|N_GALFIT_BAND_B| & FLOAT & 4.0 &  \galfitm Bulge \sersic index (identical in each band).\\
& & & ==4 for catalogue with fixed $n_{\rm B}$ values, as in the example.\\
\verb|[...]| & & & \\
\verb|MAG_GALFIT_BAND_D_[X]| &  FLOAT(9) &  (18.9752,[...],16.4439) &  \galfitm Disk magnitudes in each band\\
\verb|[...]| & & & \\
\hline
\label{tab_catalogue}
\end{tabular}
\tablefoot{This table summarises the released tables. Format with \lq (x)\rq\ indicate that these are several parameters, indicating the values either for the individual bands [X], or the individual Chebyshev parameters [C]. Some value examples are shortened to restrict the width of the table. A total of 263 columns are released. Some rows are omitted to shorten the table, but trivially follow the same naming convention. All sizes are given in arcseconds, with fit values being constrained from $\sim0.102 \arcsec$ (0.3 pixels) to $\sim135.6 \arcsec$ (400 pixels). Fit position angles are defined as follows: \lq 0\rq\ is \lq up\rq (north in the images used), counting anticlockwise (note a 90 degrees offset to the definition used by \sex).}
\end{table*}

\end{appendix}

\end{document}